\documentclass{article}

\usepackage{amssymb}
\usepackage{amsmath}

\usepackage{graphicx}

\newtheorem{theorem}{Theorem}[section]
\newtheorem{prop}[theorem]{Proposition}
\newtheorem{lem}[theorem]{Lemma}
\newtheorem{cor}[theorem]{Corollary}

\newtheorem{conj}[theorem]{Conjecture}
\newtheorem{remark}[theorem]{Remark}

\begin{document}
\title{Birkhoff strata of Sato Grassmannian and algebraic curves}
\date{\today}
\author{B.G. Konopelchenko \\
Dipartimento di Matematica e Fisica " Ennio de Giorgi", Universit\`{a} del
Salento \\
INFN, Sezione di Lecce, 73100 Lecce, Italy \\
{konopel@le.infn.it} \\
\mbox{} \\
G. Ortenzi \\
Dipartimento di Matematica Pura ed Applicazioni,\\
Universit\`{a} di Milano Bicocca, 20125 Milano, Italy\\
{giovanni.ortenzi@unimib.it} }
\maketitle

\begin{abstract}
\noindent Algebraic and geometric structures associated with
Birkhoff strata of Sato Grassmannian are analyzed. It is shown
that each Birkhoff stratum $\Sigma_S$ contains a subset
$W_{\widehat{S} }$ of points for which each fiber of the corresponding tautological subbundle 
$TB_{W_S}$ is closed with respect to multiplication. Algebraically $TB_{W_S}$  is an infinite family of infinite-dimensional
commutative associative algebras and geometrically it is an
infinite tower of families of algebraic curves. For the big cell
the subbundle $TB_{W_\varnothing}$ represents the tower of families of
normal rational (Veronese) curves of all degrees. For $W_1$ such tautological subbundle  is
the family of coordinate rings for elliptic curves. For higher
strata, the subbundles $TB_{W_{1,2,\dots,n}}$ represent families of
plane $(n+1,n+2)$ curves (trigonal curves at $n=2$) and space
curves of genus $n$. Two methods of regularization of singular curves contained in $TB_{W_{\widehat{S}}}$,
namely, the standard blowing-up and transition to higher strata
with the change of genus are discussed. 
\end{abstract}

%
%
 \tableofcontents

\section{Introduction}

Algebraic curves in finite-dimensional Grassmannians is a classical subject
in algebraic geometry and other branches of mathematics (see e.g. \cite{HP}-%
\cite{Sot}). In contrast, the story of interplay between
infinite-dimensional Grassmannians and algebraic curves was quite
non-classical.

Interest of recent years to infinite-dimensional Grassmannians is mainly due
to the Sato's papers \cite{Sat,SS}. He demonstrated that the
Kadomtsev-Petviashvilii (KP) hierarchy and hierarchies of other nonlinear
partial differential equations integrable by the inverse scattering
transform method discovered in \cite{GGKM} (see e.g. \cite{ZMNP,AS}), have a
beautiful geometrical interpretation in terms of special
infinite-dimensional Grassmannian (Sato Grassmannian). Since the papers \cite%
{Sat,SS} Sato Grassmannian became a powerful tool in many branches of
mathematics and theoretical and mathematical physics from algebraic geometry
to quantum field theory, string theory, and theory of integrable equations
(see e.g. \cite{DJKM}-\cite{KO2}).

Significance of algebraic curves in the construction of solutions of
integrable equations has been understood in the middle of seventies (see
e.g. \cite{DMN,BBEIM}). In particular Krichever \cite{Kri1,Kri2}
demonstrated that for any complex algebraic curve (with some additional
data) one can construct a (Baker-Akhiezer) function $\psi$ which obeys a
compatible pair of linear differential equations. Compatibility of these
equations is equivalent to a nonlinear integrable equation, for instance, to
the KP equation. After the Sato's results \cite{Sat,SS} on the
identification of solutions of integrable equations with subspaces $W$ in
Grassmannian it became clear that the correspondence discovered in \cite%
{Kri1,Kri2} can be extended to a map between algebraic curves and subspaces
in Sato Grassmannian \cite{SW}. This paper of Segal and Wilson and the paper 
\cite{AdCKP} coined the name of Krichever map (correspondence) for such a
map.

Since then the Krichever map, its inversion and extensions have been studied
within different contexts in a number of papers (see e.g. \cite{Mul2,Mul1},%
\cite{Mul3}-\cite{PreSpe}). In particular, in the papers \cite{Mul2,Mul1,Mul3} 
it was shown that, the so-called Schur pair plays a
central role in the construction and analysis of Krichever map. With all the
diversity of the results obtained, the constructions associated with the
Krichever map share a common feature, namely, an algebraic curve, though
closely connected, is, essentially, an object exterior to Sato Grassmannian except, probably
the rather formal papers \cite{MP,HMP}.
It seems that there are very few results concerning the study of algebraic
curves in Sato Grassmannian itself. We note the study of rational curves in
Gr$_{1}$ and in Schur cells of Gr$^{(2)}$  (section $7$ of \cite{SW}) and brief
analysis of hyperelliptic curves in Birkhoff strata of Gr$^{(2)}$ \cite{KK},%
\cite{KO}.

In the present paper we will follow a classical way adopted in \cite{SW,KK}
and we will look for algebraic curves inside the tautological bundle of Sato Grassmannian itself. We
shall use elementary methods only and shall maximally avoid the involvement
of any additional structures. Our main result is that each Birkhoff stratum $%
\Sigma _{S}$ of the Sato Grassmannian Gr contains a subset $W_{S}$ \ of
points such that for each of these points the corresponding
infinite-dimensional linear space ( fiber of the tautological subbundle $%
TB_{W_{S}}$ associated with \ $W_{S}$ ) is closed with respect to pointwise
multiplication. Algebraically all \ $TB_{W_{S}}$ are infinite families of
infinite dimensional associative commutative algebras. Geometrically each
fiber of  $TB_{W_{S}}$ is an algebraic variety and the whole  $TB_{W_{S}}$
is an algebraic $ind$-variety with each finite-dimensional subvariety being
a family of algebraic curves. For the big cell $\Sigma _{\varnothing }$ the $%
TB_{W_{\varnothing }}$ is the collection of families of normal rational
curves (Veronese curves) of the all degrees $2,3,4,\dots $. For the stratum $%
\Sigma _{1}$, each fiber of $TB_{W_{1}}$ is the coordinate ring of the
elliptic curve and $\ $ $TB_{W_{1}}$ is the infinite family of such rings
and \ the index $(\overline{\partial }_{W_{1}})=-1$. \ For the set $W_{1,2}$
the $TB_{W_{1,2}}$ is equivalent to the families of coordinate rings of a
special space curve with pretty interesting properties. This family of
curves in $\ TB_{W_{1,2}}$ contains plane trigonal curve of genus two and \
index$(\overline{\partial }_{W_{1,2}})=-2.$We conjecture that the $%
TB_{W_{1,2,...,n}}$  in higher strata $\Sigma _{1,2,\dots ,n}$ $%
(n=3,4,5,\dots )$ have similar properties. In particular,  $%
TB_{W_{1,2,...,n}}$ contains plane $(n+1,n+2)$ curve of genus $n$ \ and \
index$(\overline{\partial }_{W_{1,2,\dots ,n}})=-n$.

It is shown, that the projections of basic algebraic curves in each stratum
to lower dimensional subspaces are given by singular higher degree curves.
Two ways of their regularization are discussed. The first is the standard
blow-up by quadratic transformation within the same stratum without change
of genus. The second way consists in transition to the higher stratum. In
such a regularization procedure genus of a curve increases.

It should be noted that the main scope of the paper is a search of concrete 
algebraic curves and corresponding families in the specific Birkhoff strata 
of Sato Grassmannian in contrast to the general and formal description addressed in 
the papers \cite{Mul1,MP,HMP,Mul2,Mul3}.

The paper is organized as follows. In section 2 the basic facts on the
Birkhoff strata in Sato Grassmannian are reported. Associative algebra and
family of normal rational curves appearing in the tautological subbundle of the big cell are considered in
section 3. Family of the centered normal rational curves \ in the stratum $%
\Sigma _{0}$ are discussed in section 4. \ The stratum $\Sigma _{1}$ and the
corresponding  $TB_{W_{1}}$ which contains family of elliptic curves are
analyzed in section 5. \ The Weierstrass function reduction of these curves
are studied in the next section. Trigonal curves appearing in the stratum $%
\Sigma _{1,2}$ are studied in section 7. Some observations about algebraic
curves in higher strata are presented in section 8. Different ways of
resolution of singularities and transition between strata are discussed in
section 10. Appendices contain some explicit expressions for coefficients of
algebraic curves.

\section{Birkhoff strata and I\lowercase{ndex}$\mathbf{(\overline{\partial }%
_{W_{\protect\widehat{S}}})}$}

Here we recall briefly basic facts about Sato Grassmannian and its
stratifications (see e.g. \cite{SW,PS}).

Let $H=\mathbb{C}((z))$ be the set of all formal Laurent series with
coefficients in $\mathbb{C}$ and $H_{+}=\mathbb{C}[z]$ is the set of all
formal polynomials in $z$. Sato Grassmannian Gr is by definition the
parameter space for the totality of closed vector subspaces $W\subset H$
such that the projection $W\rightarrow H_{+}$ is Fredholm. Each $W\subset Gr$
possesses an algebraic basis ($w_{0}(z),w_{1}(z),\dots $) with the basis
elements 
\begin{equation}
w_{n}=\sum_{i=-\infty }^{n}a_{i}z^{i}  \label{basiselem}
\end{equation}%
of finite degree $n$. A point of Sato Grassmannian
represents an infinite-dimensional linear space, formed by these series with
\ various $\gamma _{k}$ and fixed $a_{i}$, which is naturally attached  to
this point. So, the linear bundle which is  the disjoint union of all
these linear spaces (fibers) is a particularly natural one. As in the
well-known case of finite-dimensional Grassmannians (see e.g. \cite{GH}) 
such bundle is referred as the tautological bundle (TB) over Sato
Grassmannian. For any subset W of  Sato Grassmannian one has the
corresponding tautological subbundle $TB_{W}$. 

Grassmannian Gr is a connected Banach manifold which exhibits a stratified
structure \cite{PS}. To describe this structure one introduces the set $%
\mathcal{I}$. It is the family of all sets $S\subset \mathbb{Z}$ which are
bounded from below and contain all sufficiently high integers. The canonical
form of such $S$ of virtual cardinality zero is 
\begin{equation}
S=\{s_{0},s_{1},s_{2}\dots \}
\end{equation}%
such that $s_{0}<s_{1}<s_{2}<\dots $ and $s_{n}=n$ for large $n$. Then for
the subspace $W\subset Gr$ one defines 
\begin{equation}
S_{W}=\{s\in \mathcal{I}:\ W\mathrm{contains\ elements\ of\ degree\ s}\}.
\label{setS}
\end{equation}%
Given $S\in \mathcal{I}$ the subset $\Sigma _{S}$ of Gr defined by 
\begin{equation}
\Sigma _{S}=\{W\in Gr:\ S_{W}=S\}
\end{equation}%
is called the Birkhoff stratum associated with the set $S$. The closure of $%
\Sigma _{S}$ (Birkhoff variety) is an infinite-dimensional irreducible $ind$%
-variety of the finite codimension $l(s)=\sum_{k\geq 0}(k-s_{k})$. In
particular, if $S=\{0,1,2,\dots \}$ the corresponding stratum has
codimension zero and it is a dense open subset of Gr which is called the
principal stratum or big cell. Lower Birkhoff strata correspond to the sets $%
S$ of type (\ref{setS}) different from $\{0,1,2,\dots \}$. For instance, the
set $S=\{-1,0,2,3,4,\dots \}$ corresponds to stratum $\Sigma _{1}$, while
the set $\{-2,-1,0,3,4,\dots \}$ is associated with $\Sigma _{1,2}$. Here
and below, for convenience, we will use also the notation $\Sigma _{\widehat{%
S}}$ fir the Birkhoff strata where $\widehat{S}=\{\mathbb{N}-S\}$ denotes a
set of holes in the positive part of $S$ with respect to $\mathbb{N}$. Note
that Grassmannian Gr has also the Schubert or Bruhat decomposition which is
dual to Birkhoff stratification. Schubert cells $C_{S}$ are subsets of the
elements of the form $\sum_{k=-N}^{N}b_{k}z^{k}$ numerated by the same sets $%
S$ as Birkhoff strata and have finite dimensions $l(s)$. Schubert cell $C_{S}
$ and Birkhoff stratum $\Sigma _{S}$ intersect transversally in a single
point. Birkhoff stratification of Sato Grassmannian induces the
stratification of the tautological bundle TB into subbubdles $TB_{\Sigma
_{S}}$ . 

Schubert varieties in finite and infinite dimensional Grassmannians have
been studied pretty well while it seems that the Birkhoff varieties have
attracted considerable interest mainly within the theory of integrable
systems (with few exceptions (see e.g. \cite{GM})). It was shown in \cite%
{SS,AvM} that the flows generated by the standard KP hierarchy belong to the
big cell. On the other hand, singular solutions of the KP hierarchy for
which the $\tau $-function and its derivatives vanish, are associated with
higher strata. A method of desingularization of wave functions near blowup
locus (Birkhoff strata) has been proposed in \cite{AvM}. In the papers \cite%
{MMM,KMAM0,KMAM} it was demonstrated that there are infinite hierarchies of
integrable equations associated with each Birkhoff strata.

In addition to algebraic and geometrical aspects the Birkhoff stratification
exhibits also an interesting analytic structure. It was observed in \cite{SW}
(section 7.3) that the Laurent series (\ref{basiselem}) are the boundary
values of certain functions $\Omega =\mathbb{C}-\mathcal{D}_{\infty }$ where 
$\mathcal{D}_{\infty }$ is a small disk around the point $z=\infty $.
Formalizing these observations Witten \cite{Wit} suggested to view Sato
Grassmannian as the space of boundary conditions for the $\overline{\partial 
}$ operator. 
Let $\overline{\partial }_{W}$ denotes the $\overline{\partial }$ operator
acting on the domain $\mathcal{D}_{\infty }$. The index of this operator is
usually defined as (see e.g. \cite{Wit}) 
\begin{equation}
\mathrm{index}\ \overline{\partial }_{W}=\mathrm{dim}(\mathrm{ker}\overline{%
\partial }_{W})-\mathrm{dim}(\mathrm{coker}\overline{\partial }_{W}).
\label{dbarindex}
\end{equation}%
Taking into account that for given $S_{W}$ one has $S_{\widetilde{W}%
}=\{-n|n\notin S_{W}\}$, one finds \cite{KMAM} 
\begin{equation}
\mathrm{index}\ \overline{\partial }_{W}=\mathrm{card}(S_{W}-\mathbb{N})-%
\mathrm{card}(S_{\widetilde{W}}-\mathbb{N}).
\end{equation}%
where $\mathbb{N}=\{0,1,2,3,\dots \}$.

For the hidden KP hierarchies associated with the Birkhoff strata the index
of the $\overline{\partial }$ operator has been calculated in \cite{KMAM}. 

\section{Big cell $\Sigma_{\varnothing}$. Families of normal rational
(Veronese) curves}

\label{sect-bigcell} We begin with the principal stratum $\Sigma_\varnothing$%
. Since in this case the basis (\ref{basiselem}) is composed by the Laurent
series of all positive degree $n=0,1,2,3,\dots$ there exists a canonical
basis $\{ p_0,p_1,p_2,\dots \}$ in $\Sigma_\varnothing$ with the basis
elements of the form 
\begin{equation}  \label{bigcellbasis}
p_i(z)=z^i+\sum_{k=1}^\infty \frac{H^i_k}{z^k}, \qquad i=0,1,2,\dots.
\end{equation}
Basis elements (\ref{bigcellbasis}) are parameterized by the infinite set of
arbitrary $H^i_k \in \mathbb{C}$.

Points of $\Sigma _{\varnothing }$ are represented by the
infinite-dimensional linear subspaces which are spans of $%
\{p_{0}(z),p_{1}(z),p_{2}(z),\dots \}$ with fixed all $H_{k}^{i}$. $\Sigma
_{\varnothing }$ itself is a family of such subspaces parameterized by $%
H_{k}^{j}$.

In this paper we will be interested by particular points in $\Sigma
_{\varnothing }$ with \ rather special property. We will look for the points
for which the corresponding subspaces ( fibers) have a specific algebraic
property, namely, when they admit a multiplication of elements. The
following Lemma is the starting point of the analysis.

\begin{lem}
\label{lembigcell} Laurent series $p_i(z)$ (\ref{bigcellbasis}) with fixed $%
H^i_k$ obey the equations 
\begin{equation}  \label{bigcellalg}
p_j(z)p_k(z)=\sum_{l=0}C_{jk}^lp_l(z), \qquad j,k=0,1,2,\dots
\end{equation}
if and only if the parameters $H^i_k$ satisfy the constraints 
\begin{equation}  \label{bigcell0coeff}
H^0_j=0, \qquad j=0,1,2,\dots
\end{equation}
and 
\begin{equation}  \label{bigcellsercoeff}
H^{j+k}_m-H^{j}_{m+k}-H^k_{j+m}+\sum_{l=1}^{j-1}
H^k_{j-l}H^l_m+\sum_{l=1}^{k-1} H^j_{k-l}H^l_m -\sum_{l=1}^{m-1}
H^k_{m-l}H^j_l=0, \qquad j,k,m=1,2,3,\dots.
\end{equation}
\end{lem}

\textbf{Proof} Let us require that Laurent series (\ref{bigcellbasis}) obey
the condition (\ref{bigcellalg}). Comparing the coefficients in front of
positive powers of $z$ in both sides of (\ref{bigcellalg}), one concludes
that 
\begin{equation}  \label{bigcellstructcoeff}
C^l_{jk}=\delta^l_{j+k}+H^k_{j-l}+H^j_{k-l}, \qquad j,k=0,1,2,3,\dots
\end{equation}
and all $H^0_j=0$. Counting of negative powers of $z$ gives the relations (%
\ref{bigcellsercoeff}). The conditions (\ref{bigcell0coeff}) and (\ref%
{bigcellsercoeff}) obviously are also the sufficient one. $\square$

Note that in this case $p_0=1$ and the conditions $p_0p_i(z)=p_i(z)$ are
identically satisfied.

Lemma (\ref{lembigcell}) has an immediate consequence.

\begin{prop}
\label{W0W0CW0} The big cell $\Sigma _{\varnothing }$ contains the subset $%
W_{\varnothing }$ of points such that the corresponding tautological
subbundle $\ TB_{W_{\varnothing }}$ \ is formed by linear spaces (fibers)
closed with respect to pointwise multiplication.
\end{prop}

\textbf{Proof} \ Indeed, take a point in $\Sigma _{\varnothing }(H)$ for
which all \ $H_{k}^{i}$ represent fixed solution of the system \ (\ref%
{bigcell0coeff}) and (\ref{bigcellsercoeff}). The conditions (\ref%
{bigcellalg}) guarantees that for any two elements $q_{1}=\sum_{j=0}^{\infty
}\alpha _{j}p_{j}(z,H)$ and $q_{2}=$ $\sum_{j=0}^{\infty }\beta
_{j}p_{j}(z,H)$ \ of the same fiber with given H the product $q_{1}q_{2}$ is
of the form $\sum_{j=0}^{\infty }\gamma _{l}p_{l}(z,H)$ with $\gamma _{l}=$ $%
\sum_{j,k=0}^{\infty }a_{j}b_{k}C_{jk}^{l}$, i.e. it belongs to the same
fiber. The collection of such points form the subset $W_{\varnothing }.$ The
properties of these points clearly are quite special with respect to the
generic points of \ $\Sigma _{\varnothing }$

In a different approach, namely in association with the Krichever map, the
subsets of points of Sato Grassmannian closed with respect to multiplication
has been discussed in \cite{MP} (Theorem 6.4).

\begin{prop}
\label{W0W0CW0 copy(1)} The subbundle  $TB_{W_{\varnothing }}$ is an
infinite family of infinite-dimensional commutative associative algebras.
\end{prop}

.Equations (\ref{bigcellalg}) for fixed $H_{k}^{i}$ \ represent the table of
multiplication for a commutative algebra with the basis $(1,p_{1},p_{2},%
\dots )$ and the structure coefficients $C_{kl}^{j}$ (\ref%
{bigcellstructcoeff}) for each $z$. It is a direct check that the conditions
(\ref{bigcell0coeff}) and (\ref{bigcellsercoeff}) are equivalent to the
associativity condition 
\begin{equation}
\sum_{l=0}^{\infty }C_{jk}^{l}C_{lm}^{p}=\sum_{l=0}^{\infty
}C_{mk}^{l}C_{lj}^{p},\qquad j,k,m,p=0,1,2,\dots \ .  \label{bigcelass}
\end{equation}%
for the structure coefficients $C_{jk}^{l}$. So, at fixed $H_{k}^{j}$, the
span of $\{p_{0}(z),p_{1}(z),p_{2}(z),\dots \}$, i.e. the fiber attached to
the point from $W_{\varnothing }$ , is the infinite-dimensional associative
commutative algebra $A_{\Sigma _{\varnothing }}$ with the basis $\langle
p_{0},p_{1},p_{2},\dots \rangle $ and structure constants (\ref%
{bigcellstructcoeff}) . The subbundle  $TB_{W_{\varnothing }}$ is a family
of such algebras. $\square $

In a different context the formulae (\ref{bigcellalg})-(\ref{bigcelass})
have appeared first in the paper \cite{KM} devoted to the coisotropic
deformations of associative algebras. In the rest of the paper we will refer
to the conditions (\ref{bigcell0coeff}) and (\ref{bigcellsercoeff}) and
similar one as the associativity conditions.

Algebra $A_{\Sigma_0}$ at fixed $H^j_k$ described above is the polynomial
algebra $\mathbb{C}[p_1]$ in the basis of Fa\`a di Bruno polynomials \cite%
{KM}. Indeed, it is easy to see that the relations (\ref{bigcellalg}) and (%
\ref{bigcellstructcoeff}) are equivalent to the following 
\begin{equation}  \label{bigcellcurr}
\begin{split}
p_2=&{p_1}^2-2H^1_1, \\
p_3=& {p_1}^3-3H^1_1p_1-3H^1_2, \\
p_4=& {p_1}^4-4H^1_1{p_1}^2-4H^1_2p_1-4H^1_3+2{H^1_1}^2, \\
p_5=& {p_1}^5-5H^1_1{p_1}^3-5H^1_2{p_1}^2- \left( 5\,H^1_{{3}}-5\,{H^1_{{1}}}%
^{2} \right)p_1-5\,H^1_{{4}}+5\,H^1_{{1}}H^1_{{2}}, \\
\dots & \\
p_n=& {p_1}^n+\sum_{k=0}^{n-2}u_{nk}{p_1}^k, \qquad n=6,7,8,\dots
\end{split}%
\end{equation}
where $u_{nk}$ are certain polynomials of $H^1_m$, $m=1,2,\dots,n-1$. The
polynomials in the r.h.s. of (\ref{bigcellcurr}) have been called Fa\`a di
Bruno polynomials in \cite{KM}.

The pointwise constraints (\ref{bigcellalg}) and (\ref{bigcellcurr}) for the
basis elements $p_{i}(z)$ have simple geometrical interpretation. Indeed, if
one treats $p_{1}(z),p_{2}(z),p_{3}(z),\dots $, for given $H_{k}^{i}$ and
variable $z$, as the local affine coordinates, then the conditions (\ref%
{bigcellalg}) become the constraints on coordinates $p_{1},p_{2},p_{3},\dots 
$ of the form 
\begin{equation}
f_{jk}=p_{j}p_{k}-\sum_{l=0}^{j+k}C_{jk}^{l}p_{l}=0.
\label{bigcellconstrcoord}
\end{equation}%
These relations define an algebraic variety for the fixed $H_{k}^{j}$. So,
under the constraint (\ref{bigcellconstrcoord}) one has an algebraic variety
at each fiber of  $TB_{W_{\varnothing }}$. Varying $H_{k}^{j}$, one gets

\begin{prop}
\label{towerbigcell} For the big cell $\Sigma _{\varnothing }$ the subbundle 
$TB_{W_{\varnothing }}$ contains an infinite family $\Gamma _{\infty }$ of
algebraic varieties which are intersections of the quadrics 
\begin{equation}
f_{jk}=p_{j}p_{k}-p_{j+k}-\sum_{l=1}^{j}H_{l}^{k}p_{j-l}-%
\sum_{l=1}^{k}H_{l}^{j}p_{k-l}=0,\qquad j,k=1,2,3,\dots 
\label{towerVeronese}
\end{equation}%
and parameterized by the variables $H_{k}^{j}$ obeying the algebraic
equations (\ref{bigcell0coeff}) and (\ref{bigcellsercoeff}). The family $%
\Gamma _{\infty }$ is the $ind$-variety with $\Gamma _{2}\subset \dots
\subset \Gamma _{d-1}\subset \Gamma _{d}\subset \Gamma _{d+1}\subset \dots $
where subvarieties $\Gamma _{d}\ (d=2,3,\dots )$ are isomorphic to a family
of rational normal curves (Veronese curves) of degree $d$.
\end{prop}

\textbf{Proof} In virtue of the equivalence of the set of equations (\ref%
{bigcellcurr}) to the set 
\begin{equation}  \label{bigcellcurrcurv}
\begin{split}
h_2=&p_2- {p_1}^2+2H^1_1=0, \\
h_3=&p_3- {p_1}^3+3H^1_1p_1+3H^1_2=0, \\
\dots& \\
h_n=&p_n- {p_1}^n+\sum_{k=0}^{n-2}u_{nk}{p_1}^k=0, \qquad n=4,5,6, \dots
\end{split}%
\end{equation}
the variety $\Gamma_{\infty}$ has dimension $1$ for each fixed $H^1_m$, $%
m=1,2,3,\dots$ . The ideal of $\Gamma_{\infty}$ is $I(\Gamma_{\infty})=%
\langle h_2,h_3,h_4,\dots\rangle$. For each finite-dimensional subspace with
coordinates $p_1,p_2,\dots,p_d$ and fixed $H^1_m$, $m=1,2,\dots,d-1$ the
corresponding variety $\Gamma_d$ is a rational normal curve of degree $d$.
For instance, $\Gamma_3$ is the twisted cubic. Formulae (\ref%
{bigcellcurrcurv}) represent the canonical parameterization of rational
normal curve (Veronese curve) (see e.g. \cite{Har}). Due to the
associativity conditions (\ref{bigcellsercoeff}) and their consequence $%
nH^i_n=iH^n_i$ all $H^i_k$ are polynomial functions of $H^1_m$, $%
m=1,2,3,\dots$. For example $H^n_1=nH^1_n$, $H^2_2={H^1_1}^2+2H^1_3$. Thus,
the family of algebraic varieties $\Gamma_d$ parameterized by $%
H^1_1,H^1_2,\dots, H^1_{d-1}$, i.e. the family of rational normal curves of
the degree $d$, is itself the affine algebraic variety in the $2d-1$%
--dimensional space. Finally, the variety $\Gamma_{\infty}$ is the $ind$%
-variety since $\Gamma_2 \subset \dots \subset \Gamma_{d-1} \subset \Gamma_d
\subset \Gamma_{d+1} \subset \dots$. We will refer such an $ind$-variety as
a tower of families of algebraic curves. $\square$

We note that within the theory of schemes (see e.g. \cite{Sha}-\cite{Don})
one can define algebraic variety associated with the fibers  of the
subbundle  $TB_{W_{\varnothing }}$  as the Spectrum Spec($R_{A}$) of the
ring $R_{A}$ corresponding to the algebra $A_{\Sigma _{\varnothing }}$.

In the theory of ideals and algebraic geometry, the so called canonical
basis, generated by elements of the form $q_n-a_n$ with some $a_n$ play a
distinguished role (see e.g. \cite{Har}, Lecture 5). For an ideal $%
I(\Gamma_\infty)$ such basis can be found in the following way. First from
the constraint $h_2=0$, one has $H^1_1=\frac{1}{2}\left({p_1}^2-p_2\right)$.
Substituting this expression for $H^1_1$ into $h_3$, one gets 
\begin{equation}
\tilde{h}_3=p_3+\frac{1}{2}{p_1}^3-\frac{3}{2}p_1p_2+3H^1_2=0.
\end{equation}
From this relation one obtains $H^1_2$ in terms of $p_1,p_2,p_3$ and then
substitutes into $h_4$ getting $\tilde{h}_4$. Continuing this procedure, one
finds (see also \cite{KM2}) 
\begin{equation}
\tilde{h}_n=-n\left(P_n(\tilde{p})-H^1_{n-1}\right), \qquad n=2,3,4, \dots
\end{equation}
where $\tilde{p_k}=-\frac{1}{k}p_k$ and $P_n(\tilde{p})$ are standard Schur
polynomials defined by the formula 
\begin{equation}
e^{\sum_{n=1}^\infty z^n t_n}=\sum_{m=0}^\infty z^m P_m(t_1,t_2,t_3,\dots).
\end{equation}
Thus one has

\begin{prop}
\label{propII=I} Canonical basis for the ideal $I(\Gamma_{\infty})$ is
composed by the elements 
\begin{equation}  \label{bigcell-h*}
h^*_n=p^*_n-H^1_{n-1}, \qquad n=2,3,4,\dots
\end{equation}
where $p^*_n=P_n\left(-p_1,-\frac{1}{2}p_2,-\frac{1}{3}p_3,\dots\right)$.
\end{prop}

This observation reveals that the variables $H^1_k$, $k=1,2,3,\dots$ play
the distinguished role in the parameterization of the associativity
conditions (\ref{bigcellsercoeff}).

The proposition \ref{propII=I} has an obvious

\begin{cor}
In the variables $p_n^*$ and $u_n=H^1_n$, $n=1,2,3,\dots$ the variety $%
\Gamma_{\infty}$ is given by intersection of the hyperplanes (\ref%
{bigcell-h*}).
\end{cor}

As far as the $\overline{\partial}-$operator is concerned one easily shows
that for the big cell 
\begin{equation}  \label{indexGr0}
\mathrm{index}\ \overline{\partial}_{W_\varnothing}=0.
\end{equation}

Rational normal curves in $\Gamma_\infty$ defined by (\ref{bigcellcurrcurv})
are smooth curves for all $d=2,3,4,\dots$. Their projection to lower
dimensional subspaces are singular algebraic curves of different types.

For the twisted cubic defined by the first two equations (\ref%
{bigcellcurrcurv}) the projection along the axis $p_1$ to the subspace with
coordinates $p_2,p_3$ is given by 
\begin{equation}  \label{singcubcurv}
\mathcal{F}^0_{23}={p_3}^2-{p_2}^3+6H^1_2p_3+3{H^1_1}^2p_2+9{H^1_2}^2-2{H^1_1%
}^3=0
\end{equation}
or in the standard form 
\begin{equation}  \label{singcubcurv-stand}
\mathcal{F}^0_{23}={\tilde{p}_3}^2-{p_2}^3+3{H^1_1}^2p_2-2{H^1_1}^3=0
\end{equation}
where ${\tilde{p}_3}=p_3+3H^1_2$. Since the discriminant of the curve
vanishes 
\begin{equation}  \label{Discr-0}
\Delta=\frac{1}{1728}\left(\left( \frac{{H^1_1}^2}{9} \right)^3-\left(- 
\frac{{H^1_1}^3}{27} \right)^2 \right)=0
\end{equation}
it has an ordinary double point and zero genus.

Curves (\ref{singcubcurv}) belong to the ideal $\mathcal{I}$ and 
\begin{equation}
\mathcal{F}^0_{23}= \left(h_3 +b_3 \right) h_3+ \left(-{h_2}^2 + b_4 h_2
+b_2 \right) h_2
\end{equation}
with 
\begin{equation}
\begin{split}
\ b_4 =& -3\,{p_1}^{2}+6\,H^1_{{1}}=-3p_2, \\
b_3 =& 2\,{p_1}^{3}-6\,H^1_{{1}}p_1=2(p_3+3H^1_2), \\
b_2 =& -9\,{H^1_{{1}}}^{2}+12\,H^1_{{1}}{p_1}^{2}-3\,{p_1}^{4}=-3{p_2}^2+3{%
H^1_1}^2
\end{split}%
\end{equation}
and 
\begin{equation}
h_3+b_3=p_3+3H^1_2-\left({p_1}^2-3H^1_1 \right)p_1=p_3+3H^1_2-\left({p_2}%
-H^1_1 \right)p_1-{h_2}^2+b_4h_2+b_2=-3{p_2}^2+6{H^1_1}^2.
\end{equation}
We note that in terms of $h_2$ and $h_3$ the projected twisted cubic is the
plane cubic too.

The projection parallel to the axis $p_1$ of the fourth degree rational
curve defined by first three equations (\ref{bigcellcurrcurv}) into the two
dimensional space $(p_3,p_4)$ is represented by the singular trigonal curve 
\begin{equation}  \label{singtrigcurv}
\begin{split}
\mathcal{F}_{34}^0={p_{{3}}}^{4}&-{p_{{4}}}^{3}-12\,H^1_{{3}}{p_{{4}}}^{2}
-12\,H^1_{{1}}H^1_{{2}}p_{{3}}p_{{4}} - \left( 6\,(H^1_{{2}})^{2}+4\,(H^1_{{1%
}})^{3} \right) {p_{{3}}}^{2} \\
&- \left( 48\,(H^1_{{3}})^{2}-12\,H^1_{{1}}(H^1_{{2}})^{2}-3\,(H^1_{{1}%
})^{4} \right) p_{{4}} - \left( 48\,H^1_{{1}}H^1_{{2}}H^1_{{3}}-8\,(H^1_{{2}%
})^{3}-12\,(H^1_{{1}})^{3}H^1_{{2}} \right) p_{{3}} \\
&+2\,(H^1_{{1}})^{6}-64\,(H^1_{{3}})^{3}+12\,H^1_{{3}}(H^1_{{1}%
})^{4}-24\,(H^1_{{1}})^{3}(H^1_{{2}})^{2} -3\,(H^1_{{2}})^{4}+48\,(H^1_{{2}%
})^{2}H^1_{{3}}H^1_{{1}}=0.
\end{split}%
\end{equation}
For the curve (\ref{singtrigcurv}) one has 
\begin{equation}  \label{singtrigcurv-0-ideal}
\mathcal{F}_{34}^0=\left({h_4}^2+a_8h_4+a_4\right)h_4+\left(-{h_3}^3+a_9{h_3}%
^2+a_7{h_4}+a_6{h_3}+a_3\right)h_3
\end{equation}
where the coefficients $a_3,a_4,a_6,a_7,a_8,a_9$ are given in the Appendix %
\ref{App-bigcell}.

Finally, let us consider the fifth degree rational normal curve defined by
first four equations (\ref{bigcellcurrcurv}). Its projection into the plane $%
(p_2,p_5)$ is the singular genus zero quintic 
\begin{equation}  \label{0F25}
\begin{split}
\mathcal{F}^0_{25}=& {p_{{5}}}^{2}-{p_{{2}}}^{5}+10\,H^1_{{2}}p_{{2}}p_{{5}%
}- \left( -10\,H^1_{ {3}}-5\,{H^1_{{1}}}^{2} \right) {p_{{2}}}^{3}- \left(
-10\,H^1_{{2}}H^1_{{1} }-10\,H^1_{{4}} \right) p_{{5}} \\
&- \left( -10\,H^1_{{1}}H^1_{{3}}-25\,{H^1_{{2}} }^{2} \right) {p_{{2}}}%
^{2}- \left( 25\,{H^1_{{3}}}^{2}-50\,H^1_{{2}}H^1_{{4 }}+5\,{H^1_{{1}}}%
^{4}+30\,{H^1_{{1}}}^{2}H^1_{{3}}-50\,H^1_{{1}}{H^1_{{2}}}^{2} \right) p_{{2}%
} \\
&+25\,{H^1_{{2}}}^{2}{H^1_{{1}}}^{2}+25\,{H^1_{{4}}}^{2}-2\,{\ H^1_{{1}}}%
^{5}+50\,H^1_{{2}}H^1_{{1}}H^1_{{4}}-20\,H^1_{{3}}{H^1_{{1}}}^{3}-50\,{H^1 _{%
{3}}}^{2}H^1_{{1}}=0.
\end{split}%
\end{equation}
A projection of the fifth degree rational curve into the two dimensional
space $(p_4,p_5)$ defined by first four equations (\ref{bigcellcurrcurv}) is
the plane singular $(4,5)$ curve in terminology of \cite{BEL,BLE2} 
{\scriptsize 
\begin{equation}  \label{0F45}
\begin{split}
\mathcal{F}^0_{45}=&p_5^4-p_4^5 -20\,H^1_{{4}}{p_5}^3 +\left( 20\,H^1_{{1}%
}H^1_{{3}}+10\,{H^1_{{2}}}^{2} \right) {p_4}{p_5}^2 +\left( 20\,H^1_{{2}}{%
H^1_{{1}}}^{2}+20\,H^1_{{2}}H^1_{{3}}\right) {p_4}^2{p_5} \\
& +\left( 5\,{H^1_{{1}}}^{4}+10\,{H^1_{{3}}}^{2}+20\,{H^1_{{2}}}^{2}H^1_{{1}%
}\right) {p_4}^3 +\left( 150\,{H^1_{{4}}}^{2}-20\,{H^1_{{2}}}^{2}H^1_{{3}%
}-30\,{H^1_{{2}}}^{2}{H^1_{{1}}} ^{2}-20\,H^1_{{1}}{H^1_{{3}}}^{2}-4\,{H^1_{{%
1}}}^{5}\right) {p_5}^2 \\
& +\left( -20\,H^1_{{2}}{H^1_{{1}}}^{4}-60\,{H^1_{{1}}}^{2}H^1_{{2}}H^1_{{3}%
}+200\,H^1_{{3}} H^1_{{4}}H^1_{{1}}-40\,H^1_{{2}}{H^1_{{3}}}^{2}+100\,{H^1_{{%
2}}}^{2}H^1_{{4}}-40\, {H^1_{{2}}}^{3}H^1_{{1}}\right) {p_4}{p_5} \\
& +\left( -40\,{H^1_{{1}}}^{3}{H^1_{{2}}}^{2}-20\,H^1_{{3}}{H^1_{{1}}}%
^{4}-20\,{H^1_{{3}}} ^{3}+5\,{H^1_{{2}}}^{4}+50\,{H^1_{{3}}}^{2}{H^1_{{1}}}%
^{2}-40\,{H^1_{{2}}}^{2} H^1_{{1}}H^1_{{3}}\right. \\
& \left. +100\,H^1_{{2}}{H^1_{{1}}}^{2}H^1_{{4}}+100\,H^1_{{4}}H^1_{{2}%
}H^1_{ {3}}\right) {p_4}^2+c_5p_5+c_4p_4+c_0=0
\end{split}%
\end{equation}
} where the coefficients $c_5,c_4,c_0$ are given in Appendix \ref%
{App-bigcell}.

In the next sections we will see that curves (\ref{singcubcurv}), (\ref%
{singtrigcurv}) are singular limits of smooth algebraic curves from the
strata $\Sigma_1$ and $\Sigma_{1,2}$.

The system (\ref{bigcellalg})-(\ref{bigcellstructcoeff}) admits infinitely
many simple reductions. The first $p_{1}=z$ is obviously trivial. The
constraint $p_{2}=z^{2}$, i.e. all $H_{n}^{2}=0$, due to (\ref%
{bigcellsercoeff}) implies that $H_{n}^{2m}=0$, $m=1,2,3,\dots ,\
n=1,2,3,\dots $, i.e. $p_{2n}={p_{2}}^{n}=z^{2n}$, $n=1,2,3,\dots $. It is
easy to show that the system (\ref{bigcellalg})-(\ref{bigcellstructcoeff})
admits the reductions $p_{ln}={p_{l}}^{n}=z^{ln}$, $l=1,2,3,\dots $, $%
n=2,3,4,\dots $. 

\bigskip

\section{Stratum $\Sigma _{0}$. Family of centered normal rational curves}

The first stratum different from $\Sigma_\varnothing$ is associated with $%
S=\{-1,1,2,\dots \}$. In the absence of zero degree element the positive
degree elements of the canonical basis are 
\begin{equation}  \label{modbigcell-curr}
p_i=z^i+H^i_0+\sum_{k \geq 1} \frac{H^i_k}{z^k}, \qquad i=1,2,3,\dots\ .
\end{equation}
Since $(p_{-1})^2 \notin \langle p_i\rangle_{i=-1,1,2,\dots}$ the element $%
p_{-1}$ cannot belong to a fiber of the subbundle $TB_{W_0}$ closed with
respect to multiplication. Considering only $p_j$ of positive degrees one has

\begin{lem}
Laurent series (\ref{modbigcell-curr}) obey the equations 
\begin{equation}  \label{alg-modbig}
p_j(z)p_k(z)=\sum_{l=1}^\infty C_{jk}^lp_l(z), \qquad j,k=1,2,3,\dots
\end{equation}
if and only if the parameters $H^j_k$, $k=0,1,2,\dots$ obey the constraints 
\begin{equation}  \label{algcoeff-modbig}
H^j_{k+m}+H^k_{j+m}+\sum_{s=0}^m
H^k_sH^j_{m-s}=H^{j+k}_m+\sum_{s=0}^{j-1}H^k_sH^{j-s}_m+%
\sum_{s=0}^{k-1}H^j_sH^{k-s}_m.
\end{equation}
\end{lem}

\textbf{Proof} Proof is similar to that of Lemma \ref{lembigcell}. The
constants $C^l_{jk}$ are given by (\ref{bigcellstructcoeff}) with $%
j,k,l=1,2,3,\dots$ . $\square$

As the consequence of this Lemma one gets

\begin{prop}
\label{tower0strat} The stratum $\Sigma _{0}$ with $S=\{-1,1,2,3,\dots \}$
contains the subset $W_{0}$ of points for which the corresponding subbundle $%
TB_{W_{0}}$ formed by fibers closed with respect to pointwise
multiplication.  The fibers are vector spaces with basis $\langle
p_{i}\rangle _{i}$ with $H_{j}^{i}$ obeying the constraints (\ref%
{algcoeff-modbig}). Algebraically the subbundle  $TB_{W_{0}}$ is the
infinite family of infinite-dimensional commutative associative algebras $%
A_{\Sigma _{0}}$ without unity element.
\end{prop}

\textbf{Proof } Proof is analogous to that of Proposition \ref{W0W0CW0}. $%
\square$

Algebra $A_{\Sigma_0}$ with fixed $H^i_j$ is a polynomial algebra since 
\begin{equation}  \label{pi-modbig}
\begin{split}
p_2=&{p_1}^2-2H^1_0p_1, \\
p_3=&{p_1}^3-3H^1_0{p_1}^2-3(H^1_1-{H^1_0}^2)p_1, \\
\dots& \\
p_n=&{p_1}^n-\sum_{k=1}^{n-1}u_k{p_1}^k, \qquad n=4,5,6,\dots
\end{split}%
\end{equation}
Geometrically interpretation of the subspace $W_0$ is similar to that given
in Section \ref{sect-bigcell}.

\begin{prop}
For the stratum $\Sigma _{0}$ the subbundle $TB_{W_{0}}$ contains an
infinite family of algebraic varieties defined as intersection of the
quadrics 
\begin{equation}
\widetilde{\mathcal{F}}_{jk}=p_{j}+p_{k}-p_{j}p_{k}+%
\sum_{l=0}^{j-1}H_{l}^{k}p_{j-l}+\sum_{l=0}^{k-1}H_{l}^{j}p_{k-l}=0,\qquad
j,k=1,2,3,\dots 
\end{equation}%
which is parameterized by $H_{k}^{j}$ obeying the equations (\ref%
{algcoeff-modbig}). This family is an infinite tower of normal rational
curves of all degrees passing through the origin.
\end{prop}

\textbf{Proof } The ideal of this family of algebraic varieties is $%
I_0(\Gamma_\infty)=\langle \tilde{h}_2, \tilde{h}_3,\tilde{h}_4,\dots
\rangle $ where 
\begin{equation}
\begin{split}
\tilde{h}_2 =& p_2 -{p_1}^2+2H^1_0p_1, \\
\tilde{h}_3 =& p_3 - {p_1}^3+3H^1_0{p_1}^2+3(H^1_1-{H^1_0}^2)p_1, \\
\dots & \\
\tilde{h}_n =& p_n- {p_1}^n+\sum_{k=1}^{n-1}u_k{p_1}^k, \qquad
n=4,5,6,\dots\ .
\end{split}%
\end{equation}
In contrast to the big cell all these normal rational curves pass through
the origin $p_1=p_2=p_3=\dots=0$. $\square$

Since $S_{\widetilde{W}_0}=\{0,1,2,\dots\}$ for the subspace $W_0$ one has
index$(\overline{\partial}_{W_0})=0$.

Similar to the big cell all normal rational curves given by (\ref{pi-modbig}%
) are smooth and have zero genus while their projections to the lower
dimensional subspaces are singular algebraic curves. For instance, the
projection of the Veronese curve of the degree $3$ defined by the first two
equations (\ref{pi-modbig}) onto the subspace with coordinates $(p_2,p_3)$
is the singular plane cubic 
\begin{equation}  \label{degEll-modbig}
\begin{split}
\mathcal{F}_{23}^{(0)}=&{p_{{3}}}^{2}-{p_{{2}}}^{3}- \left( 3\,{H^1_{{0}}}%
^{2} -6\,H^1_{{1}} \right) {p_{{2}}}^{2}- \left( -6\,H^1_{{0}}H^1_{{1}} +2\,{%
H^1_{{0}}}^{3} \right) p_{{3}} \\
&- \left( -12\,H^1_{{1}}{H^1_{{0}}}^{2} +9\,{H^1_{{1}}}^{2}+3\,{H^1_{{0}}}%
^{4} \right) p_{{2}}=0.
\end{split}%
\end{equation}
In the standard form it is 
\begin{equation}
{\tilde{p}_3}^2-{\tilde{p}_2}^2+3\,{H^1_{{1}}}^{2}\tilde{p}_2-2{H^1_1}^3=0
\end{equation}
where $\tilde{p}_3=p_{{3}}+3\,H^1_{{0}}H^1_{{1}}-{H^1_{{0}}}^{3}$ and $%
\tilde{p}_2=p_{{2}}+{H^1_{{0}}}^{2}-2\,H^1_{{1}}$. Analogously the
projection of Veronese curves to subspaces $(p_2,p_5)$, $(p_3,p_4)$, $%
(p_4,p_5)$ are singular algebraic curves of genus zero.

Comparing the formulas of this and previous sections, one observes that they
are pretty close to each other and the algebraic curves in the big cell an
stratum $\Sigma_0$ are essentially of the same type. Moreover one can easily
see that they are transformed to each other by the simple change of
``coordinates'' 
\begin{equation}
p_i^{\mathrm{big\ cell}} = p_i^{\Sigma_0}-H^i_0, \qquad i=1,2,3,\dots
\end{equation}
where 
\begin{equation}
\begin{split}
H^2_0=& 2\,H^1_{{1}}-{H^1_{{0}}}^{2}, \\
H^3_0=& -3\,H^1_{{0}}H^1_{{1}}+{H^1_{{0}}}^{3}+3\,H^1_{{2}}, \\
H^4_0=& -2\,{H^1_{{1}}}^{2}+4\,H^1_{{1}}{H^1_{{0}}}^{2} -{H^1_{{0}}}%
^{4}-4\,H^1_{{2}}H^1_{{0}}+4\,H^1_{{3}}, \\
\dots & \ .
\end{split}%
\end{equation}
So all the results for the stratum $\Sigma_0$ are easily obtainable from
those for the big cell. Formally one can consider these two cases as two
special reductions of a more general family of normal rational curves
defined by the equations 
\begin{equation}
\begin{split}
p_2=& {p_1}^2+u_{21}{p_1}+u_{20}, \\
p_3=& {p_1}^3+u_{32}{p_1}^2+u_{31}{p_1}+u_{30}, \\
\dots & \\
p_n=& {p_1}^n+\sum_{k=0}^{n-1}u_{nk}{p_1}^k, \qquad n=4,5,6,\dots
\end{split}%
\end{equation}
where $u_{nm}$ are parameters. Such class of normal rational curves is
invariant under the shifts 
\begin{equation}
p_n\to p_n+\alpha_n, \qquad n=1,2,3,\dots
\end{equation}
where $\alpha_n$ are arbitrary parameters. This invariance allows us to fix
the infinite (countable) set of parameters $u_{nm}$. The gauge in which $%
u_{n\ n-1}=0$ corresponds to normal rational curves from the big cell. In
the gauge $u_{n0}=0$ one has the normal rational curves from $\Sigma_0$.

Similar situation takes place for other strata for which $0\notin S$. By
this reason in the rest of the paper we will consider only strata for which $%
0\in S$.

I

\section{Stratum $\Sigma_1$. Elliptic curve and its coordinate ring.}

\label{firststrat} For the stratum $\Sigma_1$ one has $S=\{-1,0,2,3,\dots\}$
and the element of the first degree $z+O(z^{-1})$ is absent in the basis.
Hence, positive degree elements of the canonical basis in $\Sigma_1$ have
the form 
\begin{equation}  \label{1stratser}
\begin{split}
p_0(z)=&1+\sum_{k=1}^{\infty} \frac{H^0_k}{z^k}, \\
p_i(z)=&z^i+H^i_{-1}z+\sum_{k=1}^{\infty} \frac{H^i_k}{z^k}, \qquad
i=2,3,4,\dots.
\end{split}%
\end{equation}
Similar to stratum $\Sigma_0$ the element $p_{-1}$ cannot belong a fiber of the subbundle $TB_{W_1}$
 closed with respect to multiplication.

\begin{lem}
\label{lem1strat} Laurent series (\ref{1stratser}) obey the equations 
\begin{equation}  \label{1stratalg}
p_i(z) p_j(z) = \sum_{l=0,2,3,\dots} C^l_{ij} p_l(z), \qquad
i,j=0,2,3,4,\dots
\end{equation}
if and only if the parameters $H^i_k$ satisfy 
\begin{equation}  \label{1strat0coeff}
H^0_k=0, \qquad k=1,2,3,\dots
\end{equation}
and 
\begin{equation}  \label{1stratsercoeff}
\begin{split}
H^i_{j+l}&+H^j_{i+l}+H^j_{-1}H^i_{l+1}+H^i_{-1}H^j_{l+1}+%
\sum_{n=1}^{l-1}H^j_n H^i_{l-n} = \\
&
H^{i+j}_l+H^j_{-1}H^{i+1}_l+H^i_{-1}H^{j+1}_l+%
\sum_{n=2}^{i-1}H^{j}_{i-n}H^n_l+\sum_{n=2}^{j-1}H^{i}_{j-n}H^n_l+
H^i_{-1}H^j_{-1}H^2_l+ \\
&(H^i_j+H^j_i+H^i_{-1}H^j_1+H^i_{1}H^j_{-1})\delta^l_0, \qquad
j,k=2,3,4,\dots, \ l=-1,1,2,3,\dots.
\end{split}%
\end{equation}
\end{lem}

\textbf{Proof} Proof is similar to the case of $\Sigma_\varnothing$.
Considering positive powers of $z$ in both sides of (\ref{1stratalg}), one
gets 
\begin{equation}  \label{1stratstructcoeff}
C^l_{ij}=\delta^l_{i+j}+ H^j_{-1} \delta^l_{i+1} + H^i_{-1}
\delta^l_{j+1}+H^j_{i-l}+H^i_{j-l}+H^i_{-1}H^j_{-1}\delta^l_2 +\left( H^i_j+
H^j_i + H^i_{-1} H^j_1+H^i_1 H^j_{-1} \right)\delta^l_0 .
\end{equation}
Comparison of negatives powers gives formula (\ref{1stratsercoeff}). $%
\square $

As a consequence of the Lemma \ref{lem1strat} one has

\begin{prop}
\label{prop1strat} The stratum $\Sigma _{1}$ contains the subset $W_{1}$ of
points for which the corresponding subbundle $TB_{W_{1}}$ is formed by
fibers closed with respect to pointwise multiplication. The fibers are
vector spaces with basis $\langle p_{i}\rangle _{i}$ with $H_{k}^{i}$
satisfying the conditions (\ref{1strat0coeff}) and (\ref{1stratsercoeff}).
Moreover Codim$(W_{1})=$card$(\mathbb{N}-S_{W_{1}})=1$. The subbundle  $%
TB_{W_{1}}$ is the infinite family of infinite-dimensional
commutative associative algebras $A_{\Sigma _{1}}$ with the basis $%
(1,p_{2},p_{3},p_{4},\dots )$ and corresponding structure constants $%
C_{ij}^{l}$ given by (\ref{1stratstructcoeff}).
\end{prop}

An analysis of the multiplication table (\ref{1stratalg}), i.e. 
\begin{eqnarray} 
\label{1strat22}
  {p_2}^2&=&p_4+2H^2_{-1}p_3+(H^2_{-1})^2p_2 +2H^2_2+2H^2_1H^2_{-1}, \\
\label{1strat23}
  {p_2}p_3&=&p_{{5}}+H^2_{{-1}}p_{{4}}+H^3_{{-1}}p_{{3}}+ \left( H^2
_{{-1}}H^3_{{-1}}+H^2_{{1}} \right) p_{{2}}+H^2_{{-1}}H^3_{{1}}+H^2_{{1}}H^3_{{-1}}
+H^2_{{3}}+H^3_{{2}},\\
\label{1strat24}
  {p_2}p_4&=&p_{{6}}+H^2_{{-1}}p_{{5}}+ \left( H^2_{{1}}+H^4_{{-1}}
 \right) p_{{3}}+ \left( H^2_{{-1}}H^4_{{-1}}+H^2_{{2}}
 \right) p_{{2}}+H^2_{{4}}+H^2_{{-1}}H^4_{{1}}\\&&+H^2
_{{1}}H^4_{{-1}}+H^4_{{2}}, \nonumber\\
\label{1strat33}
  {p_3}^2&=&p_{{6}}+2\,H^3_{{-1}}p_{{4}}+ \left( {H^3_{{-1}}}^{2}+2\,H^3_{{1}} \right) p_{{2}}
+2\,H^3_{{3}}+2\,H^3_{{-1}}H^3_{{1}}, \\
{p_2}p_5 &= & p_{{7}}+H^2_{{-1}}p_{{6}}+H^2_{{1}}p_{{4}}+
 \left( H^2_{{2}}+H^5_{{-1}} \right) p_{{3}}
+ \left( H^2_{{3}}+H^2_{{-1}}H^5_{{-1}} \right) p_{{2}}+H^2_{{-1}}
H^5_{{1}}+H^5_{{2}}    \label{1strat25} \\ &&
+H^2_{{5}}+H^2_{{1}}H^5_{{-1}}, \nonumber \\
{p_3}p_4 &= & p_{{7}}+H^3_{{-1}}p_{{5}}+H^4_{{-1}}p_{{4}}+
H^3_{{1}}p_{{3}}+ \left( H^3_{{2}}+H^3_{{-1}}H^4_{{-1}}
+H^4_{{1}} \right) p_{{2}}+H^4_{{3}}+H^3_{{4}}\label{1strat34} \\&&
+H^3_{{-1}}H^4_{{1}}+H^3_{{1}}H^4_{{-1}} \nonumber
\end{eqnarray}
and so on shows that the algebra $A_{\Sigma_1}$ at fixed $H^j_k$ is the
polynomial algebra generated by $p_0=1,p_2,p_3$. However the formulae and (%
\ref{1strat22})-(\ref{1strat33}) immediately indicate that they are not
free. Indeed, subtracting (\ref{1strat24}) from (\ref{1strat33}), one first
eliminates $p_6$ then, using (\ref{1strat22}) and (\ref{1strat23}), gets 
\begin{equation}  \label{1stratellcurv}
\begin{split}
\mathcal{C}_6=& {p_3}^{2}-{p_2}%
p_4+H^2_{-1}p_5+2H^3_{-1}p_4+(H^2_{1}+H^4_{-1})p_3 +(-{H^3_{{-1}}}%
^{2}-2\,H^3_{{1}}+H^2_{{-1}}H^4_{{-1}}+H^2_{{2}})p_2 \\
& -2\,H^3_{{-1}}p_{{4}}-2\,H^3_{{3}} -2\,H^3_{{-1}}H^3_{{1}}+H^2_{{-1}}p_{{5}%
}+H^2_{{4}} +H^2_{{-1}}H^4_{{1}}+H^2_{{1}}H^4_{{-1}}+H^4_{{2}} \\
=&{p_3}^{2}-{p_2}^{3}+3\,H^2_{-1}p_3\,p_2-2\, H^3_{{-1}}{p_2}^{2}+ \left( {%
H^2_{{-1}}}^{3}+3\,H^2 _{{1}}+H^2_{{-1}}H^3_{{-1}} \right) p_3 \\
&- \left( {H^3_{{-1}}}^{2}+2\,H^3_{{1}}-3\,H^2_{{-1}}H^2_{{1}} -3\,H^2_{{2}%
}+H^3_{{-1}}{H^2_{{-1}}}^{2} \right) {p_2} \\
&+3\,H^2_{{4}}+3\,H^2_{{3}}H^2_{{-1}} +3\,H^2_{{2}}{H^2_{{-1}}}^{2}+3\,{H^2_{%
{1}}}^{2}-2\,H^3_{{-1}}H^3 _{{1}}-2\,H^3_{{3}}-3\,H^2_{{-1}}H^3_{{2}} -3\,{%
H^2_{{-1}}}^{2}H^3_{{1}} \\
& +H^2_{{-1}}H^3_{{-1}}H^2_{{1}}+4\,H^2_{{2}}H^3_{{-1}} =0.
\end{split}%
\end{equation}
This constraint, due to (\ref{1stratalg}), leads to the following
constraints on $H^2_i$ and $H^3_i$ 
\begin{equation}  \label{1strat23rel}
\begin{split}
H^3_2=&\frac{3}{2}\,H^2_{{3}}-\frac{1}{2}\,H^2_{{-1}}H^3_{{1}}+\frac{1}{2}%
\,H^3_{{-1}}H^2_{{1}}, \\
H^3_4=& \frac{1}{2}\,H^2_{{-1}}H^2_{{2}}H^3_{{-1}} +\frac{1}{2}\,H^3_{{-1}%
}H^2_{{3}}+\frac{1}{4}\,{H^2_{{-1}}}^{3}H^3_{{1}} -\frac{1}{2}\,H^3_{{1}%
}H^2_{{1}}+\frac{3}{2}\,H^2_{{4}}H^2_{{-1}} +\frac{3}{2}\,H^2_{{2}}H^2_{{1}}+%
\frac{3}{2}\,H^2_{{5}} \\
&-\frac{3}{4}\,H^2_{{3}}{H^2_{{-1}}}^{2}-\frac{3}{2}\,H^2_{{-1}}H^3_{{3}} -%
\frac{1}{4}\,{H^2_{{-1}}}^{2}H^3_{{-1}}H^2_{{1}}, \\
H^3_5=& \frac{3}{4}\,H^2_{{1}}H^2_{{3}}-\frac{3}{4}\,H^2_{{-1}}H^2_{{5}} +%
\frac{3}{2}\,H^2_{{6}}-\frac{3}{4}\,H^2_{{4}}{H^2_{{-1}}}^{2} +2\,H^3_{{-1}%
}H^2_{{4}}+\frac{1}{4}\,H^3_{{-1}}{H^2_{{1}}}^{2} +\frac{1}{2}\,{H^3_{{-1}}}%
^{2}H^2_{{2}}-H^3_{{-1}}H^3_{{3}} \\
&-\frac{1}{4}\,H^2_{{-1}}H^2_{{1}}{H^3_{{-1}}}^{2} -\frac{1}{4}\,H^3_{{-1}%
}H^2_{{2}}{H^2_{{-1}}}^{2} -H^3_{{-1}}H^2_{{-1}}H^2_{{3}}+\frac{1}{4}\,H^3_{{%
-1}}{H^2_{{-1}}}^{2}H^3_{{1}} -\frac{3}{4}\,H^2_{{1}}H^2_{{-1}}H^2_{{2}} \\
&+\frac{1}{8}\,{H^2_{{-1}}}^{3}H^3_{{-1}}H^2_{{1}} +\frac{3}{4}\,{H^2_{{-1}}}%
^{2}H^3_{{3}}-\frac{1}{2}\,{H^3_{{1}}}^{2} -\frac{1}{8}\,{H^2_{{-1}}}%
^{4}H^3_{{1}}+\frac{3}{8}\,H^2_{{3}}{H^2_{{-1}}}^{3} +H^2_{{2}}H^3_{{1}}, \\
&\dots
\end{split}%
\end{equation}
It is not difficult to see that the conditions (\ref{1strat23rel}) form a
subset of the system (\ref{1stratsercoeff}). The coefficient $H^3_2$ appear
in the curve (\ref{1stratellcurv}) itself which, after the substitution,
becomes 
\begin{equation}  \label{1stratellcurv-red}
\begin{split}
\mathcal{F}^1_{23}=&{p_3}^{2}-{p_2}^{3}+3\,H^2_{-1}p_3\,p_2-2\, H^3_{{-1}}{%
p_2}^{2}+ \left( {H^2_{{-1}}}^{3}+3\,H^2 _{{1}}+H^2_{{-1}}H^3_{{-1}} \right)
p_3 \\
&- \left( {H^3_{{-1}}}^{2}+2\,H^3_{{1}}-3\,H^2_{{-1}}H^2_{{1}} -3\,H^2_{{2}%
}+H^3_{{-1}}{H^2_{{-1}}}^{2} \right) {p_2} \\
&-2\,H^3_{{3}}-2\,H^3_{{-1}}H^3_{{1}}+3\,H^2_{{4}}+ 3\,H^2_{{2}}{H^2_{{-1}}}%
^{2}-\frac{3}{2}\,H^2_{{-1}}H^2_{{\ 3}}+3\,{H^2_{{1}}}^{2}-\frac{3}{2}\,{%
H^2_{{-1}}}^{2}H^3_{{1}} \\
&-\frac{1}{2}\,H^2_{{-1}}H^2_{{1}}H^3_{{-1}}+4\,H^3_{{-1}} H^2_{{2}} =0.
\end{split}%
\end{equation}
The constraint (\ref{1stratellcurv}) implies also that any element $p_n \in
A_{\Sigma_1}$ has the form 
\begin{equation}  \label{1stratunique}
p_n=\alpha_n(p_2)+\beta_n(p_2)p_3, \qquad n=2,3,4,5,6,\dots
\end{equation}
where $\alpha_n$ and $\beta_n$ are certain polynomials of degrees $\left[%
\frac{n}{2}\right]$ and $\left[\frac{n-3}{2}\right]$, respectively.

The system of equations (\ref{1stratsercoeff}) gives rise to infinitely many
other constraints between $p_2$ and $p_3$. For instance, subtracting the
formulae (\ref{1strat34}) and (\ref{1strat25}), expressing $p_4,p_5,p_6$ via 
$p_2$ and $p_3$, one gets 
\begin{equation}  \label{1stratC7}
\begin{split}
\mathcal{C}_7=& p_2p_5-p_3p_4 +\dots \\
=& H^2_{-1}{p_3}^2-H^2_{-1}{p_2}^3-3\,{H^2_{{-1}}}^{2}p_2p_3 +2\,H^2_{{-1}%
}H^3_{{-1}}{p_2}^2+ (-3\,H^2_{{1}}H^2_{{-1}}-{H^2_{{-1}}}^{2}H^3_{{-1}} -{%
H^2_{{-1}}}^{4})p_3 \\
& +(2\,H^2_{{-1}}H^3_{{1}}-3\,H^2_{{-1}}H^2_{{2}} -3\,H^2_{{1}}{H^2_{{-1}}}%
^{2}+{H^2_{{-1}}}^{3}H^3_{{-1} }+H^2_{{-1}}{H^3_{{-1}}}^{2})p_2+ 2\,H^2_{{-1}%
}H^3_{{-1}}H^3_{{1}} \\
&-3\,{H^2_{{-1}}}^{ 2}H^2_{{3}}+3\,{H^2_{{-1}}}^{3}H^3_{{1}}+3\,{H^2_{ {-1}}}%
^{2}H^3_{{2}}-3\,H^2_{{2}}{H^2_{{-1}}}^{3} -{H^2_{{-1}}}^{2}H^3_{{-1}}H^2_{{1%
}}-3\,H^2_{{4}}H^2 _{{-1}} \\
&+2\,H^2_{{-1}}H^3_{{3}}-3\,H^2_{{-1}}{H^2_{ {1}}}^{2}-4\,H^2_{{-1}}H^2_{{2}%
}H^3_{{-1}} =0.
\end{split}%
\end{equation}
It is easy to see that 
\begin{equation}
\mathcal{C}_7=H^2_{-1}\mathcal{F}^1_{23},
\end{equation}
i.e. the constraint (\ref{1stratC7}) is satisfied due to the constraint (\ref%
{1stratellcurv}).\newline
One observes that other simple constraints obtained in such a way have
similar properties 
\begin{equation}  \label{strat1-C8C9}
\begin{split}
\mathcal{C}_8=& {p_4}^2-p_{3}p_5+\dots={p_2}^4-{p_3}^2p_2+\dots = -p_2 
\mathcal{F}^1_{23}, \\
\mathcal{C}_9=& {p_3}p_6-p_{4}p_5+\dots={p_3}^3-{p_2}^3{p_3}+\dots = \left(
p_{{3}}-3\,H^2_{{-1}}p_{{2}}-3\,H^2_{{1}}-{H^2 _{{-1}}}^{3}-H^2_{{-1}}H^3_{{%
-1}} \right) \mathcal{F}^1_{23}.
\end{split}%
\end{equation}
For other examples see Appendix \ref{App-1strat}.

In general one has the following

\begin{lem}
\label{lemC6} For any constraint 
\begin{equation}  \label{effe}
f(p_2,p_3)=0
\end{equation}
arising from the system (\ref{1stratsercoeff}) the polynomial $f(p_2,p_3)$
is in the ideal generated by $\mathcal{F}^1_{23}$.
\end{lem}

\textbf{Proof} Let us assume that $f(p_2,p_3)$ is not in the ideal generated
by $\mathcal{F}^1_{23}$, i.e. 
\begin{equation}  \label{effeqG}
f(p_2,p_3)=q(p_2,p_3) \mathcal{F}^1_{23}+R(p_2,p_3)
\end{equation}
where $q(p_2,p_3)$ is certain polynomial and the rest $R(p_2,p_3)$ is not
identically zero. Since $R(p_2,p_3)=f(p_2,p_3)|_{\mathcal{F}^1_{23}=0}$ the
rest $R(p_2,p_3)$ has the form 
\begin{equation}  \label{Gpol}
R(p_2,p_3)=A(p_2)+B(p_2)p_3
\end{equation}
where $A$ and $B$ are certain polynomials. So our assumption, due to (\ref%
{effe},\ref{effeqG},\ref{Gpol}), is equivalent to the existence of nonzero $%
A $ and $B$ such that 
\begin{equation}  \label{restozero}
A(p_2)+B(p_2)p_3=0.
\end{equation}
The point is that such polynomials $A$ and $B$ do not exist. Indeed the
l.h.s. of (\ref{restozero}), is a polynomial in $p_2$ and $p_3$ of certain
degree and, hence, can be written as $\sum_{k=0,2,3,\dots}^n \gamma_k p_k$.
Since $p_0,p_2,p_3,p_4,\dots$ are elements of a basis in ${\Sigma_1}$ then
the condition (\ref{restozero}) is satisfied iff all $\gamma_k=0$. One
arrives to the same conclusion considering the representation of $p_2$ and $%
p_3$ as Laurent series (\ref{1stratser}). $\square$

This lemma leads to

\begin{prop}
Algebra $A_{\Sigma_1}$ at fixed $H^j_k$ is equivalent to the algebra $%
\mathbb{C}[p_2,p_3]\slash \mathcal{F}^1_{23}$.
\end{prop}

Similar to $\Sigma_\varnothing$ one can treat $p_2(z),p_3(z),...$ for given $%
H^i_k$ and variable $z$ as the local coordinates in $\Sigma_1$. In such
interpretation the condition (\ref{1stratalg}) becomes constraints on the
coordinates and one has

\begin{prop}
\label{tower-1strat} The stratum $\Sigma _{1}$ contains the subset $W_{1}$
for which the corresponding \ subbundle \ $TB_{W_{1}}$ is an infinite family 
$\Gamma _{\infty }^{1}$ of infinite dimensional varieties which are
intersection of the quadrics 
\begin{equation}
\begin{split}
f_{ij}^{(1)}=&
p_{i}p_{j}-p_{i+j}-H_{-1}^{j}p_{i+1}-H_{-1}^{i}p_{j+1}-%
\sum_{l=2}^{i-1}H_{i-l}^{j}p_{l}-\sum_{l=2}^{j-1}H_{j-l}^{i}p_{l} \\
& -H_{-1}^{i}H_{-1}^{j}p_{2}-\left(
H_{j}^{i}+H_{i}^{j}+H_{-1}^{i}H_{1}^{j}+H_{1}^{i}H_{-1}^{j}\right) =0,\quad
i,j=2,3,4,\dots .
\end{split}
\label{1stratquadcurv}
\end{equation}%
and parameterized by the variables $H_{k}^{j}$ $(j=2,3,\dots )$ obeying the
algebraic equations (\ref{1stratsercoeff}). This family $\Gamma _{\infty
}^{1}$ is the infinite tower of algebraic curves of genus $1$ with the
elliptic curve in the base.
\end{prop}

\textbf{Proof} As it was shown above the relations (\ref{1stratquadcurv})
are equivalent to the following 
\begin{equation}  \label{1strat-h1n}
\begin{split}
\mathcal{F}^1_{23}&=0, \qquad h^{(1)}_n=p_n-\alpha_n(p_2)-\beta_n(p_2)p_3
=0, \qquad n=4,5,6,\dots .
\end{split}%
\end{equation}
So, in the subspace $(p_2,p_3)$ one has, for given $H^i_j$ an elliptic curve
which generically has genus $1$. In the three dimensional space $p_2,p_3,p_4$
one has a curve which is the intersection on the cylindrical surface
generated by the elliptic curve and the quadric 
\begin{equation}
h^{(1)}_4=p_4-{p_{{2}}}^{2}+2\,H^2_{{-1}}p_{{3}}+{H^2_{{-1}}}^{2}p_{{2}%
}+2\,H^2_{{-1}}H^2_{{1}}+2\,H^2_{{2}}.
\end{equation}
In the $d$-dimensional subspace one has the curve with the ideal 
\begin{equation}
I(\Gamma_d^1)=\langle \mathcal{F}^1_{23},h^{(1)}_4,h^{(1)}_5,%
\dots,h^{(1)}_{d+1}\rangle.
\end{equation}
$\square$

Moduli $g_2$, $g_3$ (see e.g. \cite{HC,Sil}) of the elliptic curves (\ref%
{1stratellcurv}) are equal to 
\begin{equation}  \label{1stratg2g3}
\begin{split}
g_2=& \frac{3}{2}\,H^2_{{1}}H^2_{{-1}}-\frac{1}{3}\,{H^3_{{-1}}}^{2}-3\,H^2_{%
{2}} -\frac{1}{2}\,{H^2_{{-1}}}^{2}H^3_{{-1}}+2\,H^3_{{1}}-\frac{3}{16}\,{%
H^2_{{-1}}}^{4}, \\
g_3=&2\,H^3_{{3}}-3\,H^2_{{4}}-\frac{3}{4}\,{H^2_{{1}}}^{2} +\frac{2}{3}%
\,H^3_{{-1}}H^3_{{1}}+\frac{3}{2}\,H^2_{{-1}}H^2_{{3}} +H^2_{{-1}}H^2_{{1}%
}H^3_{{-1}}-2\,H^3_{{-1}}H^2_{{2 }}-\frac{3}{4}\,H^2_{{2}}{H^2_{{-1}}}^{2} \\
&-{\frac {2}{27}}\,{H^3_{{-1}}}^{3}-\frac{1}{8}\,{H^2_{{-1}}}^{4}H^3_{{-1}}-%
\frac{1}{6}\,{H^2_ {{-1}}}^{2}{H^3_{{-1}}}^{2}+\frac{3}{8}\,{H^2_{{-1}}}%
^{3}H^2_{{\ 1}}-\frac{1}{32}\,{H^2_{{-1}}}^{6}
\end{split}%
\end{equation}
and the $J$-invariant is $J=1728\frac{4{g_2}^3}{\Delta}$ where the
discriminant $\Delta=-16\,(4{g_{{2}}}^{3}+27\,{g_{{3}}}^{2})$ is given in
the Appendix \ref{App-1strat}.

It follows from equations (\ref{1stratsercoeff}) that all $H^i_j$ can be
expressed (polynomially) in terms of $H^2_i$ $(i=-1,1,2,3,\dots)$, $%
H^3_{-1},H^3_{1}$, and $H^3_{3}$. For instance 
\begin{equation}
H^3_2=\frac{3}{2}\,H^2_{{3}}-\frac{1}{2}\,H^2_{{-1}}H^3_{{1}}+\frac{1}{2}%
\,H^3_{{-1}}H^2_{{1}}.
\end{equation}
Thus the family of curves $\Gamma^1_{\infty}$ is parameterized by $%
H^2_{-1},H^2_1,H^2_2,\dots, H^3_{-1},H^3_1,$ and $H^3_3$.

We emphasize that the elliptic curve $\mathcal{F}_{23}^{1}$ and its
coordinate ring correspond to a point in $W_{1}$. Hence, in the stratum $%
\Sigma _{1}$ one may refer to such a point as an \textit{elliptic point} and 
$W_{1}$ as an \textit{elliptic subset} of $\Sigma _{1}$.

\begin{prop}
Index$({\overline{\partial}_{W_1}})=-1$.
\end{prop}

\textbf{Proof } Since $S_{W_1}=\mathbb{N}-\{1\}$, then $S_{\widetilde{W}%
_1}=\{-1,1,2,3,\dots \}$. Hence, Index$({\overline{\partial}_{W_1}})=$card$%
(\{\varnothing\})$-card$(\{-1\})$=-1. $\square$

Coordinate ring of the elliptic curve (\ref{1stratellcurv-red}) contains
various higher degree singular algebraic curves. One of the examples is the
singular hyperelliptic curve 
\begin{equation}  \label{1F25}
\begin{split}
\mathcal{F}^1_{25}=&{p_5}^2-{p_2}^5+5\,H^2_{{-1}}{p_2}^2p_5+\left(5\,H^2_{{1}%
}+5\,{H^2_{{-1}}}^{3}\right)p_2p_5+\left(-2\,{H^2_{{-1}}}^{2}H^3_{{-1}%
}+11\,H^2_{{-1}}H^2_{ {1}}+2\,{H^3_{{-1}}}^{2}\right. \\
&\left.-2\,{H^2_{{-1}}}^{4}+3\,H^2_{{2 }}-2\,H^3_{{1}} \right){p_2}%
^3+\left(-H^2_{{-1}}{H^3_{{-1}}}^{2}+{H^2_{{-1}}}^{3}H^3_{{\ -1}}+2\,H^2_{{1}%
}{H^2_{{-1}}}^{2}-4\,H^2_{{2}}H^2_ {{-1}}\right. \\
&\left.+H^2_{{-1}}H^3_{{1}}+2\,{H^2_{{-1}}}^{5} +5\,H^2_{{3}%
}\right)p_5+D_4\,p_4+D_2\,p_2 +D_0=0
\end{split}%
\end{equation}
whose coefficients are given in Appendix \ref{App-1strat}. This plane
quintic has genus $1$ and 
\begin{equation}
\begin{split}
&\mathcal{F}^1_{25}=\left(p_{{5}}+4\,H^2_{{-1}}{p_{{2}}}^{2} + \left( -H^2_{{%
-1}}H^3_{{-1}}+p_{{3}}+6\,{H^2_{{-1}}}^{3}+4\,H^2_{{1}} \right) p_{{2}}+%
\frac{1}{2}\,H^2_{{-1}}H^3_{{1}}-H^2_{{-1}}{\ H^3_{{-1}}}^{2}\right. \\
&\left.+{H^2_{{-1}}}^{3}H^3_{{-1}}+4\,H^2_ {{1}}{H^2_{{-1}}}^{2}-2\,H^2_{{2}%
}H^2_{{-1}}+\frac{5}{2}\,\mathit{H2}_{{3}}-p_{{3}}H^3_{{-1}}-\frac{3}{2}%
\,H^3_{{-1}}H^2_{{1}}+ 2\,p_{{3}}{H^2_{{-1}}}^{2}+2\,{H^2_{{-1}}}%
^{5}\right)h_5^{(1)} \\
& +\left({p_{{2}}}^{2}+ \left( -2\,H^3_{{-1}}+4\,{H^2_{{-1}}}^{2} \right) p_{%
{2}}-4\,{H^2_{{-1}}}^{2}H^3_{{-1}}+{H^3_{ {-1}}}^{2}+4\,{H^2_{{-1}}}%
^{4}\right)\mathcal{F}^1_{23}.
\end{split}%
\end{equation}
Different type of curve contained in the family $\Gamma^1_{\infty}$ is given
,for instance, by the trigonal curve 
\begin{equation}  \label{1stratTrig}
\begin{split}
&\mathcal{F}^{1}_{34}= \\
&{p_4}^3-{p_3}^4 +4\,H^3_{{-1}}{p_3}{p_4}^2 +\left(3\,{H^2_{{-1}}}%
^{3}+6\,H^2_{{-1}}H^3_{{-1}} -6\,H^2_{{1}}\right){p_3}^3+\left(-2\,{H^3_{{-1}%
}}^{2}+4\,H^3_{{1}}\right){p_4}^2 \\
& +\left(-6\,H^2_{{-1}}H^2_{{2}}+3\,H^2_{{1}}{H^2_{{-1}}}^{2} +4\,H^2_{{-1}%
}H^3_{{1}}+12\,H^3_{{-1}}H^2_{{1}}- 5\,H^3_{{-1}}{H^2_{{-1}}}^{3}-10\,H^2_{{%
-1}}{H^3_{{-1}}}^{2}\right){p_3}p_4 \\
& +\left(-3\,{H^2_{{-1}}}^{6}-12\,{H^2_{{-1}}}^{4}H^3_{{-1}} +12\,{H^2_{{-1}}%
}^{3}H^2_{{1}} -12\,{H^2_{{-1}}}^{2}{H^3_{{-1}}}^{2}+27\,H^2_{{-1}}H^2_{{1}%
}H^3_{{-1}} -15\,{H^2_{{1}}}^{2} \right. \\
&\left. +3\,H^2_{{3}}H^2_{{-1}}-6\,H^2_{{4}} -3\,{H^2_{{-1}}}^{2}H^3_{{1}%
}+3\,H^2_{{2}}{H^2_{{-1}}}^{2} +4\,H^3_{{-1}}H^3_{{1}}+4\,H^3_{{3}}\right){%
p_3}^2 \\
& +A_4p_4\, +A_3 p_3\, +A_0=0
\end{split}%
\end{equation}
where $A_4,A_3$ and $A_0$ are given in the Appendix \ref{App-1strat}. This
plane curve has genus $1$. For this curve one has 
\begin{equation}  \label{1F34-ideal}
\mathcal{F}^1_{34} =\left({p_4}^2+ap_4p_3+b{p_{3}}^2+cp_4+dp_3+f
\right)h_4^{(1)}+ \left(-{p_3}^2+h p_3 + j \right)\mathcal{F}^1_{23}
\end{equation}
where the coefficients $a,b,c,d,f,h,j$ are given in Appendix \ref{App-1strat}%
.

Finally a projection of the curve in the four dimensional space $%
(p_2,p_3,p_4,p_5)$ defined by the equation 
\begin{equation}
\mathcal{F}^1_{23}=0, \qquad h^{(1)}_4=0, \qquad h^{(1)}_5=0
\end{equation}
to the subspace $(p_4,p_5)$ is given by (4,5) curve 
\begin{equation}  \label{1F45}
\mathcal{F}^1_{45}={p_5}^4-{p_4}^5+\dots=0
\end{equation}
where the coefficients are too long for writing them here. They can be
easily computed by an algebraic manipulator. The curve (\ref{1F45}) is
singular and has genus one.


\section{Stratum $\Sigma_1$. Weierstrass function reduction}

Here we will study the reduction of the system (\ref{1stratalg})-(\ref%
{1stratsercoeff}) associated with the celebrated Weierstrass $\wp$-function
given by the series (see e.g. \cite{HC,Sil}) 
\begin{equation}
\wp(u)=\frac{1}{u^2}+\sum_{n=2}^\infty c_n u^{2n-2}
\end{equation}
where the coefficients $c_n$ are defined by the recurrence relation 
\begin{equation}
c_n=\frac{1}{(n-3)(2n+1)}\sum_{k=2}^{n-2}c_kc_{n-k}, \qquad n=4,5,6,\dots\ .
\end{equation}
The Weierstrass function $\wp(u)$ and its derivative $\wp^{\prime }(u)$ obey
the equation 
\begin{equation}  \label{WPeqn}
\wp^{\prime 2}=4\wp(u)^3-g_2 \wp (u)-g_3.
\end{equation}
where $g_2=20 c_2$ and $g_3=28c_3$. Equation (\ref{WPeqn}) clearly indicates
that equation (\ref{1stratellcurv-red}) $\mathcal{F}^1_{23}=0$ should admit
the reduction for which $p_2$ and $p_3$ are connected with $\wp(u)$ and $%
\wp^{\prime }(u)$, respectively. It is indeed the case and for such a
reduction 
\begin{equation}  \label{p23wp}
p_2(z)=\wp(1/z)\qquad \mathrm{and} \qquad p_3(z)=-\frac{1}{2}\wp^{\prime
}(u)|_{u=z^{-1}},
\end{equation}
i.e. $H^2_{2n}=c_{n+1}$, $n=1,2,3,\dots$, $H^2_{2m+1}=0$, $m=-1,0,1,\dots$, $%
H^{3}_{k}=-\frac{k}{2}H^{2}_{k}$, $k=-1,0,1,\dots$ . Then it is a
straightforward check that the whole system (\ref{1stratalg})-(\ref%
{1stratsercoeff}) admits the reduction 
\begin{equation}  \label{WPred}
\begin{split}
p_{n+2}=-\frac{1}{(n+1)!}\partial_u^n \wp(u)|_{u=z^{-1}}, \qquad n \ odd \\
p_{n+2}=\frac{1}{(n+1)!}\partial_u^n \wp(u)|_{u=z^{-1}}-\frac{1}{n+1}%
c_{n/2+1}, \qquad n \ even .
\end{split}%
\end{equation}
Under this reduction equations (\ref{1strat22})-(\ref{1strat34}) take the
form 
\begin{equation}
\begin{split}
p_2p_2=& p_4 +\frac{1}{10}g_2, \\
p_2p_3=& p_5, \\
p_3p_3=& p_6 -\frac{g_2}{10}p_2-\frac{g_3}{7}, \\
p_2p_4=& p_6 +\frac{g_2}{20}p_2+\frac{3g_3}{28}, \\
\dots &.
\end{split}%
\end{equation}
Due to (\ref{WPred}) they are nothing else than the classical equations (see
e.g. \cite{HC,Sil}) for the Weierstrass function $\wp(u)$ 
\begin{equation}  \label{wprel}
\begin{split}
\wp^{\prime \prime 2}-\frac{g_2}{2}, \\
\wp^{\prime \prime \prime }=&12\wp \wp^{\prime }, \\
\wp^{\prime \prime \prime \prime }=&30 {\wp^{\prime }}^2+12g_2\wp+18 g_3, \\
\wp^{\prime \prime \prime \prime }=& 20 \wp^{\prime \prime }\wp -8 g_2 \wp
-12 g_3, \\
\dots\ . &
\end{split}%
\end{equation}
In particular, the last two equations (\ref{wprel}) and the first one give
equation (\ref{WPeqn}). One can observe also that the formula (\ref%
{1stratunique}) under this reduction becomes the well known expression (see
e.g. \cite{HC}) 
\begin{equation}  \label{WPpoly}
\alpha_n(\wp(u))+\beta_n(\wp(u))\wp^{\prime }(u)
\end{equation}
for an entire elliptic function.

Pure algebraic characterization of the reduction (\ref{WPred}) is an
interesting open problem.

For the Weierstrass reduction 
\begin{equation}  \label{WP2P5}
p_2=\wp(u)|_{u=1/z}, \qquad p_5=-\frac{1}{24}\wp^{\prime \prime \prime
}(u)|_{u=1/z}
\end{equation}
the hyperelliptic curve (\ref{1F25}) has the form 
\begin{equation}  \label{WP125}
{\wp^{\prime \prime \prime }(u)}^2-576{\wp(u)}^5+144g_2{\wp(u)}^3+144g_3{%
\wp(u)}^2=0,\qquad u=1/z
\end{equation}
which is a consequence of the formula (\ref{WPeqn}). It has, obviously,
genus one. The formula (\ref{WP2P5}) reproduces the well known
parametrization of the fifth degree hyperelliptic curve of genus one in
terms of the Weierstrass-$\wp$ function.

Weierstrass reduction (\ref{WPred}) of the trigonal curve (\ref{1stratTrig})
is given by 
\begin{equation}
\begin{split}
p_3=&-\frac{1}{2}\wp^{\prime }(u)\Big{|}_{u=z^{-1}}, \\
p_4=&\frac{1}{6} \wp^{\prime \prime }(u)\Big{|}_{u=z^{-1}}-\frac{g_2}{60}
=\wp^2(u)\Big{|}_{u=z^{-1}}+\frac{g_2}{10}
\end{split}%
\end{equation}
while the curve (\ref{1stratTrig}) takes the form 
\begin{equation}  \label{trigwp}
\left({\wp^{\prime }(u)}^2+g_3\right)^2=\frac{2}{27}\left(\wp^{\prime \prime
}(u)-g_2\right)^2\left(\wp^{\prime \prime }(u)+\frac{g_2}{2}\right).
\end{equation}
Finally, for the Weierstrass reduction, i.e. for 
\begin{equation}  \label{p4p5WP}
\begin{split}
p_4=&\frac{1}{6}\wp^{\prime \prime }(u)|_{u=z^{-1}}-\frac{g_2}{10}, \\
p_5=&-\frac{1}{24}\wp^{\prime \prime \prime }(u)|_{u=z^{-1}}
\end{split}%
\end{equation}
the curve (\ref{1F45}) is 
\begin{equation}
\begin{split}
&3(\wp^{\prime \prime \prime 4 }- 128(\wp^{\prime \prime 5}+64\,g_{{2}%
}(\wp^{\prime \prime 4}+144\,g_{{3}}(\wp^{\prime \prime \prime 2}\wp^{\prime
\prime }(u) +160\,{g_{{2}}}^{2}(\wp^{\prime \prime 3} \\
&+72\,g_{{2}}g_{{3}}(\wp^{\prime \prime \prime 2}+(-16\,{g_{{2}}}^{3} +1728\,%
{g_{{3}}}^{2})(\wp^{\prime \prime 2}+(-64\,{g_{{2}}}^{4}+1728\,{g_{{3}}}%
^{2}g_{{2}})\wp^{\prime \prime }(u) \\
&+432\,{g_{{3}}}^{2}{g_{{2}}}^{2}-16\,{g_{{2}}}^{5}=0.
\end{split}%
\end{equation}
It is a straightforward check that this equation is satisfied due to
equations (\ref{wprel}) and (\ref{1stratTrig}). The formula (\ref{p4p5WP})
gives us a parameterization of the genus one $(4,5)$ curve (\ref{1F45}) in
terms of Weierstrass $\wp$-function.

In a similar manner one can get Weierstrass function parameterization of $%
(n,n+1)$ curves $n=5,6,7,\dots$ of genus one. 

\section{Stratum $\Sigma_{1,2}$. Trigonal curve of genus two}

Now we will consider the stratum $\Sigma_{1,2}$ which corresponds to the set 
$S=(-2,-1,0,3,4,\dots)$. First and second degree elements are absent and,
hence, the positive degree elements of the canonical basis are 
\begin{equation}  \label{2stratLaurbas}
\begin{split}
p_0&=1+\sum_{k=1}^{\infty} \frac{H^0_{k}}{z^k}, \\
p_i&=z^i+H^i_{-2}z^2+H^i_{-1}z+\sum_{k=1}^{\infty}\frac{H^i_{k}}{z^k},
\qquad i=3,4,5,\dots\ .
\end{split}%
\end{equation}
In this case $(p_{-2})^2,(p_{-1})^2 \notin \langle
p_i\rangle_{i=-2,-1,2,3,\dots}$, and, hence, only the elements of the
positive degree can be involved in the subset closed with respect to
pointwise multiplication.

\begin{lem}
\label{lem2stratalg} Laurent series (\ref{2stratLaurbas}) obey the equations 
\begin{equation}  \label{2stratalg}
p_n(z) p_m(z)=\sum_{l=0,3,4,\dots}C_{nm}^l p_l(z), \qquad n,m=0,3,4,5,\dots
\end{equation}
if and only if 
\begin{equation}
H^0_k=0,\qquad k=1,2,\dots
\end{equation}
and 
\begin{equation}  \label{alg2-comp}
\begin{split}
&H^m_{n+l}+H^n_{m+l}+\sum_{i=1}^2
(H^n_{-i}H^m_{l+i}+H^m_{-i}H^n_{l+i})+\sum_{i=1}^{l-1}
(H^n_{i}H^m_{l-i}+H^m_{i}H^n_{l-i}) \\
&=H^{n+m}_l+\sum_{i=m+1}^{m+2} H^n_{m-i} H^{i}_l+\sum_{i=3}^{m-1} H^n_{m-i}
H^{i}_l +\sum_{i=n+1}^{n+2}H^m_{n-i} H^{i}_l+\sum_{i=3}^{n-1} H^m_{n-i}
H^{i}_l+H^n_m+H^m_n \\
&+H^n_{-2}H^m_{-2}H^{4}_l+(H^n_{-1}H^m_{-2}+H^n_{-2}H^m_{-1})H^{3}_l+%
\sum_{i=1}^2(H^n_{-i}H^m_i+H^m_{-i}H^n_i)\delta^0_l, \qquad n,m=3,4,5,\dots\
.
\end{split}%
\end{equation}
The constants $C^l_{ij}$ have the form 
\begin{equation}  \label{2stratstructconst}
\begin{split}
C^l_{nm}=&\delta_{n+m}^l+\sum_{i=m+1}^{m+2} H^n_{m-i}
\delta_i^l+\sum_{i=3}^{m-1} H^n_{m-i} \delta_i^l
+\sum_{i=n+1}^{n+2}H^m_{n-i} \delta_i^l+\sum_{i=3}^{n-1} H^m_{n-i}
\delta_i^l +H^n_{-2}H^m_{-2}\delta_4^l \\
&+(H^n_{-1}H^m_{-2}+H^n_{-2}H^m_{-1})\delta_3^l+(H^n_m+H^m_n)\delta_0^l+%
\sum_{i=1}^2(H^n_{-i}H^m_i+H^m_{-i}H^n_i) \delta_0^l, \\
& n,m=3,4,5,\dots\ .
\end{split}%
\end{equation}
\end{lem}

An analog of Proposition \ref{prop1strat} is

\begin{prop}
\label{prop2strat} The stratum $\Sigma _{1,2}$ contains the subset $W_{1,2}$
of points of codimension $2$ for which the corresponding subbundle $%
TB_{W_{1,2}}$ is formed by fibers closed with respect to the pointwise
multiplication$.$ These fibers  are vector spaces with basis $\langle
p_{i}\rangle _{i}$ with $H_{k}^{i}$ obeying the condition (\ref{alg2-comp}).
The subbundle $TB_{W_{1,2}}$ is the infinite family of infinite-dimensional
associative algebras $A_{\Sigma _{1,2}}$ with the basis $%
(1,p_{3},p_{4},p_{5},\dots )$ and the structure constants given by (\ref%
{2stratstructconst}).
\end{prop}

Analysis of the equations (\ref{2stratalg}), i.e. the equations 
\begin{equation}
\begin{split}
{p_3}^2=&p_{{6}}+2\,H^3_{{-2}}p_{{5}} + \left( 2\,H^3_{{-1}}+{H^3_{{-2}}}%
^{2} \right) p_{{4}}+2\,H^3_{{-1}}H^3 _{{-2}}p_{{3}}+2\,H^3_{{2}}H^3_{{-2}}
+2\,H^3_{{1}}H^3_{{-1}}+2\,H^3_{{3}}, \\
p_{3}p_{4}=&p_{{7}}+H^3_{{-2}}p_{{6}}+ \left( H^4_{{-2}} +H^3_{{-1}} \right)
p_{{5}}+ \left( H^3_{{-2}}H^4_{{-2} }+H^4_{{-1}} \right) p_{{4}}+ \left(
H^3_{{1}} +H^3_{{-1}}H^4_{{-2}}+H^3_{{-2}}H^4_{{-1}} \right) p_{{3}} \\
& +H^3_{{2}}H^4_{{-2}}+H^3_{{1}}H^4_{{-1}}+H^3_{{-1}} H^4_{{1}}+H^3_{{-2}%
}H^4_{{2}}+H^3_{{4}}+H^4_{{3}}, \\
\dots
\end{split}%
\end{equation}
shows that $A_{\Sigma_{1,2}}$ for fixed $H^j_k$ is the polynomial algebra
generate by four elements $1,p_3,p_4,p_5$.

Analogously to $A_{\Sigma_1}$ these generators are not free and obey certain
constraints. Considering equations (\ref{2stratalg}) with $i+j=8$, $i+j=9$,
and $i+j=10$, one gets the following constraints 
\begin{equation}  \label{2stratC8}
\begin{split}
\mathcal{C}_8=&p_3p_5-{p_4}^2- H^3_{{-2}}p_3p_4 - \left( H^3_{{-1}}-2\,H^4_{{%
-2}}-{H^3_{{-2}}}^{2} \right) {p_3}^{2} \\
&- \left( H^5_{{-2}}-2\,H^4_{{-1}}+3\,H^3_{{-2}}H^4 _{{-2}}-3\,H^3_{{-2}%
}H^3_{{-1}}+2\,{H^3_{{-2}}}^{3} \right) p_5 \\
&- \left( H^3_{{-2}}{H^5}_{{-2}}+H^5_{{-1}}+H^3_{{1}}-{H^4_{{-2}}}^{2}+{\
H^3_{{-2}}}^{2}H^4_{{-2}}-H^3_{{-2}}H^4_{{-1}}-2\,{\ H^3_{{-1}}}^{2}+H^3_{{-1%
}}{H^3_{{-2}}}^{2} \right. \\
&\left.+4\,H^4_{{-2}}H^3_{{-1}}+{H^3_{{-2}}}^{4} \right) p_4- \left( H^3_{{-1%
}}H^5_{{-2}}+H^3_{{-2}}H^5_ {{-1}}+H^3_{{2}}-2\,H^4_{{1}}-2\,H^4_{{-1}}H^4_{{%
- 2}}-H^3_{{-2}}H^3_{{1}}\right. \\
&\left.+3\,H^3_{{-2}}H^3_{{-1}}{\ H^4}_{{-2}}-{H^3_{{-2}}}^{2}H^4_{{-1}%
}-2\,H^3_{{-2}} {H^3_{{-1}}}^{2}+2\,{H^3_{{-2}}}^{3}H^3_{{-1}} \right) p_3 \\
&+2\,H^3_{{-1}}H^3_{{-2}}H^3_{{2}}+H^3_{{-2 }}H^3_{{1}}H^4_{{-1}}+H^3_{{-2}%
}H^3_{{-1}}H^4 _{{1}}-3\,H^3_{{-2}}H^3_{{2}}H^4_{{-2}} -H^3_{{5}}-H^3_{{-1}%
}H^5_{{1}}-H^3_{{2}}H^5_{{-2}} \\
&-H^3_{{1}}H^5_{{-1}} +{H^3_{{-2}}}^{2}H^4_{{2}}+H^3_{{-2}}H^3_{{4}}+H^3_{{-2%
}}H^4_{{3}} -H^5_{{3}}+2\,H^4_{{4}}-H^3_{{-2}}{\ H^5}_{{2}}-2\,{H^3_{{-2}}}%
^{2}H^3_{{3}}+2\,H^4_{{2}} H^4_{{-2}} \\
&+2\,H^4_{{1}}H^4_{{-1}}+2\,H^3_{{1}}{{\ H^3}_{{-1}}}^{2} +2\,H^3_{{3}}H^3_{{%
-1}}-4\,H^4_{{-2} }H^3_{{3}}-2\,{H^3_{{-2}}}^{3}H^3_{{2}}-4\,H^4_{{- 2}}H^3_{%
{1}}H^3_{{-1}} \\
&-2\,{H^3_{{-2}}}^{2}H^3_{{1} }H^3_{{-1}}=0
\end{split}%
\end{equation}
and 
\begin{equation}  \label{2stratC9}
\begin{split}
\mathcal{C}_9=&p_3p_6-p_4p_5+\dots \\
=&{p_3}^{3}-{p_4}\,{p_5}-3\,H^3_{{-2}}{p_3}\,{p_5}- \left( 3\,H^3_{{-1}%
}-H^4_{{-2}} \right) {p_3}\,{p_4} - \left( {H^3_{{-2}}}^{3}-H^5_{{-2}}-H^4_{{%
-1}}+{H^3 }_{{-2}}H^4_{{-2}} \right) {p_3}^{2} \\
&- \left( 3\,H^3_{{1}}+3 \,H^3_{{-1}}{H^3_{{-2}}}^{2}-H^5_{{-1}}-H^3_{{-2}}
H^5_{{-2}}-2\,H^4_{{-2}}H^3_{{-1}}+{H^4_{{-2}}}^{2 }-2\,{H^3_{{-2}}}%
^{4}+2\,H^3_{{-2}}H^4_{{-1}}\right. \\
&\left.-2\,{{\ H^3}_{{-2}}}^{2}H^4_{{-2}} \right) {p_5} +N_4\, p_4+N_3\,
p_3+N_0=0
\end{split}%
\end{equation}
and 
\begin{equation}  \label{2stratC10}
\begin{split}
\mathcal{C}_{10}=&{p_5}^2-p_4p_6+\dots \\
=& {p_{{5}}}^{2}-p_{{4}}{p_{{3}}}^{2}+2\,H^3_{{-2}}{p_{{3}}}^{3}- \left(
-H^4_{{-2}}-2\,H^3_{{-1}}+5\,{H^3_{{-2}}}^{2} \right) p_{{3}}p_{{5}} \\
&- \left( 2\,H^5_{{-2}}+6\,H^3_{{-1}}H^3_{{-2}}-H^4_{{-1}}-H^4_{{-2}}H^3_{{-2%
}}+{\ H^3_{{-2}}}^{3} \right) p_{{3}}p_{{4}} \\
&- \left( 2\,H^5_{{-1}}-H^4_{{-1}}H^3_{{-2}}-2\,H^3_{{1}}+{H^3_{{-1}}}^{2}-
H^4_{{-2}}H^3_{{-1}}+{H^3_{{-2}}}^{2}H^3_{{-1}}+{\ H^3_{{-2}}}^{4}-2\,H^3_{{%
-2}}H^5_{{-2}} \right) {p_{{3}} }^{2} \\
&- \left( -H^4_{{1}}+H^4_{{-2}}H^4_{{-1}}-4\,H^3_{{-2}}H^5_{{-1}}+8\,H^3_{{1}%
}H^3_{{-2}}-H^5_{{\ -2}}{H^3_{{-2}}}^{2}-H^5_{{-2}}H^4_{{-2}}-5\,H^4_{ {-2}%
}H^3_{{-1}}H^3_{{-2}}\right. \\
&\left.-2\,H^3_{{2}}-2\,{H^3_{{-2}}}^{5}-2\,{H^3_{{-1}}}^{2}H^3_{{-2}}+{H^4_{%
{-2}}}^{2} H^3_{{-2}}-H^4_{{-2}}{H^3_{{-2}}}^{3}-H^4_{{-1}} H^3_{{-1}}+H^4_{{%
-1}}{H^3_{{-2}}}^{2}\right. \\
& \left. +3\,{H^3_{{-2} }}^{3}H^3_{{-1}} \right) p_{{5}} \\
& +B_4\, p_4+B_3\, p_3+B_0=0
\end{split}%
\end{equation}
where the coefficients $N_i$ and $B_i$ are given in Appendix \ref{App-2strat}%
. These three constraints are not independent since 
\begin{equation}
\begin{split}
&(p_{{3}}+2\,H^4_{-1}-H^5_{-2}-2\,{H^3_{-2}}^{3}-3\,H^3_{-2}H^4_{-2}+3%
\,H^3_{-1}H^3_{-2})\mathcal{C}_{10} \\
&+ (p_4-2\,H^3_{-2}p_{3} +2\,H^5_{-1}+3\,{H^3_{-2}}^{4}-2\,H^3_{1}-2\,{%
H^3_{-1}}^{2}-3\,H^3_{-2}H^4_{-1}-2\,{H^4_{-2}}^{2}+6\,H^4_{-2} H^3_{-1} \\
&+2\,H^3_{-2}H^5_{-2}+3\,H^4_{-2}{H^3_{-2}}^{2}-2\,H^3_{-1}{H^3_{-2}}^{2} )%
\mathcal{C}_9 \\
&+(-p_5+(-3\,H^3_{-1}+H^4_{-2})p_3 +(H^4_{-2}H^5_{-2}+3\,H^3_{-2}H^5_{-1}-3\,%
{H^3_{-1}}^{2}H^3_{-2}+2\,{H^3_{-2}}^{3}H^3_{-1} \\
&-2\,H^3_{-1}H^5_{-2}+{H^3_{-2}}^{5}+H^4_{1}-3\,H^3_{2}+5%
\,H^3_{-1}H^3_{-2}H^4_{-2} +{H^3_{-2}}^{3}H^4_{-2}+2\,H^5_{-2}{H^3_{-2}}%
^{2}-H^4_{-1}H^4_{-2} \\
&-{H^3_{-2}}^{2}H^4_{-1}-H^3_{-2}{H^4_{-2}}^{2}-3%
\,H^3_{1}H^3_{-2}+H^4_{-1}H^3_{-1}))\mathcal{C}_8 =0 .
\end{split}%
\end{equation}
There are infinitely many other constraints between $p_3,p_4$, and $p_5$.
Two of them are given by 
\begin{equation}
\begin{split}
\tilde{\mathcal{C}}_{10}=&p_3p_7-p_4p_6+\dots =-2H^3_{-2}\mathcal{C}%
_9+(2H^3_{-1}+{H^3_{-1}}^2)\mathcal{C}_8, \\
{\mathcal{C}}_{11}=&p_3p_8-p_5p_6+\dots =2\,H^3_{-2}\mathcal{C}_{10}+ \left(
-2H^3_{-1}-{H^3_{-2}}^{2} \right) \mathcal{C}_9.
\end{split}%
\end{equation}
An important constraint is given by 
\begin{equation}  \label{2strattrigcurv}
\begin{split}
\mathcal{C}_{12}=& p_4p_8-{p_6}^2+\dots \\
=&\mathcal{F}^2_{34}={p_4}^3-{p_3}^4 +4\,H^3_{{-2}}{p_3}{p_4}^2
+\left(-3\,H^4_{{-2}}+4\,H^3_{{-1}}+2\,{H^3_{{-2}}}^{2}\right){p_3}^2p_4 \\
& +\left(-3\,H^4_{{-1}}-2\,H^3_{{-2}}H^4_{{-2}}\right){p_3}^3 +Q_8 {p_{4}}^2
+Q_7p_4p_3+Q_6 {p_{3}}^2 +Q_4 p_4+Q_3p_3+Q_0=0.
\end{split}%
\end{equation}
The coefficients $Q_i$ of this trigonal curve are given in Appendix \ref%
{App-2strat}. One has 
\begin{equation}
\begin{split}
\mathcal{F}^2_{34} =&(p_{{4}}+3\,H^3_{{-2}}p_{{3}}+3\,H^3_{{-1}}{H^3_{{-2}}}%
^{2}+3\, H^3_{{1}}-H^5_{{-1}}-H^3_{{-2}}H^5_{{-2}}-2\,H^4_{{-2}}H^3_{{-1}} +{%
H^4_{{-2}}}^{2}-2\,{H^3_{{-2}}}^{4} \\
&+2\,H^3_{{-2}}H^4_{{-1}}-2\,H^4_{{-2}}{H^3_{{-2}}}^{2})\mathcal{C}_8 \\
& +(-p_{{3}}-2\,H^4_{{-1}}+3\,H^3_{{-2}}H^4_{{-2}}+2\,{H^3_{{-2}}}^{3} +H^5_{%
{-2}}-3\,H^3_{{-1}}H^3_{{-2}})\mathcal{C}_9.
\end{split}%
\end{equation}
i.e. $\mathcal{F}^2_{34}$ belongs to the ideal $\langle \mathcal{C}_8,%
\mathcal{C}_9 \rangle$.

Constraints (\ref{2stratC8}),(\ref{2stratC9}) and (\ref{2stratC10}) show
that any element of the algebra $A_{\Sigma_{1,2}}$ can be represented in the
form 
\begin{equation}  \label{2stratpn}
p_n=a_n(p_3)+b_n(p_3)p_4+c_n(p_3)p_5, \qquad n=3,4,5,\dots
\end{equation}
where $a_n,b_n,$ and $c_n$ are polynomials.

This observation allows us to prove the following

\begin{prop}
\label{propAS2} Algebra $A_{\Sigma_{1,2}}$ with fixed $H^j_k$ is equivalent
to the polynomial algebra $\mathbb{C}[p_3,p_4,p_5] / \langle \mathcal{C}_8,%
\mathcal{C}_9,\mathcal{C}_{10} \rangle $.
\end{prop}

The proof is based on the

\begin{lem}
For any constraint 
\begin{equation}
f(p_3,p_4,p_5)=0
\end{equation}
arising from the system of equations (\ref{2stratalg}), the polynomials $%
f(p_3,p_4,p_5)=0$ belong to the ideal generated by $\mathcal{C}_8,\mathcal{C}%
_9$ and $\mathcal{C}_{10}$.
\end{lem}

\textbf{Proof} The proof is similar to that of Lemma \ref{lemC6}. Indeed we
assume that $f$ does not belongs to the ideal $\langle \mathcal{C}_8,%
\mathcal{C}_9,\mathcal{C}_{10} \rangle$. Hence 
\begin{equation}
f(p_3,p_4,p_5)=q_8(p_3,p_4,p_5)\mathcal{C}_8+q_9(p_3,p_4,p_5)\mathcal{C}%
_9+q_{10}(p_3,p_4,p_5)\mathcal{C}_{10}+R(p_3,p_4,p_5)
\end{equation}
where $q_8,q_9,q_{10}$ are some polynomials and $R(p_3,p_4,p_5)$ is not
identically zero. Since 
\begin{equation}
R(p_3,p_4,p_5)=f(p_3,p_4,p_5)\Big{|}_{\mathcal{C}_8=\mathcal{C}_9=\mathcal{C}%
_{10}=0}
\end{equation}
the rest $R(p_3,p_4,p_5)$ has the form 
\begin{equation}
R(p_3,p_4,p_5)=A(p_3)+B(p_3)p_4+C(p_3)p_5
\end{equation}
where $A,B$ and $C$ are certain polynomials in $p_3$. Our assumption due to (%
\ref{2stratC8}-\ref{2stratC10}) is equivalent to the existence of nonzero $%
A,B$ and $C$ such that 
\begin{equation}  \label{2stratcondrest}
A(p_3)+B(p_3)p_4+C(p_3)p_5=0.
\end{equation}
Since the numbers $3n$,$3m+4$, and $3l+5$ for positive integers $n,m,l$
never coincide, the count of gradation or power of Laurent series shows that
the three terms in (\ref{2stratcondrest}) always have different degrees.
Consequently equation (\ref{2stratcondrest}) has no nontrivial solutions. $%
\square$

Similar to the previous section one can treat $p_{3}$,$p_{4}$,$p_{5},\dots $
for given $H_{k}^{i}$ and $z$ as the local affine coordinates in fibers of .$%
TB_{W_{1,2}}.$ Thus one has

\begin{prop}
\label{prop2tower} For the stratum $\Sigma _{1,2}$ \ the subbundle $%
TB_{W_{1,2}}$ contains an infinite family $\Gamma _{\infty }^{2}$ of
infinite-dimensional algebraic varieties defined by the quadrics 
\begin{equation}
\begin{split}
f_{nm}=&
p_{n+m}+\sum_{i=m+1}^{m+2}H_{m-i}^{n}p_{i}+\sum_{i=3}^{m-1}H_{m-i}^{n}p_{i}+%
\sum_{i=n+1}^{n+2}H_{n-i}^{m}p_{i}+%
\sum_{i=3}^{n-1}H_{n-i}^{m}p_{i}+H_{m}^{n}+H_{n}^{m} \\
&
+H_{-2}^{n}H_{-2}^{m}p_{4}+(H_{-1}^{n}H_{-2}^{m}+H_{-2}^{n}H_{-1}^{m})p_{3}+%
\sum_{i=1}^{2}(H_{-i}^{n}H_{i}^{m}+H_{-i}^{m}H_{i}^{n})=0, \\
& n,m=3,4,5,\dots 
\end{split}%
\end{equation}%
and varying with parameters $H_{m}^{n}$ obeying the algebraic equation (\ref%
{alg2-comp}). The prime ideal $I(\Gamma _{\infty }^{2})$ of $\Gamma _{\infty
}^{2}$ is generated by $\mathcal{C}_{8},\mathcal{C}_{9},\mathcal{C}_{10}$
and 
\begin{equation}
h_{n}^{(2)}=p_{n}-a_{n}(p_{3})-b_{n}(p_{3})p_{4}-c_{n}(p_{3})p_{5},\qquad
n=6,7,8,\dots \ .
\end{equation}
\end{prop}

\textbf{Proof } Proof is based on the equivalence of the set of equations (%
\ref{2stratalg}) to the equations (\ref{2stratC8}-\ref{2stratC10}) and (\ref%
{2stratpn}). The constraint $\mathcal{C}_{10}=0$ is necessary to guarantee
the irreducibility of varieties. For the discussion of this point in the
case of all $H^n_m=0$ see e.g. \cite{CLOS}. $\square$

The family $\Gamma^2_{\infty}$ of algebraic curves contains the plane
trigonal curve given by the equation $\mathcal{F}^2_{34}=0$. The polynomial $%
\mathcal{F}^2_{34}$ given by (\ref{2strattrigcurv}) is the standard $(3,4)$
trigonal polynomial defining trigonal curve \cite{BLE2}. But $\mathcal{F}%
^2_{34}$ is not a generic $(3,4)$ trigonal polynomial. Computer calculation
show that the curve $\mathcal{F}^2_{34}=0$ has genus two. The family $%
\Gamma^2_{\infty}$ includes also the plane $(4,5)$ curve $\mathcal{C}_{20}=%
\mathcal{F}^2_{45}=0$. It is too complicated to be presented here. In the
three dimensional space with coordinates $p_3,p_4,p_5$ equations (\ref%
{2stratC8}-\ref{2stratC10}) define an irreducible algebraic curve $\Gamma$.
It is the intersection of well-known surfaces. For instance, the surface
defined by the equation $\mathcal{C}_{10}=0$ is the celebrated Whitney
umbrella (see e.g. \cite{CLOS}). On the other hand equation $\mathcal{F}%
^2_{34}=0$ defines the cylindrical surface generated by the trigonal curve.
So, the curve $\Gamma$ is the intersection of the Whitney umbrella surface
(see e.g. \cite{Har}) and the cylindrical surfaces generated by the trigonal
curve. So one expects that the curve $\Gamma$ has genus two.

\begin{prop}
Index$(\overline{\partial}_{W_{1,2}})=-2$.
\end{prop}

\textbf{Proof } $S_{\widetilde{W}_{1,2}}=\{-2,-1,1,2,3,\dots\}$ and hence
card$\{S_{W_{1,2}}-\mathbb{N}\}=0$ and card$\{S_{\widetilde{W}_{1,2}}-%
\mathbb{N}\}=2$ $\square$

We note that similar to the first stratum for a generic curve $\Gamma$ one
has 
\begin{equation}  \label{indexconj}
\mathrm{genus}(\Gamma)+\mathrm{index}(\overline{\partial}_{W_{1,2}})=0.
\end{equation}
Now let us consider the stratum $\Sigma^*_2$ with $S_{\Sigma^*_2}=%
\{-1,0,1,3,\dots\}$. It has the codimension $3$, i.e. one half of codim$%
(\Sigma_{1,2})$. On the other hand a basis in $\Sigma^*_2$ contains an
element of the first degree. Hence in $\Sigma^*_2$ there is the canonical
basis given by formulae (\ref{2stratLaurbas}) with $H^i_{-1}$, $%
i=3,4,5,\dots $. Thus, the results formulated above in Lemmas (\ref%
{lem2stratalg}), (\ref{lemC6}) and Propositions (\ref{prop2strat}),(\ref%
{propAS2}) (\ref{prop2tower}) with $H^i_{-1}=0$, $i=3,4,5,\dots$ are valid
for the stratum $\Sigma^*_2$ too. In contrast, codim$(W^*_2)=1$ and index$(%
\overline{\partial}_{W^*_2})$= card$(\varnothing)$-card$(\{-2\})=-1$. 

\section{Higher strata. Plane $(n+1,n+2)$ curves}

For the higher strata all calculations and formulas become much more
involved. For the stratum $\Sigma_{1,2,3}$ with $S=\{-3,-2,-1,0,4,5,6,\dots%
\} $ the canonical basis have the form 
\begin{equation}  \label{3stratLaurser}
\begin{split}
p_0=&1+\sum_{k=1}^\infty\frac{H^0_k}{z^k}, \\
p_i=&z^i+\sum_{k=-3}^{-1}\frac{H^i_k}{z^k}+\sum_{k=1}^{\infty}\frac{H^i_k}{%
z^k}, \qquad i=4,5,6,\dots\ .
\end{split}%
\end{equation}
Again only the elements of positive degrees may be involved in 
$TB_{W_{1,2,3}}$.

The Laurent series (\ref{3stratLaurser}) obey the analogue of equations (\ref%
{2stratalg}) if $H_{k}^{0}=0$ and $H_{k}^{i},\ i=4,5,6,\dots $ satisfy a
system of quadratic equations analogue to (\ref{alg2-comp}). As a
consequence $\Sigma _{1,2,3}$ contains the subbundle $TB_{W_{1,2,3}}$ with
fibers closed with respect to multiplication.  This subbundle is the infinite
dimensional algebra $A_{\Sigma _{1,2,3}}$ with the basis $%
(1,p_{4},p_{5},p_{6},\dots )$. The algebra $A_{\Sigma _{1,2,3}}$ is
generated by four elements $p_{4},p_{5},p_{6},p_{7}$. These elements are not
free and obey several constraints. They can be obtained exactly in the same
manner as for $\Sigma _{1,2}$. The corresponding expressions are pretty
long. To give an idea of their form and number we will present them in the
case when all $H_{k}^{i}=0$, i.e. $p_{i}=z^{i}$. Since $p_{i}p_{j}=p_{i+j},$ 
$i,j=4,5,6,\dots $, the simplest constraints are 
\begin{equation}
\begin{split}
& C_{10}={p_{5}}^{2}-p_{4}p_{6},\quad C_{11}={p_{5}}p_{6}-p_{4}p_{7},\quad
C_{12}={p_{6}}^{2}-p_{4}p_{7},\quad \widetilde{C}_{12}=p_{7}p_{5}-{p_{4}}%
^{3}, \\
& C_{13}={p_{6}}p_{7}-{p_{4}}^{2}p_{5},\quad C_{14}={p_{7}}^{2}-{p_{4}}%
^{2}p_{6},\quad \widetilde{C}_{14}={p_{7}}^{2}-{p_{5}}^{2}p_{4}.
\end{split}
\label{3stratcurv}
\end{equation}%
First three of the constraints (\ref{3stratcurv}) are independent. Others
are not since 
\begin{equation}
\begin{split}
p_{4}\widetilde{C}_{12}=& p_{4}C_{12}+p_{6}C_{10}-p_{5}C_{11}, \\
p_{4}{C}_{13}=& p_{5}C_{12}-p_{6}C_{11}, \\
p_{4}C_{14}=& -p_{7}C_{11}+p_{6}\widetilde{C}_{12}={p_{6}}%
^{2}C_{10}-(p_{7}+p_{5}p_{6})C_{11}+p_{4}p_{6}C_{12}, \\
\widetilde{C}_{14}=& C_{14}-p_{4}C_{10}.
\end{split}%
\end{equation}%
It is easy to show also that the constraints (\ref{3stratcurv}) imply that 
\begin{equation}
C_{20}={p_{4}}^{5}-{p_{5}}^{4}=0.  \label{3stratC20}
\end{equation}%
Constraints (\ref{3stratcurv}) imply that the general element of the algebra 
$A_{\Sigma _{1,2,3}}$ in this case has the form 
\begin{equation}
p_{k}=A_{k}(p_{4})+B_{k}(p_{4})p_{5}+C_{k}(p_{4})p_{6}+D_{k}(p_{4})p_{7}
\end{equation}%
where $A_{k},B_{k},C_{k},D_{k}$ are certain polynomials. This observation
allows us to prove that for any constraint $f(p_{4},p_{5},p_{6},p_{7})=0$
arising from the equations $p_{i}p_{j}=p_{i+j}$ the polynomial $%
f(p_{4},p_{5},p_{6},p_{7})$ belongs to the ideal generated by $C_{10},C_{11}$%
, and $C_{12}$. Indeed, since $%
f(p_{4},p_{5},p_{6},p_{7})|_{C_{10}=C_{11}=C_{12}=0}=A(p_{4})+B(p_{4})p_{5}+C(p_{4})p_{6}+D(p_{4})p_{7}
$ with some polynomials $A,B,C,D$, the r.h.s. of this formula is identically
zero, due to the fact that the integers $4n,4m+5,4l+6,4k+7$ are always
distinct.

So the algebra $A_{\Sigma_{1,2,3}}(H^i_k=0)$ is equivalent to the polynomial
algebra $\mathbb{C}[p_4,p_5,p_6,p_7]/ \langle C_{10},C_{11},C_{12} \rangle$.
Geometrically the subspace $W_{1,2,3}$ is the infinite family of the
algebraic varieties with the $(4,5)$ curve (\ref{3stratC20}) in the basis.

Taking into account these observations it is natural to conjecture that in
general case $H^i_k\neq 0$ one has similar results for the algebra $%
A_{\Sigma_j}$ and affine algebraic variety.

In the general case one has

\begin{prop}
Index$(\overline{\partial}_{W_{1,2,3}})=-3$.
\end{prop}

\textbf{Proof } $S_{W_{1,2,3}}-\mathbb{N}=\varnothing$, $S_{\widetilde{W}_3}-%
\mathbb{N}=\{-3,-2,-1\}$. $\square$

This result suggests to conjecture that the curve $\mathcal{F}^3_{45}$ and
the basic curve $\Gamma$ in the four dimensional space with the coordinates $%
(p_4,p_5,p_6,p_7)$ defined by the equations $C_{10}=C_{11}=C_{12}=0$ have
genus $3$.

For n-th stratum $\Sigma_{1,2,\dots,n}$ associated with the set $%
S=\{-n,-n+1,\dots,-1,0,n+1,n+2,\dots \}$ the closed subspace $%
W_{1,2,\dots,n} $ ($W_{1,2,\dots,n}\cdot W_{1,2,\dots,n} \subset
W_{1,2,\dots,n}$) has the basis $(1,p_{n+1},p_{n+2},\dots)$ with 
\begin{equation}
p_i=z^i+\sum_{k=1}^iH^i_{-k}z^k+\sum_{k=1}^\infty \frac{H^i_{k}}{z^k}
\end{equation}
and $H^i_j$ obeying the system of quadratic algebraic equations analogue to (%
\ref{alg2-comp}). Algebraically $W_{1,2,\dots,n}$ is the infinite family of
infinite-dimensional algebra generated by $n+1$ elements $%
(p_{n+1},p_{n+2},\dots,p_{2n+1})$ modulo $n$ independent constraints 
\begin{equation}  \label{generalC}
\begin{split}
C_{2n+4}=& p_{n+1}p_{n+3}-{p_{n+2}}^2+\dots=0, \\
C_{2n+5}=& p_{n+1}p_{n+4}-{p_{n+2}}p_{n+3}+\dots=0, \\
\dots& \\
C_{3n+3}=& p_{n+1}p_{2n+2}-{p_{n+2}}p_{2n+1}+\dots=0.
\end{split}%
\end{equation}
These constraints imply that any element of the algebra $A_{\Sigma_{1,2,%
\dots,n}}$ can be represented as 
\begin{equation}
p_n=\alpha_{n0}(p_{n+1})+\sum_{k=n+2}^{2n+1}\alpha_{nk}(p_{n+1})p_k
\end{equation}
where $\alpha_{nk}$ are certain polynomials. Geometrically $W_{1,2,\dots,n}$
is the infinite-dimensional algebraic varieties varying with parameters $%
H^j_k$ $(i=n+1,n+2,\dots,\ k=1,2,\dots)$. In the base of this family there
is an algebraic curve in $n+1$-dimensional subspace with coordinates $%
(p_{n+1},p_{n+2},\dots,p_{2n+1})$ defined by $n$ constraints mentioned
above. An ideal of this curve contains the element 
\begin{equation}  \label{n+1n+2curve}
\mathcal{F}^n_{n+1,n+2}={p_{n+1}}^{n+2}-{p_{n+2}}^{n+1}+\dots
\end{equation}
which defines a $(n+1,n+2)$ curve in the two dimensional subspace with
coordinates $(p_{n+1},p_{n+2})$.

These statements are easily provable in the trivial case when $p_i=z^i$.
General case will be considered elsewhere. In general case one has

\begin{prop}
Index$(\overline{\partial}_{W_{1,2,\dots,n}})=-n$.
\end{prop}

Proof is straightforward.

This observation and results obtained in the previous section suggests to
formulate the following

\begin{conj}
\label{Theconjecture} The stratum $\Sigma _{1,2,\dots ,n}$ contains the
subset $W_{1,2,\dots ,n}$ of codimension $n$ for which the subbundle $%
TB_{W_{1,2,...,n}}$ is formed by fibers closed with respect to pointwise
multiplication. Algebraically $TB_{W_{1,2,...,n}}$ $\ $is the infinite
family of polynomial algebras equivalent to the coordinate ring 
\begin{equation}
\mathbb{C}[p_{n+1},p_{n+2},p_{n+3},\dots ,p_{2n+1}]/C_{2n+4},C_{2n+5},\dots
C_{3n+3}  \label{nstrat-conj}
\end{equation}%
with $C_{2n+4},C_{2n+5},\dots C_{3n+3}$ given by (\ref{generalC}).
Geometrically $\ $ $TB_{W_{1,2,...,n}}$is the infinite family of algebraic
curves with the basic algebraic curve $\Gamma $ in the $n+1$-dimensional
subspace with the local affine coordinates ($p_{n+1}$, $p_{n+2}$, $\dots $, $%
p_{2n+1}$) defined by equations (\ref{generalC}). $W_{1,2,\dots ,n}$ at
fixed $H_{k}^{j}$ contains the plane $(n+1,n+2)$ curves given by the
equation $\mathcal{F}_{n+1,n+2}^{n}=0$ of genus $n$. Curves $\Gamma $ have
genus $n$ too. Moreover Index$(\overline{\partial }_{W_{1,2,\dots ,n}})=-n$.
\end{conj}

So we conjecture that for strata $W_{1,2,\dots,n}$ one has the relation 
\begin{equation}
\mathrm{genus\ of\ } \Gamma + \mathrm{index} \overline{\partial}=0
\end{equation}
that could be useful for an analysis of the interrelations between various
geometric and analytic objects in Birkhoff strata of Sato Grassmannian. We
will analyze this conjecture and we will study other Birkhoff strata in
separate paper. 

\section{Resolution of singularities and transitions between strata}

\label{sect-desing} In the previous sections it was shown that each Birkhoff
strata contains subsets of points for which the corresponding tautological
subbundles contain infinite towers of families of algebraic curves.
Generically these curves are smooth. On the other hand it was also noted
that the projection of these smooth curves on the lower dimensional
subspaces in the same subbundle  are represented by the higher degree
singular curves which appear in the higher strata as smooth curve. This
observation clearly indicates that there is intimate interconnection between
the curves of the same type in different strata. It suggests also to adopt
wider approach in analyzing the possible mechanisms of resolution of
singularities of such curves.

Let us begin with the simplest example of the twisted cubic in the big cell
defined by the equations 
\begin{equation}
\begin{split}
q_{2}=& {q_{1}}^{2}-2H_{1}^{1}, \\
q_{2}=& {q_{1}}^{3}-3H_{1}^{1}q_{1}-3H_{2}^{1}.
\end{split}
\label{bigcellparam}
\end{equation}%
To avoid confusion we denote here the coordinate in $\ TB_{W_{\varnothing }}$
by $(q_{1},q_{2},q_{3})$. Its general projection on the two dimensional
subspace $\ TB_{W_{\varnothing }}$ with coordinates $(k_{2},k_{3})$ is given
by the plane cubic 
\begin{equation}
\left( k_{{3}}+\frac{3}{2}\,\alpha \,k_{{2}}+3\,H_{{2}}^{1}+\frac{3}{2}\,H_{{%
1}}^{1}\alpha +\frac{1}{2}\,{\alpha }^{3}+\frac{1}{2}\,\beta \,\alpha
\right) ^{2}=\left( k_{{2}}+2\,H_{{1}}^{1}+\frac{1}{4}\,{\alpha }^{2}\right)
\left( k_{{2}}-H_{{1}}^{1}+\beta +{\alpha }^{2}\right) ^{2}
\label{degEllzeros}
\end{equation}%
The nodal cubic (\ref{degEllzeros}) has polynomial parameterization 
\begin{equation}
\label{k2k3}
\begin{split}
k_{2}=& {k_{1}}^{2}+\alpha k_{1}-2H_{1}^{1}, \\
k_{3}=& {k_{1}}^{3}+(\beta -3H_{1}^{1})k_{1}-3H_{2}^{1}.
\end{split}%
\end{equation}%
It has ordinary double point at $k_{{2}}=H_{{1}}^{1}-\beta -{\alpha }^{2}$, $%
k_{3}=-3\,H_{{2}}^{1}-{3}\,H_{{1}}^{1}\alpha +{\alpha }^{3}+\,\beta \,\alpha 
$ and zero genus.

A standard way to resolve this singularity is to blow-up it by quadratic
transformation (see e.g. \cite{Har,Wal}). For simplicity we will consider
the case $\alpha=\beta=0$. An appropriate quadratic transformation in this
case is of the form 
\begin{equation}  \label{00blowup}
k_3=\tilde{k}\left(k_2-H^1_1\right)-3H^1_2.
\end{equation}
In virtue of equation (\ref{degEllzeros}) with $\alpha=\beta=0$ the new
variable $\tilde{k}$ obeys also the equation 
\begin{equation}  \label{newtildeq00}
\tilde{k}^2-\left(k_2+2H^1_1\right)=0.
\end{equation}
The system of equations (\ref{00blowup}) and (\ref{newtildeq00}) defines the
curve in the three-dimensional space $(\tilde{k},k_2,k_3)$. This system is
equivalent to the system 
\begin{equation}  \label{equiv00sys}
\begin{split}
k_2-H^1_1=&\tilde{k}^2-3H^1_1, \\
k_3+3H^1_2=&\tilde{k}^3-3H^1_1\tilde{k}.
\end{split}%
\end{equation}
So two points $\left(3\sqrt{H^1_1},H^1_1,-3H^1_2\right)$ and $\left(-3\sqrt{%
H^1_1},H^1_1,-3H^1_2\right)$ on the three dimensional curve defined by (\ref%
{equiv00sys}) correspond to the ordinary double point $\left(H^1_1,-3H^1_2%
\right)$ of the plane curve (\ref{degEllzeros}). Moreover, comparing (\ref%
{equiv00sys}) and (\ref{bigcellparam}), one concludes that the three
dimensional curve is nothing but the twisted cubic (\ref{bigcellparam}).
Twisted cubic is regular. So the transformation (\ref{00blowup}) represents
a resolution of singularity (blowing-up) of the curve (\ref{degEllzeros}).
This observation is very natural and almost trivial, since the curve (\ref%
{degEllzeros}), has been obtained as the projection of the original twisted
cubic.

Important features of such regularization is that the regularized curve (\ref%
{degEllzeros}) is the curve in three-dimensional space. It belongs to the
same fiber of  $TB_{W_{\varnothing }}$ and the genus of the regularized
curve remains to be zero.

The presence of the elliptic curve (\ref{1stratellcurv-red}) in the $%
TB_{W_{1}}$  indicates the existence of a different regularization
procedure. Generically the curve (\ref{1stratellcurv-red}) has genus one and 
$p_{2},p_{3}$ are full Laurent series (\ref{1stratser}) with $H_{k}^{i}$
obeying to the constraints (\ref{1stratsercoeff}). An important property of
these constraints is that the system (\ref{1stratsercoeff}) does not have
reductions for which $H_{m}^{j}=0$ for $m\geq n$. It is a well known fact
for the Weierstrass reduction (\ref{WPred}). So $p_{2}$ and $p_{3}$ are
either full Laurent series or polynomials $p_{2}^{s}=z^{2}+H_{-1}^{2}z$, $%
p_{3}^{s}=z^{3}+H_{-1}^{3}z$. In the latter case the cubic curve (\ref%
{1stratellcurv-red}) is singular and has the form 
\begin{equation}
\left( p_{{3}}+\frac{3}{2}\,H_{{-1}}^{2}p_{{2}}+\frac{1}{2}\,H_{{-1}}^{2}H_{{%
-1}}^{3}+\frac{1}{2}\,{H_{{-1}}^{2}}^{3}\right) ^{2}-\left( p_{2}+\frac{1}{4}%
{H_{-1}^{2}}^{2}\right) \left( p_{2}+{H_{-1}^{3}}+{H_{-1}^{2}}^{2}\right)
^{2}=0.  \label{singpolyn-cubcurve}
\end{equation}%
Now let us compare singular curves (\ref{singpolyn-cubcurve}) and 
(\ref{degEllzeros}). Taking into account (\ref{k2k3}), one readily concludes that
they represent the same curve with correspondence 
\begin{equation}
\alpha \leftrightarrow H_{-1}^{2},\quad \beta -3H_{1}^{1}\leftrightarrow
H_{-1}^{3},\quad p_{1}\leftrightarrow z,\quad p_{2}\leftrightarrow
k_{2}+2H_{1}^{1},\quad p_{3}\leftrightarrow k_{3}+3H_{2}^{1}.
\end{equation}%
So, the boundary form of the elliptic curve (\ref{1stratellcurv-red}) on the
boundary $\Delta _{01}$ between strata $\Sigma _{1}$ and $\Sigma
_{\varnothing }$ coincides with the projection of the twisted cubic (\ref%
{bigcellparam}) on this boundary from the side of $\Sigma _{\varnothing }$.

This observation suggests the following mechanism for transition between the
fibers of  $TB_{W_{\varnothing }}$ and \ $TB_{W_{1}}$ for the strata $\Sigma
_{\varnothing }$ and $\Sigma _{1}$. Inside $TB_{W_{\varnothing }}$ one has
the twisted cubic (\ref{bigcellparam}). Its form on boundary $\Delta _{01}$
from the side of $\Sigma _{\varnothing }$ is given by the nodal cubic (\ref%
{degEllzeros}). It coincides (under the identification with the boundary
form (\ref{singpolyn-cubcurve})) with the elliptic curve (\ref%
{1stratellcurv-red}). In order to move inside $\Sigma _{1}$ the polynomials (%
\ref{k2k3}) should become the Laurent series 
\begin{equation}
\begin{split}
k_{2}+2H_{1}^{1}\rightarrow & {k_{1}}^{2}+\alpha k_{1}+\sum_{n=1}^{\infty }%
\frac{H_{n}^{2}}{{k_{1}}^{n}}, \\
k_{3}+3H_{2}^{1}\rightarrow & {k_{1}}^{3}+\left( \beta -3H_{1}^{1}\right)
k_{1}+\sum_{n=1}^{\infty }\frac{H_{n}^{3}}{{k_{1}}^{n}},
\end{split}
\label{01tailgrowth}
\end{equation}%
where $H_{j}^{2}$ and $H_{j}^{3}$ should obey the constraints (\ref%
{1stratsercoeff}). We emphasize that in the transition $W_{\varnothing
}\rightarrow W_{1}$ the variable $k_{1}=p_{1}$ in $TB_{W_{\varnothing }}$ 
becomes the formal variable $z$ in $\Sigma _{1}$.

The transition from the elliptic curve (\ref{1stratellcurv-red}) to the
twisted cubic in  $TB_{W_{\varnothing }}$ is just the inverse process. The
boundary form (\ref{singpolyn-cubcurve}) of the elliptic genus one curve (%
\ref{1stratellcurv-red}) on $\Delta _{01}$ from the side of $\Sigma _{1}$ is
obtained by cutting the Laurent tails of $p_{2}$ and $p_{3}$. Passing to $%
\Sigma _{\varnothing }$ one has the curve (\ref{degEllzeros}). Then
blowing-up the singularity, one gets the twisted cubic.

In such a transition mechanism both generic curves in fibers of the
subbundles $TB_{W_{\varnothing }}$ and $TB_{W_{1}}$, i.e. the twisted cubic (%
\ref{bigcellparam}) and elliptic curve (\ref{1stratellcurv-red}) are
regular, but have different genus. This mechanism provides us with the
method of regularization of the nodal cubic (\ref{singpolyn-cubcurve}) by
instantaneous growing-up the full Laurent tail according to (\ref%
{01tailgrowth}).

This mechanism is valid also for the transition between entire infinite-
dimensional $TB_{W_{\varnothing }}$ and $TB_{W_{1}}$ .

Now let us consider the quintic (\ref{0F25}). It has two ordinary double
points. Complete resolution of these singularities without changing the
genus (zero) is performed by quadratic transformations in two steps. In the
final form it is given by the first four equations (\ref{bigcellcurr}) and
the fifth degree Veronese curve in the space with coordinates $%
(p_1,p_2,p_3,p_4,p_5)$ is the corresponding regularized curve.

The regularization of the quintic (\ref{0F25}) by increasing genus to one is
provided by the transition to the quintic (\ref{1F25}) via the procedure of
rising the Laurent tail of the type (\ref{01tailgrowth}). Cutting the
Laurent tail one passes from the stratum $\Sigma_1$ to $\Sigma_\varnothing$.

Similar procedure of resolution of singularities take place for trigonal
curve (\ref{singtrigcurv}) and $(4,5)$ curve (\ref{0F45}). In the big cell
their genus one regularized version are given by the curves (\ref{1stratTrig}%
) and (\ref{1F45}).

Singularities of trigonal curve (\ref{1stratTrig}) in the first stratum can
be resolved again in two ways . The first way consists in performing
quadratic transformations from $p_{4}$ to the new variable $p_{2}$ defined
by 
\begin{equation}
p_{4}={p_{2}}^{2}-2H_{-1}^{2}p_{3}-{H_{-1}^{2}}%
^{2}p_{2}-2H_{-1}^{2}H_{1}^{2}-2H_{2}^{2}.  \label{p4p2-1strat}
\end{equation}%
The corresponding regularized curve in the three dimensional space $%
(p_{2},p_{3},p_{4})$ is the genus one curve which is the intersection of the
cylindrical surfaces generated by the elliptic curve (\ref{1stratellcurv-red}%
) and the surface defined by equation (\ref{p4p2-1strat}). Second
regularization is provided by the transition from the curve (\ref{1stratTrig}%
) to the genus two trigonal curve (\ref{2strattrigcurv}) in $TB_{W_{1,2}}$ .

Analogous mechanism of resolution of singularities take place for other
algebraic curves in Sato Grassmannian.

\section{Conclusion}

The results presented in this paper are essentially the observations about
the existence of special points in the Birkhoff strata of Sato Grassmannian.
For these points and sets of such points the corresponding tautological
subbundles contain associative algebras and families of algebraic curves
with interesting and important properties. In our analysis we have tried to
deal only with the Birkhoff strata itself and to avoid the use of any
auxiliary and additional structure. \ This was done on a purpose by two
reasons. \ The first reason was to demonstrate the richness and peculiarity
of \ Birkhoff strata of Sato Grassmannian \ themselves independently of their
known connections with other algebro-geometric structures like integrable
systems and auxiliary vector bundles. The second one is to leave completely
open the way for possible interpretations of these properties of Birkhoff
strata. \ Our approach is apparently different from those discussed earlier,
in particular, from the methods of Krichever \cite{Kri1,Kri2}, Segal-Wilson 
\cite{SW}, Mulase \cite{Mul1,Mul2,Mul3} and Takasaki \cite{Tak2}. The
possible principal differences (if so) or eventual interrelation between our
method and those mentioned above will be discussed elsewhere. 

\subsubsection*{Acknowledgments}

We thank Marco Pedroni and Andrea Previtali for many fruitful
discussions. We are very grateful to some readers of the manuscript
for the useful remarks. This work has been partially supported by PRIN grant no
28002K9KXZ and by FAR 2009 (\emph{Sistemi dinamici Integrabili e Interazioni
fra campi e particelle}) of the University of Milano Bicocca. 
\appendix

\section[Curves in Birkhoff strata]{Curves in Birkhoff strata}


\begin{center}
\begin{tabular}{|c|c|c|c|}
\hline
&  &  &  \\ 
\textbf{Birkhoff stratum} & \textbf{\quad associated algebra \quad} & 
\textbf{plane curve} & \textbf{genus} \\ 
&  &  &  \\ \hline
&  &  &  \\ 
$\Sigma_\varnothing$ & $\mathbb{C}[p]$ & Parabola & 0 \\ 
&  &  &  \\ \hline
&  &  &  \\ 
$\Sigma_1$ & $\mathbb{C}[p_2,p_3]/ \mathcal{F}^1_{23}$ & Elliptic & 1 \\ 
&  &  &  \\ \hline
&  &  &  \\ 
$\Sigma_{1,2}$ & $\mathbb{C}[p_3,p_4,p_5]/ \mathcal{C}_{8},\mathcal{C}_{9},%
\mathcal{C}_{10}$ & Trigonal & 2 \\ 
&  &  &  \\ \hline
&  &  &  \\ 
$\Sigma_{1,2,\dots,n}$ & $\dfrac{\mathbb{C}[p_{n+1},p_{n+2},\dots,p_{2n+1}]}{%
\{C_{i}\}_{i=2n+4,\dots,3n+3}}$ & $(n+1,n+2)$ curve & n\ ? \\ 
&  &  &  \\ \hline
\end{tabular}
\end{center}

\section[Big cell $\Sigma_{\varnothing}$]{Big cell $\Sigma_{\varnothing}$}

\label{App-bigcell} The coefficients in the formula (\ref%
{singtrigcurv-0-ideal}) are 
\begin{equation}
\begin{split}
a_9=& -4\,{p_{{1}}}^{3}+12\,H^1_1p_{{1}}+12\,H^1_2\ , \\
a_8=& 3\,{p_{{1}}}^{4}-12\,H^1_1{p_{{1}}}^{2}-12\,H^1_2p_{{1}}+6\,{H^1_1}%
^{2}\ , \\
a_7=& 12\,H^1_1H^1_2\ , \\
a_6=& -6\,{p_{{1}}}^{6}+36\,H^1_1{p_{{1}}}^{4}+36\,H^1_2{p_{{1}}}^{3}-54 \,{%
H^1_1}^{2}{p_{{1}}}^{2}-108\,H^1_1p_{{1}}H^1_2-48\,{H^1_2}^ {2}+4\,{H^1_1}%
^{3}\ , \\
a_4=& 9\,{H^1_1}^{4}+3\,{p_{{1}}}^{8}-48\,{H^1_2}^{2}H^1_1-24\,H^1_1{p_{{1} }%
}^{6}-24\,{p_{{1}}}^{5}H^1_2+60\,{p_{{1}}}^{4}{H^1_1}^{2}- 48\,{H^1_1}^{3}{%
p_{{1}}}^{2}+48\,{H^1_2}^{2}{p_{{1}}}^{2} \\
&+108\,H^1_1{p_{{1}}}^{3}H^1_2-84\,H^1_2p_{{1}}{H^1_1}^{2}\ , \\
a_3=& 64\,{H^1_2}^{3}-4\,{p_{{1}}}^{9}-12\,{H^1_1}^{3}H^1_2-204\,{p_{{\ 1}}}%
^{4}H^1_1H^1_2+276\,{H^1_1}^{2}{p_{{1}}}^{2}H^1_2 +240\,H^1_1p_{{1}}{H^1_2}%
^{2}+36\,{p_{{1}}}^{6}H^1_2 \\
&-108\,{p_{{1}}}^{5}{H^1_1}^{2}-96\,{p_{{1}}}^{3}{H^1_2}^{2}+116\,{H^1_1}^{3}%
{p_{{1}}} ^{3}+36\,{p_{{1}}}^{7}H^1_1-24\,{H^1_1}^{4}p_{{1}}\ .
\end{split}%
\end{equation}
The coefficients in the formula (\ref{0F45}) are 
\begin{equation}
\begin{split}
c_5=& 20\,H^1_{{2}}{H^1_{{3}}}^{3}-200\,{H^1_{{2}}}^{2}H^1_{{3}}H^1_{{4}%
}-200\,H^1_{{4}} H^1_{{1}}{H^1_{{3}}}^{2}-20\,H^1_{{2}}{H^1_{{1}}}%
^{6}-40\,H^1_{{4}}{H^1_{{1}}}^{5} +16\,{H^1_{{2}}}^{5} \\
& +140\,H^1_{{2}}{H^1_{{1}}}^{2}{H^1_{{3}}}^{2}-300\,{H^1_{{2} }}^{2}{H^1_{{1%
}}}^{2}H^1_{{4}}+100\,H^1_{{2}}{H^1_{{1}}}^{4}H^1_{{3}}+120\,{H^1_{{\ 2}}}%
^{3}H^1_{{3}}H^1_{{1}}+500\,{H^1_{{4}}}^{3}\ , \\
c_4=& -100\,{H^1_{{1}}}^{2}{H^1_{{3}}}^{3}+50\,{H^1_{{1}}}^{4}{H^1_{{3}}}%
^{2} -300\,H^1_{{2}}{H^1_{{1}}}^{2}H^1_{{4}}H^1_{{3}}+500\,{H^1_{{4}}}%
^{2}H^1_{{1}}H^1_{{3}}-200 \,H^1_{{4}}{H^1_{{3}}}^{2}H^1_{{2}} \\
& -50\,{H^1_{{2}}}^{4}{H^1_{{1}}}^{2}+15\,{H^1_{{\ 3}}}^{4}+250\,{H^1_{{4}}}%
^{2}{H^1_{{2}}}^{2}-5\,{H^1_{{1}}}^{8}+20\,{H^1_{{2}} }^{4}H^1_{{3}}+20\,{%
H^1_{{2}}}^{2}{H^1_{{1}}}^{5}+200\,{H^1_{{2}}}^{2}{H^1_{{1}} }^{3}H^1_{{3}}
\\
&-100\,H^1_{{2}}{H^1_{{1}}}^{4}H^1_{{4}}+120\,{H^1_{{2}}}^{2}{H^1_{{3 }}}%
^{2}H^1_{{1}}-200\,H^1_{{4}}{H^1_{{2}}}^{3}H^1_{{1}}-20\,H^1_{{3}}{H^1_{{1}}}%
^{ 6}\ , \\
c_0=&-360\,{H^1_{{2}}}^{2}{H^1_{{3}}}^{2}{H^1_{{1}}}^{3}+60\,{H^1_{{2}}}^{2}{%
H^1_{{1} }}^{5}H^1_{{3}}-100\,{H^1_{{2}}}^{2}{H^1_{{3}}}^{3}H^1_{{1}} -500\,{%
H^1_{{4}}}^{2 }{H^1_{{2}}}^{2}H^1_{{3}}-750\,{H^1_{{4}}}^{2}{H^1_{{2}}}^{2}{%
H^1_{{1}}}^{2} \\
&-500\,{H^1_{{4}}}^{2}H^1_{{1}}{H^1_{{3}}}^{2}-60\,{H^1_{{2}}}^{4}H^1_{{3}}{%
H^1_{{1}}}^ {2}+700\,{H^1_{{1}}}^{2}{H^1_{{3}}}^{2}H^1_{{2}}H^1_{{4}}-100\,{%
H^1_{{1}}}^{6}H^1_ {{2}}H^1_{{4}}-4\,{H^1_{{3}}}^{5}+2\,{H^1_{{1}}}^{10} \\
&+625\,{H^1_{{4}}}^{4}+100 \,H^1_{{2}}{H^1_{{3}}}^{3}H^1_{{4}}+600\,{H^1_{{2}%
}}^{3}H^1_{{3}}H^1_{{4}}H^1_{{1}}+ 500\,H^1_{{2}}{H^1_{{1}}}^{4}H^1_{{4}%
}H^1_{{3}}+80\,H^1_{{4}}{H^1_{{2}}}^{5}-80\,{\ H^1_{{2}}}^{6}H^1_{{1}} \\
&+20\,{H^1_{{1}}}^{6}{H^1_{{3}}}^{2}+50\,{H^1_{{1}}}^{2}{H^1 _{{3}}}%
^{4}-160\,{H^1_{{1}}}^{4}{H^1_{{3}}}^{3}+105\,{H^1_{{2}}}^{4}{H^1_{{1}} }%
^{4}+20\,{H^1_{{1}}}^{8}H^1_{{3}}-40\,{H^1_{{1}}}^{7}{H^1_{{2}}}^{2} \\
&-100\,{H^1_ {{1}}}^{5}{H^1_{{4}}}^{2}.
\end{split}%
\end{equation}

\section[Stratum $\Sigma_{1}$]{Stratum $\Sigma_{1}$}

\label{App-1strat} The first following cases in (\ref{strat1-C8C9}) are 
\begin{equation}  \label{1stratlowcasecurv}
\begin{split}
\mathcal{C}_{10}=&{p_2}^5-{p_3}^2{p_2}^2 + \dots =-{p_2}^2 \mathcal{F}%
^1_{23}\ , \\
\mathcal{C}_{11}=& {p_2}^4{p_3}-{p_3}^3{p_2}+\dots = \left( -p_{{3}}p_{{2}%
}+3\,H^2_{{-1}}{p_{{2}}}^{2}+p_{{2}}H^2_{{-1}}H^3_{{-1}}+3\,p_{{2}}H^2_{{1}%
}+p_{{2}}{H^2_{{-1}}}^{3} \right) \mathcal{F}^1_{23}\ , \\
\mathcal{C}_{12}=& {p_3}^4-{p_2}^6+ \dots = \left( \mathcal{F}^1_{23}+2p_2^3
-6\,H^2_{{-1}}p_2p_3+ (4\,H^3_{{-1}}+9\,{H^2_{{-1}}}^{2})p_2^2\right. \\
&\left.+ (-6\,H^2_{{1}}-2\,H^2_{{-1}}H^3_{{-1}}-2\,{H^2_{{-1 }}}^{3})p_3+
(12\,H^2_{{1}}H^2_{{-1}}+2\,{H^3_{{-1}}}^{2} +6\,{H^2_{{-1}}}^{4}+4\,H^3_{{1}%
}-6\,H^2_{{2}}\right. \\
&\left.+8\,{H^2_{{-1}}} ^{2}H^3_{{-1}})p_2+ 7\,H^2_{{-1}}H^3_{{-1}}H^2_{{1}%
}+3\,{H^2_{{-1}}}^{ 2}H^3_{{1}}+{H^2_{{-1}}}^{2}{H^3_{{-1}}}^{2} +3\,H^2_{{3}%
}H^2_{{-1}}-6\,H^2_{{2}}{H^2_{{-1}}}^{2}\right. \\
&\left. +4\,H^3_{{1}}H^3_{{-1}}-8\,H^3_{{-1}}H^2_{{2}}+6\,H^2_ {{1}}{H^2_{{-1%
}}}^{3}+2\,{H^2_{{-1}}}^{4}H^3_{{-1}}-6\, H^2_{{4}}+4\,H^3_{{3}}+3\,{H^2_{{1}%
}}^{2}+{H^2_{{- 1}}}^{6} \right) \mathcal{F}^1_{23}\ , \\
\dots\ .
\end{split}%
\end{equation}
The discriminant of the elliptic curve (\ref{1stratellcurv}) is \newline
{{\small 
\begin{equation}
\begin{split}
&\Delta = \\
& -396\,{H^2_{{1}}}^{2}{H^2_{{-1}}}^{2}{H^3_{{-1}}}^{2}- 243\,{H^2_{{1}}}%
^{4}+360\,{H^2_{{-1}}}^{3}H^3_{{1}}H^2_{{1}}H^3_{{-1}} +1152\,{H^3_{{-1}}}%
^{2}H^2_{{2}} H^2_{{-1}}H^2_{{1}}+432\,H^2_{{2}}{H^2_{{-1}}}^{3} H^2_{{1}%
}H^3_{{-1}} \\
& -32\,{H^3_{{-1}}}^{3}H^3_{{1}}{\ H^2_{{-1}}}^{2}+972\,{H^2_{{-1}}}^{3}H^2_{%
{3}}H^2_ {{2}}+96\,H^2_{{-1}}H^2_{{3}}{H^3_{{-1}}}^{3} +216\,{H^2_{{-1}}}%
^{3}H^2_{{3}}{H^3_{{-1}}}^{2}-432\,{H^2_ {{-1}}}^{4}H^3_{{1}}H^2_{{2}} \\
&-72\,{H^2_{{-1}}}^{4}H^3_{{1}}{H^3_{{-1}}}^{2} -432\,H^3_{{-1}}{H^2_{{2}}}%
^{2}{H^2_{{-1}}}^{2} -144\,{H^3_{{-1}}}^{3}H^2_{{2}}{H^2_{{-1}}}%
^{2}-1152\,H^3_{{3}}H^3_{{-1}}H^3_{{1}} -1944\,H^2_{{4}}H^2_{{2}}{H^2_{{-1}}}%
^{2} \\
&-864\,{H^2_{{1 }}}^{2}{H^2_{{-1}}}^{2}H^3_{{1}}-2592\,H^3_{{3}}H^2_{{-1}%
}H^2_{{3}} -{\frac {27}{2}}\,{H^2_{{-1}}}^{8}H^3_ {{1}}-81\,{H^2_{{-1}}}%
^{6}H^2_{{4}} +{\frac {81}{2}}\,{H^2_{{-1}}}^{7}H^2_{{3}}+54\,{H^2_{{-1}}}%
^{6}H^3_{{3}} \\
&-1728\,{H^3_{{3}}}^{2}+5184\,H^3_{{3}}H^2_{{4}} +1296\,H^3_{{3}}{H^2_{{1}}}%
^{2}+128\,H^3_{{3}}{H^3_{{-1}}} ^{3}-3888\,{H^2_{{4}}}^{2}-1944\,H^2_{{4}}{%
H^2_{{1}}}^{ 2}-192\,H^2_{{4}}{H^3_{{-1}}}^{3} \\
&-48\,{H^2_{{1}}}^{2}{H^3_{{-1}}}^{3} +64\,{H^3_{{-1}}}^{2}{H^3_{{1}}}%
^{2}-972 \,{H^2_{{-1}}}^{2}{H^2_{{3}}}^{2}+144\,{H^2_{{-1}}}^{4} {H^3_{{1}}}%
^{2}-1152\,{H^3_{{-1}}}^{2}{H^2_{{2}}}^{2}- 64\,{H^3_{{-1}}}^{4}H^2_{{2}} \\
&+81\,{H^2_{{2}}}^{2}{H^2_{{-1}}}^{4} +810\,{H^2_{{1}}}^{2}H^2_{{2}}{H^2_{{-1%
} }}^{2}+1728\,H^2_{{4}}H^3_{{-1}}H^3_{{1}} +648\,{H^2_{{1}}}^{3}H^2_{{-1}%
}H^3_{{-1}} +972\,{H^2_{{-1}}}^{3}H^2_{{1}}H^2_{{4}} \\
&-324\,H^3_{{-1}}H^2_{{4}}{H^2_{{-1}}}^{4}+216\,H^3_{{-1}}{H^2_{{-1}}}%
^{4}H^3_{{3}}- 54\,H^3_{{-1}}{H^2_{{-1}}}^{6}H^3_{{1}}+162\,H^3_{ {-1}}H^2_{{%
3}}{H^2_{{-1}}}^{5}-27\,{H^2_{{-1}}}^{6}H^2_{{2}}H^3_{{-1}} \\
&+216\,{H^2_{{-1}}}^{5}H^3_{{1}}H^2_{{1}}+384\,{H^3_{{-1}}}^{2}H^3_{{1}}H^2_{%
{2}}+ 432\,{H^2_{{1}}}^{2}H^3_{{-1}}H^3_{{1}}+972\,{H^2_ {{1}}}^{2}H^2_{{-1}%
}H^2_{{3}} -81\,{H^2_{{-1}}}^{5}H^2_{{2}}H^2_{{1}} \\
&+{\frac {27}{2}}\,{H^2_{{-1}}}^{7}H^3 _{{-1}}H^2_{{1}}-486\,{H^2_{{-1}}}%
^{4}H^2_{{1}}H^2 _{{3}}-648\,{H^2_{{-1}}}^{3}H^2_{{1}}H^3_{{3}} -1728\,H^3_{{%
3}}H^2_{{-1}}H^2_{{1}}H^3_{{-1}} +2592\,H^2_{{4}}H^2_{{-1}}H^2_{{1}}H^3_{{-1}%
} \\
&+3456\,H^2_{{2}}H^3_{{1}}H^2_{{-1}}H^2_{{1}} +384\,H^3_{{-1}}{H^3_{{1}}}^{2}%
{H^2_{{-1}}}^{2}-512\,{H^3_{{1}}}^{3}-864 \,H^3_{{-1}}H^3_{{1}}H^2_{{-1}%
}H^2_{{3}} -720\,H^3_{{-1}}H^3_{{1}}H^2_{{2}}{H^2_{{-1}}}^{2} \\
&-192\,{H^3_{{-1}}}^{2}H^3_{{1}}H^2_{{-1}}H^2_{{1}}+2592\, H^2_{{-1}}H^2_{{3}%
}H^3_{{-1}}H^2_{{2}} -1296\,{H^2_{{-1}}}^{2}H^2_{{3}}H^2_{{1}}H^3_{{-1}}
+1728\,{H^2_{{2}}}^{3} +54\,{H^2_{{-1}}}^{5}{H^3_{{-1}}}^{2}H^2_{{1}} \\
&-189\,{H^2_{{-1}}}^{4}H^3_{{-1}}{H^2_{{1}} }^{2}-1296\,{H^2_{{1}}}^{2}H^3_{{%
-1}}H^2_{{2}} -3456\,{H^2_{{2}}}^{2}H^3_{{1}}+2304\,H^2_{{2}}{H^3_{{1}}}^
{2}+27\,{H^2_{{-1}}}^{3}{H^2_{{1}}}^{3} \\
&-432\,H^2_{{4}}{H^2_{{-1}}}^{2}{H^3_{{-1}}}^{2}-108\,H^2_{{2}}{H^2 _{{-1}}}%
^{4}{H^3_{{-1}}}^{2}+32\,{H^3_{{-1}}}^{4}H^2_{{\ -1}}H^2_{{1}}+72\,{H^2_{{-1}%
}}^{3}H^2_{{1}}{H^3_{{\ -1}}}^{3}+3456\,H^3_{{3}}H^3_{{-1}}H^2_{{2}} \\
&+1296\,H^3_{{3}}H^2_{{2}}{H^2_{{-1}}}^{2} +3888\,H^2_{{4}}H^2_{{-1}}H^2_{{3}%
} -5184\,H^2_{{4}}H^3_{{-1}}H^2_{{2}}-2592\,{H^2_{{2}}}^{2}H^2_{{-1}}H^2_{{1}%
}-1152 \,{H^3_{{1}}}^{2}H^2_{{-1}}H^2_{{1}} \\
&+288\,H^3_{{3}}{H^2_{{-1}}}^{2}{H^3_{{-1}}}^{2}\ .
\end{split}%
\end{equation}%
}} The coefficients in equation (\ref{1F25}) are 
\begin{equation}
\begin{split}
D_4=& 7\,{H^2_{{1}}}^{2}-4\,{H^2_{{-1}}}^{6}+3\,H^2_{{4}}-2\, H^3_{{3}}+4\,{%
H^2_{{-1}}}^{2}{H^3_{{-1}}}^{2} -4\,{H^2_{{-1}}}^{4}H^3_{{-1}}+10\,{H^2_{{-1}%
}}^{3}H^2_{{1}} -2\,H^3_{{-1}}H^2_{{2}} \\
&-7\,{H^2_{{-1}}}^{2}H^3_{{1 }}+11\,H^2_{{-1}}H^2_{{3}}+5\,H^2_{{2}}{H^2_{{-1%
}} }^{2}+2\,H^3_{{-1}}H^3_{{1}}+H^2_{{-1}}H^2_{{1}}H^3_{{-1}}\ ,
\end{split}
\notag
\end{equation}
\begin{equation}
\begin{split}
D_2=& {\frac {25}{2}}\,H^2_{{1}}H^2_{{3}}+4\,H^3_{{3}} H^3_{{-1}}-5\,H^2_{{2}%
}{H^3_{{-1}}}^{2}-\frac{3}{2}\,H^3_{{-1}}{H^2_{{1}}}^{2} -6\,H^3_{{-1}}H^2_{{%
4}}-\frac{23}{2}\,{H^2_{{\ -1}}}^{4}H^3_{{1}}+\frac{13}{2}\,H^2_{{3}}{H^2_{{%
-1}}}^{3} \\
&-8\,{\ H^2_{{-1}}}^{2}H^3_{{3}}+12\,H^2_{{4}}{H^2_{{-1}}} ^{2}+14\,{H^2_{{-1%
}}}^{4}H^2_{{2}}+2\,{H^3_{{-1}}}^{2}H^3_{{1}} +2\,{H^2_{{-1}}}^{4}{H^3_{{-1}}%
}^{2}+2\,{H^2_{{-1}}}^{2}{H^3_{{-1}}}^{3} \\
&-2\,{H^2_{{-1}}}^{5}H^2_ {{1}}+8\,{H^2_{{-1}}}^{2}{H^2_{{1}}}^{2}-3\,H^3_{{%
-1}}{\ H^2_{{-1}}}^{6}+3\,H^2_{{-1}}H^2_{{3}}H^3_{{-1}}+\frac{5}{2}\,H^2_{{-1%
}}H^2_{{1}}H^3_{{1}}-2\,H^2_{{2}}{H^2_{{-1}}}^{2}H^3_{{-1}} \\
& -10\,H^2_{{-1}}H^2_{{1}}H^2_{{2}}+3\,{H^2_{{-1}}}^{2}H^3_{{-1}}H^3_{{1}} +%
\frac{7}{2}\,{H^2_{{-1}}}^{3}H^3_{{-1}}H^2_{{1}}-2\,{H^2_{{- 1}}}^{8}-{H^3_{{%
-1}}}^{4} \ ,
\end{split}%
\end{equation}
\begin{equation}
\begin{split}
D_0=& 6\,{H^2_{{-1}}}^{2}{H^3_{{-1}}}^{2}H^3_{{1}}-{H^2_ {{-1}}}^{5}H^2_{{3}%
}-11\,H^2_{{2}}{H^2_{{-1}}}^{2}{H^3_{{-1}}}^{2} -3\,{H^2_{{-1}}}^{2}H^3_{{-1}%
}{H^2_{ {1}}}^{2}+5\,{H^2_{{-1}}}^{2}H^2_{{1}}H^2_{{3}} \\
&-\frac{3}{2}\,H^3_{{-1}}{H^2_{{-1}}}^{4} H^3_{{1}}+8\,H^3_{{-1}}{H^2_{{-1}}}%
^{2}H^3_{{3}}-2\,H^2_{{-1}}{H^3_{{-1}}}^ {3}H^2_{{1}}+\frac{11}{2}\,{H^2_{{-1%
}}}^{3}{H^3_{{-1}}}^{2}H^2_{{1}}+2\,{H^2_{{-1}}}^{4}H^2_{{2}}H^3_{{-1}} \\
&-4 \,{H^2_{{-1}}}^{3}H^2_{{1}}H^2_{{2}}+{\frac {25}{4}}\,{\ H^2_{{3}}}^{2}-{%
H^2_{{-1}}}^{5}H^3_{{-1}}H^2_{{1}} +{H^2_{{-1}}}^{3}H^3_{{1}}H^2_{{1}}+\frac{%
3}{4}\,{H^3_{{-1 }}}^{2}{H^2_{{1}}}^{2}+\frac{1}{4}\,{H^2_{{-1}}}^{2}{H^3_{{1%
}}} ^{2} \\
&+8\,{H^2_{{-1}}}^{6}H^2_{{2}} +\frac{17}{2}\,H^3_{{-1}}H^2_{{3}}{H^2_{{-1}}}%
^{3}-2\,H^3_{{3}}{H^3_{{-1}}}^{2}- 5\,{H^2_{{-1}}}^{6}H^3_{{1}}-8\,{H^2_{{-1}%
}}^{4}{H^2 }_{{3}}+12\,{H^2_{{-1}}}^{4}H^2_{{4}} \\
&+4\,H^2_{{2}}{H^3_{{-1}}}^{3}+\frac{5}{2}\,H^2_{{-1}}H^2_{{3}}H^3_{{1}}+3 \,%
{H^3_{{-1}}}^{2}H^2_{{4}}-2\,{H^3_{{-1}}}^{3}H^3 _{{1}}+4\,{H^2_{{1}}}^{2}{%
H^2_{{-1}}}^{4}-4\,{H^2_{{-1} }}^{7}H^2_{{1}} \\
&+4\,{H^2_{{-1}}}^{2}{H^2_{{2}}}^{2}-12\, H^3_{{-1}}H^2_{{4}}{H^2_{{-1}}}%
^{2} -10\,H^2_{{3}}H^2_{{2}}H^2_{{-1}} -2\,H^2_{{2}}{H^2_{{-1}}}^{2}H^3_{{1}%
}-4\,H^2_{{-1}}H^2_{{3}}{H^3_{{-1}}}^{2}\ .
\end{split}
\notag
\end{equation}
The longer coefficients in the formula (\ref{1stratTrig}) are {\small
\begin{equation}
\begin{split}
A_4=&\left(-{\frac {27}{2}}\,{H^2_{{-1}}}^{3}H^3_{{-1}}H^2_{{1}}+ 12\,{H^2_{{%
-1}}}^{2}H^2_{{2}}H^3_{{-1}}-4\,{H^2_{{-1}}}^{2}H^3_{{-1}}H^3_{{1}}
-16\,H^2_{{-1}}{H^3_{{-1}}}^{2}H^2_{{1}}+6\,{H^2_{{-1}}}^{4}H^2_{{2}}
-3\,H^2_{{1}}{H^2_{{-1}}}^{5} \right. \\
&\left. +6\,{H^2_{{1}}}^{2}{H^2_{{-1}}}^{2} -4\,{H^3_{{-1}}}^{2}H^3_{{1}%
}+9\,H^2_{{4}}{H^2_{{-1}}}^{2} -6\,{H^2_{{-1}}}^{2}H^3_{{3}}+{H^3_{{-1}}}%
^{4}-6\,H^2_{{3}}H^2_{{-1}}H^3_{{-1}} +4\,{H^3_{{-1}}}^{2}{H^2_{{-1}}}%
^{4}+4\,{H^2_{{-1}}}^{2}{H^3_{{-1}}}^{3} \right. \\
&\left. -\frac{5}{2}\,{H^2_{{-1}}}^{4}H^3_{{1}}-\frac{9}{2}\,H^2_{{3}}{H^2_{{%
-1}}}^{3} +12\,H^3_{{-1}}{H^2_{{1}}}^{2}+12\,H^3_{{-1}}H^2_{{4}}+2\,H^2_{{2}}%
{H^3_{{-1}}}^{2} +4\,H^2_{{2}}H^3_{{1}}+4\,H^2_{{1}}H^2_{{-1}}H^3_{{1}}-3\,{%
H^2_{{2}}}^{2} \right. \\
&\left. -8\,H^3_{{3}}H^3_{{-1}}+4\,{H^3_{{1}}}^{2}+H^3_{{-1}}{H^2_{{-1}}}%
^{6} -6\,H^2_{{1}}H^2_{{-1}}H^2_{{2}}\right)\ , \\
\end{split}
\notag
\end{equation}
\begin{equation}
\begin{split}
A_3=&\left(18\,{H^2_{{-1}}}^{3}{H^2_{{1}}}^{2}+{H^2_{{-1}}}^{9} -6\,{H^2_{{-1%
}}}^{6}H^2_{{1}}+12\,{H^2_{{-1}}}^{5}{H^3_{{-1}}}^{2} +9\,{H^2_{{-1}}}^{3}{%
H^3_{{-1}}}^{3}+6\,{H^2_{{-1}}}^{7}H^3_{{-1}} +2\,{H^3_{{-1}}}^{4}H^2_{{-1}}
\right. \\
&\left. -6\,{H^2_{{-1}}}^{5}H^2_{{2}} +8\,H^2_{{-1}}{H^3_{{1}}}^{2}+6\,H^2_{{%
-1}}{H^2_{{2}}}^{2} +\frac{9}{2}\,{H^2_{{-1}}}^{5}H^3_{{1}}-\frac{3}{2}\,{%
H^2_{{-1}}}^{4}H^2_{{3}} -2\,{H^2_{{-1}}}^{3}H^3_{{3}} +3\,{H^2_{{-1}}}%
^{3}H^2_{{4}}+12\,H^3_{{3}}H^2_{{1}} \right. \\
&\left. -18\,{H^2_{{1}}}^{3} +18\,H^2_{{4}}H^2_{{-1}}H^3_{{-1}} -16\,H^2_{{-1%
}}H^2_{{2}}H^3_{{1}}+9\,H^2_{{-1}}H^2_{{1}}H^2_{{3}} -{\frac {57}{2}}\,{H^2_{%
{-1}}}^{4}H^3_{{-1}}H^2_{{1}} -15\,{H^2_{{-1}}}^{3}H^2_{{2}}H^3_{{-1}} -{%
H^2_{{-1}}}^{2}H^3_{{1}}H^2_{{1}} \right. \\
&\left. +3\,{H^2_{{-1}}}^{2}H^2_{{2}}H^2_{{1}} +45\,H^2_{{-1}}H^3_{{-1}}{%
H^2_{{1}}}^{2} -4\,{H^3_{{-1}}}^{2}H^2_{{-1}}H^3_{{1}}+12\,H^3_{{-1}}H^3_{{1}%
}H^2_{{1}} -18\,H^2_{{1}}H^2_{{4}}-32\, {H^2_{{-1}}}^{2}{H^3_{{-1}}}^{2}H^2_{%
{1}} \right. \\
&\left. -8\,H^2_{{-1}}H^2_{{2}}{H^3_{{-1}}}^{2} +7\,H^3_{{-1}}{H^2_{{-1}}}%
^{3}H^3_{{1}} -9\,H^3_{{-1}}H^2_{{3}}{H^2_{{-1}}}^{2}-12\,H^3_{{-1}}H^2_{{-1}%
}H^3_{{3}}\right)\ , \\
\end{split}%
\end{equation}
\begin{equation}
\begin{split}
A_0=&\left( -10\,H^3_{{-1}}{H^2_{{-1}}}^{2}{H^3_{{1}}}^{2} -12\,{H^2_{{-1}}}%
^{2}{H^2_{{2}}}^{2}H^3_{{-1}} +\frac{9}{2}\,{H^2_{{-1}}}^{3}H^2_{{2}}H^2_{{3}%
} +6\,{H^2_{{-1}}}^{2}H^2_{{2}}H^3_{{3}}+6\,H^2_{{-1}}{H^2_{{2}}}^{2}H^2 _{{1%
}}-9\,{H^2_{{-1}}}^{2}H^2_{{2}}H^2_{{4}} \right. \\
&\left. -10\,{H^2_{{-1}}}^{2}H^3_{{1}}H^3_{{3}} +8\,H^2_{{-1}}{H^3_{{1}}}%
^{2}H^2_{{1}} +15\,{H^2_{{-1}}}^{2}H^3_{{1}}H^2_{{4}} -{\frac {85}{4}}\,{%
H^3_{{-1}}}^{2}{H^2_{{1}}}^{2}{H^2_{{-1}}}^{2} +12\,H^3_{{-1}}{H^2_{{1}}}%
^{2}H^3_{{1}}-\frac{15}{2}\,H^2_{{3}}{H^2_{{-1}}}^{3}H^3_{{1}} \right. \\
&\left. +9\,H^2_{{3}}H^2_{{4}}H^2_{{-1}} -\frac{3}{2}\,{H^2_{{-1}}}^{3}{H^3_{%
{-1}}}^{2}H^2_{{3}} +\frac{13}{2}\,{H^2_{{-1}}}^{5}{H^3_{{-1}}}^{2}H^2_{{1}%
}+23\,{H^2_{{-1}}}^{4}H^2_{{2}}{H^3_{{-1}}}^{2} -\frac{15}{2}\,{H^2_{{-1}}}%
^{4}{H^3_{{-1}}}^{2}H^3_{{1}} \right. \\
&\left. +\frac{15}{2}\,{H^2_{{-1}}}^{3}{H^3_{{-1}}}^{3}H^2 _{{1}}+12\,{H^2_{{%
-1}}}^{2}{H^3_{{-1}}}^{3}H^2_{{2}}-2 \,{H^2_{{-1}}}^{2}{H^3_{{-1}}}^{2}H^3_{{%
3}} +3\,{H^2_{{-1}}}^{2}{H^3_{{-1}}}^{2}H^2_{{4}}-2\,{H^2_{{-1}}}^ {2}{H^3_{{%
-1}}}^{3}H^3_{{1}} \right. \\
&\left. +2\,H^2_{{-1}}{H^3_{{- 1}}}^{4}H^2_{{1}}+8\,{H^3_{{-1}}}^{2}H^2_{{2}%
}H^3_ {{1}}+12\,H^3_{{-1}}H^2_{{4}}H^3_{{1}}+9\,H^3_{{-1 }}{H^2_{{-1}}}%
^{4}H^2_{{4}}-8\,H^3_{{-1}}H^3_{{1}} H^3_{{3}}-6\,{H^2_{{-1}}}^{3}H^3_{{3}%
}H^2_{{1}} \right. \\
&\left. -\frac{9}{2}\,{H^2_{{-1}}}^{4}H^2_{{1}}H^2_{{3}}+\frac{3}{2}\,{H^2_{{%
\ -1}}}^{7}H^3_{{-1}}H^2_{{1}}+15\,{H^2_{{-1}}}^{6} H^2_{{2}}H^3_{{-1}}-%
\frac{1}{2}\,{H^2_{{-1}}}^{5}H^3_{{1}} H^2_{{1}}-3\,{H^2_{{-1}}}^{5}H^2_{{2}%
}H^2_{{1}}-{\frac {33}{2}}\,{H^2_{{-1}}}^{4}H^3_{{-1}}{H^2_{{1}}}^{2} \right.
\\
&\left. +9\,{H^2_{{-1}}}^{3}H^2_{{1}}H^2_{{4}}+9\,H^2_{{-1}}{H^2_{{1}}}%
^{2}H^2_{{3}} +7\,{H^2_{{-1}}}^{2}H^3_{{1} }{H^2_{{1}}}^{2}-3\,{H^2_{{-1}}}%
^{2}H^2_{{2}}{H^2_ {{1}}}^{2}-6\,{H^2_{{-1}}}^{4}{H^2_{{2}}}^{2}-\frac{3}{2}%
\,{H^2 _{{-1}}}^{8}H^3_{{1}}-\frac{3}{2}\,{H^2_{{-1}}}^{7}H^2_{{3}} \right.
\\
&\left. -2 \,{H^2_{{-1}}}^{6}H^3_{{3}}+3\,{H^2_{{-1}}}^{6}H^2 _{{4}}+11\,{%
H^2_{{-1}}}^{3}{H^2_{{1}}}^{3}+12\,H^3_{{3} }{H^2_{{1}}}^{2}-18\,{H^2_{{1}}}%
^{2}H^2_{{4}} +8\,H^2_{{2}}{H^3_{{1}}}^{2}+2\,{H^2_{{2}}}^{3}-9\,{H^2_{{1 }}}%
^{4} \right. \\
&\left. -\frac{21}{2}\,{H^2_{{-1}}}^{3}H^2_{{2}}H^3_{{-1}}H^2_{{1}}
-16\,H^2_{{-1}}H^2_{{2}}H^3_{{1}}H^2_{{1} }-6\,H^2_{{-1}}H^3_{{1}}H^3_{{-1}%
}H^2_{{3}} -\frac{9}{2}\,{H^2_{{-1}}}^{3}H^3_{{1}}H^3_{{-1}}H^2_{{1}} +30\,{%
H^2_{{-1}}}^{2}H^3_{{1}}H^2_{{2}}H^3_{{-1}} \right. \\
&\left. -8\,{H^3_{{-1}}}^{2}H^2_{{1}}H^2_{{-1}}H^2_{{2}} -18\,H^3_{{-1}}H^2_{%
{1}}H^2_{{-1}}H^3_{{3}}+27\,H^3 _{{-1}}H^2_{{1}}H^2_{{4}}H^2_{{-1}} -{\frac {%
27}{2}}\,H^2_{{3}}{H^2_{{-1}}}^{2}H^3_{{-1}}H^2_{{1}} -10\,H^2_{{-1}}H^3_{{1}%
}{H^3_{{-1}}}^{2}H^2_{{1}} \right. \\
&\left. -\frac{9}{2}\,H^3_{{-1}}{H^2_{{-1}}}^{5}H^2_{{3}}-3\,{H^2_{{1}}}^ {2}%
{H^2_{{-1}}}^{6}+3\,{H^2_{{-1}}}^{8}H^2_{{2}} -8\,H^3_{{1}}{H^2_{{2}}}%
^{2}+12\,H^2_{{4}}H^3_{{3}}-6\, H^3_{{-1}}{H^2_{{-1}}}^{4}H^3_{{3}}-6\,H^2_{{%
3}}H^2_{{-1}}H^3_{{3}} \right. \\
&\left. +27\,H^3_{{-1}}H^2_{{-1}}{H^2_{{1}}}^{3}-\frac{13}{2}\,H^3_{{-1}}{%
H^2_{{-1}}}^{6}H^3_{{1} }-9\,{H^2_{{4}}}^{2}+{\frac {29}{2}}\,{H^2_{{-1}}}%
^{4}H^2_{{2}}H^3_{{1}} -4\,{H^3_{{3}}}^{2}-\frac{9}{4}\,{H^2_{{3}}}^{2}{H^2_{%
{-1}}}^{2} -{\frac {21}{4}}\,{H^2_{{-1}}}^{4}{H^3_{{1}}}^{2} \right. \\
&\left. +2\,H^2_{{2}}{H^3_{{-1}}}^{4} -4\,{H^3_{{-1}}}^{2}{H^3_{{1}}}^{2}-4\,%
{H^3_{{-1}}}^{2}{H^2_{{2}}}^{2} \right)\ .
\end{split}
\notag
\end{equation}
} The coefficients in equation (\ref{1F34-ideal}) are {\small 
\begin{equation}
\begin{split}
a=&-2\,H^2_{{-1}}\ , \\
b=&4\,H^3_{{-1}}+4\,{H^2_{{-1}}}^{2}\ , \\
c=&4\,H^3_{{1}}-2\,{H^3_{{-1}}}^{2}-{H^2_{{-1}}}^{2}\pi_{{\ 2}}-2\,H^2_{{-1}%
}H^2_{{1}}+{\pi_{{2}}}^{2}-2\,H^2_{{2}}\ , \\
d=&-4\,H^2_{{-1}}{\pi_{{2}}}^{2}+4\,{H^2_{{-1}}}^{3} \pi_{{2}}+11\,H^2_{{1}}{%
H^2_{{-1}}} ^{2}-4\,H^2_{{-1}}H^3_{{1}}+12\,H^3_{{-1}}H^2_{{1} }-5\,H^3_{{-1}%
}{H^2_{{-1}}}^{3}-6\,H^2_{{-1}}{H^3_ {{-1}}}^{2} \\
&+2\,H^2_{{-1}}H^2_{{2}} \\
f=&-2\,{H^2_{{-1}}}^{2}{\pi_{{2}}}^{3}+{H^2_{{-1}}}^{4}{\pi_{{2 }}}%
^{2}+4\,H^3_{{1}}{\pi_{{2}}}^{2}-2\,{H^3_{{-1}}}^{2}{\pi_ {{2}}}^{2}-4\,{%
\pi_{{2}}}^{2}H^2_{{2}}+4\,{H^2_{{-1}}}^{2} \pi_{{2}}H^2_{{2}}-3\,{H^2_{{-1}}%
}^{5}H^2_{{1}} \\
&+10\,{H^2_{{-1}}}^{2}{H^2_{{1}}}^{2}+4\,{H^2_{{-1}}}^{4}{H^3_{{-1}}}^{2}
+6\,{H^2_{{-1}}}^{4}H^2_{{2}}+H^3_{{-1}} {H^2_{{-1}}}^{6}-8\,H^3_{{3}}H^3_{{%
-1}}+4\,{H^3_{{\ 1}}}^{2}+4\,{H^2_{{-1}}}^{2}{H^3_{{-1}}}^{3} \\
&+{H^2_{{2}} }^{2}+{\pi_{{2}}}^{4}+{H^3_{{-1}}}^{4}-4\,H^2_{{-1}}H^2 _{{1}%
}H^3_{{1}}-6\,H^2_{{-1}}H^2_{{3}}H^3_{{-1}}+9 \,H^2_{{4}}{H^2_{{-1}}}^{2}-4\,%
{H^3_{{-1}}}^{2}H^3_{{1}} -6\,{H^2_{{-1}}}^{2}H^3_{{3}} \\
&+2\,H^2_{{1}}H^2_{{-1}}H^2_{{2}} -12\,H^2_{{-1}}{H^3_{{-1}}}^{2}H^2_{{1}%
}-4\,{H^2_{{-1}}}^{2}H^3_{{-1}}H^3_{{1}} -{\frac {27}{2}}\,{H^2_{{-1}}}%
^{3}H^3_{{-1}}H^2_{{1}} +12\,{H^2_{{-1}}}^{2}H^2_{{2}}H^3_{{-1}} \\
&+12\,H^3_{{-1}}H^2_{{4}}+12\,H^3_{{-1}}{H^2_{{1}}}^{2}-\frac{9}{2}\,H^2_{{3}%
}{\ H^2_{{-1}}}^{3}-\frac{5}{2}\,{H^2_{{-1}}}^{4}H^3_{{1}} -4\,H^2_{{2}}H^3_{%
{1}}+6\,H^2_{{2}}{H^3_{{-1}}}^{2} -4\,H^3_{{1}}{H^2_{{-1}}}^{2}\pi_{{2}} \\
&+2\,{H^3_{{-1}}}^{2}{H^2_{{-1}}}^{2}\pi_{{2}} +4\,{H^2_{{-1}}}^{3}\pi_{{2}%
}H^2_{{1}}-4\,H^2_{{-1}}H^2_{{1}}{\pi_{{2}}}^{2}\ , \\
h=&3\,H^2_{{-1}}\pi_{{2}}-4\,{H^2_{{-1}}}^{3}-3\,H^2_{{1}} -H^2_{{-1}}H^3_{{%
-1}}\ , \\
j=&-{\pi_{{2}}}^{3}-3\,H^2_{{4}}+{H^2_{{-1}}}^{6} -\pi_{{2}}{H^3_{{-1}}}%
^{2}-5\,\pi_{{2}}{H^2_{{-1}}}^{2}H^3_{{-1}}+3 \,\pi_{{2}}H^2_{{-1}}H^2_{{1}%
}+3\,\pi_{{2}}H^2_{{2}}-4 \,H^3_{{-1}}H^2_{{2}} \\
&+\frac{7}{2}\,{H^2_{{-1}}}^{2}H^3_{{1 }}+\frac{3}{2}\,H^2_{{-1}}H^2_{{3}%
}-6\,H^2_{{2}}{H^2_{{-1} }}^{2}+2\,H^3_{{-1}}H^3_{{1}}+\frac{1}{2}\,H^2_{{-1}%
}H^2_{ {1}}H^3_{{-1}}-3\,{H^2_{{1}}}^{2} +3\,{H^2_{{-1}}}^{4}H^3_{{-1}} \\
&-3\,{H^2_{{-1}}}^{3}H^2_{{1}}+{H^2_{{-1}} }^{2}{H^3_{{-1}}}^{2}-2\,\pi_{{2}%
}H^3_{{1}}+2\,H^3_{{3} }-3\,{H^2_{{-1}}}^{4}\pi_{{2}}+2\,H^3_{{-1}}{\pi_{{2}}%
}^{2}+ 3\,{H^2_{{-1}}}^{2}{\pi_{{2}}}^{2}\ .
\end{split}%
\end{equation}%
}

\section[Stratum $\Sigma_{1,2}$]{Stratum $\Sigma_{1,2}$}

\label{App-2strat} The coefficients $N_i$ in the formula (\ref{2stratC9})
are {\small 
\begin{equation}
\begin{split}
N_4=& - \left( 3\,H^3_{{-2}}H^3_{{1}}+3\,H^3_{{2}}+3\,{\ H^3}_{{-2}}{H^3_{{-1%
}}}^{2}-H^4_{{-2}}H^5_{{-2}}-{H^4 }_{{1}}-2\,{H^3_{{-2}}}^{2}H^5_{{-2}%
}-3\,H^3_{{-2}}{\ H^5}_{{-1}}-5\,H^3_{{-2}}H^3_{{-1}}H^4_{{-2}}\right. \\
&\left.+{\ H^3}_{{-2}}{H^4_{{-2}}}^{2}-H^4_{{-1}}H^3_{{-1}}+{H^4 }_{{-1}%
}H^4_{{-2}}-2\,{H^3_{{-2}}}^{3}H^3_{{-1}}-{{\ H^3}_{{-2}}}^{5}+2\,H^3_{{-1}%
}H^5_{{-2}}+{H^3_{{-2}}}^{2 }H^4_{{-1}}-{H^3_{{-2}}}^{3}H^4_{{-2}} \right) \
, \\
N_3=&- \left( -2\,{H^3_{{-2}}}^{2}{\ H^3}_{{-1}}H^4_{{-2}}-H^5_{{1}}-H^4_{{2}%
}+3\,{H^3 }_{{3}}-3\,H^4_{{-2}}{H^3_{{-1}}}^{2}+{H^4_{{-2}}}^{2}{\ H^3}_{{-1}%
}+3\,H^3_{{-2}}H^3_{{2}}+H^3_{{-2}}{H^4 }_{{-1}}H^4_{{-2}}\right. \\
&\left. -H^3_{{-2}}H^3_{{-1}}H^5_{{-2}}- 2\,{H^3_{{-2}}}^{4}H^3_{{-1}}+{H^3_{%
{-1}}}^{3}-H^4 _{{-1}}H^5_{{-2}}-H^4_{{-2}}H^5_{{-1}}-3\,{H^3_{{- 2}}}%
^{2}H^5_{{-1}}+3\,{H^3_{{-2}}}^{2}H^3_{{1}}+3\,{\ H^3}_{{1}}H^3_{{-1}}\right.
\\
& -H^3_{{-2}}H^3_{{-1}}H^4_{{\ -1}}+H^4_{{-2}}H^3_{{1}} \Big) \ , \\
N_0=&-3\,H^3_{{6}}+H^4_{{5}}+H^4_{{-2}}{\ H^5}_{{2}}+H^4_{{-1}}H^5_{{1}%
}+3\,H^3_{{-2}}H^3_{{- 1}}H^5_{{1}}+3\,H^3_{{2}}H^4_{{-2}}H^3_{{-1}}+H^3_{{1}%
}H^4_{{-1}}H^3_{{-1}} +H^4_{{1}}H^5_{{-1}}-H^3_{{-2}}{H^4 }_{{2}}H^4_{{-2}}
\\
&-2\,H^3_{{-2}}H^3_{{2}}H^4_{{-1}} +H^3_{{-2}}H^3_{{2}}H^5_{{-2}}+3\,H^3_{{-2%
}}{\ H^3}_{{1}}H^5_{{-1}}-2\,H^5_{{-2}}H^3_{{1}}H^3_{{-1 }}-H^4_{{3}}H^4_{{-2%
}}+H^5_{{4}}-H^3_{{4}}H^4 _{{-2}}+3\,{H^3_{{-2}}}^{2}H^5_{{2}} \\
& +3\,{H^3_{{-1}}}^{2 }H^4_{{1}}+3\,H^4_{{3}}H^3_{{-1}}-3\,H^3_{{4}}{{\ H^3}%
_{{-2}}}^{2}-3\,H^3_{{2}}{H^3_{{-1}}}^{2}-3\,H^3 _{{-2}}{H^3_{{1}}}^{2}+2\,{%
H^3_{{-2}}}^{4}H^3_{{2}}+2\, {H^3_{{-2}}}^{3}H^3_{{3}}-2\,H^4_{{-1}}H^3_{{3}}
\\
&-3\,H^3_{{-2}}H^3_{{5}}-2\,H^3_{{3}}H^5_{{-2}}-3\,{\ H^3}_{{4}}H^3_{{-1}%
}-6\,H^3_{{2}}H^3_{{1}}+3\,{\ H^3}_{{-2}}H^5_{{3}}+3\,H^3_{{-2}}H^3_{{-1}%
}H^4_{{2 }}-6\,H^3_{{3}}H^3_{{-2}}H^3_{{-1}} \\
&+2\,{H^3_{{-2}} }^{3}H^3_{{1}}H^3_{{-1}}+ 2\,H^3_{{-2}}H^3_{{3}}H^4_{{-2}%
}+2\,{H^3_{{-2}}}^{ 2}H^4_{{-2}}H^3_{{2}}-H^3_{{-1}}H^4_{{1}}H^4_ {{-2}%
}-H^3_{{2}}{H^4_{{-2}}}^{2}+H^4_{{2}}H^5_{{-2 }}-H^3_{{1}}H^4_ {{-1}}H^4_{{-2%
}} \\
&+2\,H^3_{{-2}}H^4_{{-2}}H^3_{{1}}{\ H^3}_{{-1}}\ ,
\end{split}%
\end{equation}
} and for the formula (\ref{2stratC10}) are {\small
\begin{equation}
\begin{split}
B_4=&9\,H^3_{{1}}{H^3_{{-2}}}^{2}-H^3_{{-2}}H^5_{{-2}}H^4_{{-2}}-3\,{H^3_{{-2%
}}}^{4}H^3_{{-1}}-7\,H^5_{{-1}}{H^3_{{-2}}}^{2}-3\,{H^3_{{-2}}}^{3}H^5_{ {-2}%
}+4\,H^3_{{-1}}H^3_{{1}}+2\,H^4_{{-1}}H^4_{{-2 }}H^3_{{-2}} \\
&+4\,H^3_{{2}}H^3_{{-2}}-2\,H^4_{{1}}H^3_{{-2}}-H^4_{{2}}-H^4_{{-2}}H^3_{{1}%
}+{H^4_ {{-2}}}^{2}{H^3_{{-2}}}^{2}-2\,{H^3_{{-1}}}^{3}-2\,H^3_ {{3}}+H^4_{{%
-2}}{H^3_{{-1}}}^{2}-{H^3_{{-2}}}^{6}+{H^4_{{-1}}}^{2} \\
&-6\,H^4_{{-1}}H^3_{{-1}}H^3_{{-2}}+2 \,H^5_{{1}}+{H^5_{{-2}}}^{2}-5\,H^4_{{%
-2}}{H^3_{{- 2}}}^{2}H^3_{{-1}}+3\,{H^3_{{-1}}}^{2}{H^3_{{-2}}}^{2}- 2\,H^5_{%
{-1}}H^3_{{-1}}+6\,H^3_{{-2}}H^5_{{-2}}H^3_{{-1}} \\
&-2\,H^4_{{-1}}H^5_{{-2}} -H^4_{{-2}}{H^3_{{-2}}}^{4}+H^5_{{-1}}H^4_{{-2}}\ ,
\\
B_3=&- \left( -H^4_{{1}}{H^3_{{-2}}}^{2}+7\,H^3_{{2}}{H^3_{{-2}}}^{2} +2\,{%
H^3_{{-1}}}^{2}H^5_{{-2}}-5\,{H^3_{{-2}}}^{3}H^5_{{-1}} -2\,H^3_{{1}}H^5_{{-2%
}}-2\,{H^3_{{-1}}}^{2}{H^3_{{-2}}}^{3}-2\,H^3_{{-1}}{H^3_{{-2}}}
^{5}+5\,H^3_{{1}}{H^3_{{-2}}}^{3}\right. \\
&\left.-H^3_{{-1}}H^5_{{\ -2}}{H^3_{{-2}}}^{2}-H^3_{{-1}}H^5_{{-2}}H^4_{{-2}
}+H^3_{{-2}}H^5_{{-1}}H^4_{{-2}}+{H^4_{{-2}}}^{2}H^3_{{-1}}H^3_{{-2}} -H^4_{{%
-2}}H^3_{{-1}}{H^3_{{-2}}}^{3}+10\,H^3_{{1}}H^3_{{-1}}H^3_{{-2}}\right. \\
&\left.-H^4_ {{-2}}H^3_{{1}}H^3_{{-2}}-H^4_{{-1}}{H^3_{{-1}}}^{ 2}-H^4_{{-1}%
}H^3_{{1}}+{H^4_{{-1}}}^{2}H^3_{{-2}}- H^4_{{-1}}{H^3_{{-2}}}^{4}-H^4_{{-2}%
}H^3_{{2}}+4\, H^3_{{3}}H^3_{{-2}}+2\,H^5_{{-1}}H^5_{{-2}}-2\,H^4_{{2}}H^3_{{%
-2}}\right. \\
&\left.+2\,H^5_{{2}} -H^4_{{3}}-2\,H^3_{{4}}-2\,H^3_{{-2}}H^5_{{-1}}H^3_{{-1}%
}-4\,H^4 _{{-2}}{H^3_{{-1}}}^{2}H^3_{{-2}}-4\,H^4_{{-1}}{H^3_{{-2}}}^{2}H^3_{%
{-1}} -2\,H^4_{{-1}}H^3_{{-2}}H^5_{{-2}}+H^4_{{-1}}{H^3_{{-2}}}^{2}H^4_{{-2}%
}\right. \\
&\left.+H^4_{ {-2}}H^3_{{-1}}H^4_{{-1}}-2\,H^4_{{1}}H^3_{{-1}} \right)\ , \\
B_0=& 2\,{H^3_{{-2}}}^{4}H^3_{{-1}}H^3_{{1}}+H^4_{{-1}}H^3_{{4}} +H^4_{{2}}{%
H^3_{{-1}}}^{2}-H^4_{{3}}H^4_{{-1}}-2\,H^3_{{2}}H^5_{{-2}}H^3_{{-1}} -H^3_{{%
-2}}H^5_{{2}}H^4_{{-2}}+2\,H^4_{{2}}H^3_{{1}} +6\,H^4_{{1}}{H^3_{{-1}}}%
^{2}H^3_{{-2}} \\
&+H^4_{{-2}}H^3_{{2}}H^5_{{-2}} +2\,H^4_{{-1}}H^3_{{1}}H^5_{{-2}}-2\,H^4_{{-1%
}}{H^3_{{-2}}}^{2}H^3_{{2}}-6\,H^3_{ {4}}H^3_{{-1}}H^3_{{-2}} -2\,H^3_{{-2}%
}H^5_{{2}}H^3_{{-1}}-2\,H^3_{{3}}H^3_{{1}}+2\,{H^3_{{-2}}}^{2 }{H^3_{{-1}}}%
^{2}H^3_{{1}} \\
&+2\,{H^3_{{-2}}}^{4}H^3_ {{3}}+2\,H^4_{{5}}H^3_{{-2}}-7\,H^3_{{5}}{H^3_{{-2 
}}}^{2}+2\,H^4_{{1}}H^3_{{2}}+2\,H^4_{{4}}H^3_{{-1 }}-2\,{H^3_{{-1}}}%
^{2}H^5_{{1}} +5\,{H^3_{{-2}}}^{3}H^5_{{2}}+2\,H^3_{{4}}H^5_{{-2}} -5\,H^3_{{%
4}}{H^3_{{-2}}}^{3} \\
&-2\,H^5_{{3}}H^3_{{-1}} +5\,H^5_{{3}}{H^3_{{-2}}}^{2}+2\,H^4_{{3}}H^5_{{-2}%
}-2\,H^5_{{2}} H^5_{{-2}}+2\,H^4_{{1}}H^3_{{-1}}H^5_{{-2}} +H^4_{{4}}{H^3_{{%
-2}}}^{2}-4\,H^3_{{6}}H^3_{{-2}} -H^5_{{3}}H^4_{{-2}} \\
&+{H^3_{{-2}}}^{2}H^5_{{-2}}H^3_ {{2}}+2\,H^3_{{1}}H^5_{{-1}}H^3_{{-1}}
+H^4_{{-2}}H^3_{{5}}-16\,H^3_{{-2}}H^3_{{2}}H^3_{{1}}-H^4 _{{2}}{H^3_{{-2}}}%
^{2}H^4_{{-2}}-H^4_{{-2}}H^3_{{2 }}H^4_{{-1}}+8\,H^4_{{3}}H^3_{{-1}}H^3_{{-2}%
}+H^4_{{6}} \\
&+2\,H^3_{{7}}+2\,{H^3_{{-1}}}^{2}H^3_{{3}}+2\,{H^3_{{-2}}}^{5}H^3_{{2}} +2\,%
{H^3_{{-1}}}^{3}H^3_{{1}}-H^4_{{1}}H^3_{{-1}}H^4_{{-2}}H^3_{{-2}}- H^4_{{-1}%
}H^3_{{1}}H^4_{{-2}}H^3_{{-2}} +4\,H^4_{{-2}}H^3_{{2}}H^3_{{-1}}H^3_{{-2}} \\
&+4\,H^4_{{-1} }H^3_{{1}}H^3_{{-1}}H^3_{{-2}}-4\,H^3_{{-2}}H^5_{{-2}}H^3_{{-1%
}}H^3_{{1}} -6\,{H^3_{{-2}}}^{2}{H^3_{{1}}}^{2}+4\,H^5_{{-1}}H^3_{{3}}
-2\,H^5_{{-1}}H^5_{{1}}+6\,H^4_{{2}}{H^3_{{-2}}}^{2}H^3_{{-1}} \\
&-10\,H^3_{{3}}H^3_{{-1}}{H^3_{{-2}}}^{2}+H^4_{{-2}}{H^3_{{1}}}^{2} +5\,H^3_{%
{1}}H^5_{{-1}}{H^3_{{-2}}}^{2}-H^4_{{3}}H^4_{{-2}}H^3_{{-2}} +2\,H^4_{{2}%
}H^3_{{-2}}H^5_{{-2}}-H^4_{{1}}H^3_{{-1}}H^4_{{-1}} +H^4_{{-2}}H^3_{{-2}%
}H^3_{{4}} \\
&-4\,H^3_{{-2}}H^3_{{2}}{H^3_{{-1}}}^{2} -4\,H^3_{{-2}}H^5_{{-2}}H^3_{{3}}-{%
H^4_{{-2}}}^{2}H^3_{{2}}H^3_{{-2}} +2\,H^4_{{-1}}H^3_{{2}}H^3_{{-1}}+2\,H^4_{%
{1}}H^3_{{1} }H^3_{{-2}}+2\,{H^3_{{-2}}}^{3}H^3_{{-1}}H^3_{{2}} \\
&+5\,H^3_{{-1}}H^5_{{1}}{H^3_{{-2}}}^{2}+H^4_{{-1}} H^3_{{1}}{H^3_{{-2}}}%
^{3}-2\,H^4_{{-2}}{H^3_{{-1}} }^{2}H^3_{{1}}+H^4_{{2}}{H^3_{{-2}}}^{4}-{H^4_{%
{-1 }}}^{2}H^3_{{1}}-H^3_{{1}}H^5_{{-1}}H^4_{{-2}} +{H^3_{{2}}}^{2}-2\,H^5_{{%
5}} \\
&+H^4_{{-2}}H^3_{{2}}{H^3_{{-2}}}^{3} +4\,H^5_{{-1}}H^3_{{2}}H^3_{{-2}}-H^3_{%
{-1}}H^5_{{1}}H^4_{{-2}} -4\,H^3_{{-1}}{H^3_{{1}}}^{2}+H^4_{{3}}{H^3_{{-2}}}%
^{3} +H^4_{{1}}H^3_{{-1}}{H^3_{{-2}}}^{3}-H^4_{{2}}H^3_{{-2}}H^4_{{\ -1}}\ .
\end{split}%
\end{equation}
} For the trigonal curve (\ref{2strattrigcurv}) the coefficients $Q_k$ are
given by {\small
\begin{equation}
\begin{split}
Q_8=&4\,H^3_{{1}}-2\,{H^3_{{-1}}}^{2}+4\,H^3_{{-1}}{H^3_{{-2}}}^{2} -{H^3_{{%
-2}}}^{4}+H^3_{{-2}}H^4_{{-1}}-{H^3_{{-2}}}^{2}H^4_{{-2}}+2\,H^4_{{-2} }H^3_{%
{-1}}\ , \\
Q_7=&-3\,H^4_{{-1}}H^4_{{-2}}-3\,H^4_{{1}}+4\,H^3_{{2}}+4\,{H^3_{{-1}}}^{2}
H^3_{{-2}}+8\,H^3_{{1}}H^3_{{-2}}+2\,H^3_{{-2}}{H^4_{{-2}}}^{2} -{H^3_{{-2}}}%
^{2}H^4_{{-1}}-4\,H^3_{{-2}}H^4_{{-2}}H^3_{{-1}} \\
&+5\,H^4_{{-1}}H^3_{{-1}}+2\,{H^3_{{-2}}}^{3}H^4_{{-2}}\ , \\
Q_6=&4\,H^3_{{3}}-3\,H^4_{{2}}+4\,H^3_{{2}}H^3_{{-2}}+4\,H^3_{{1}}H^3_{{-1}
}-{H^4_{{-2}}}^{3}-3\,{H^4_{{-1}}}^{2}-{H^3_{{-2}}}^{2}{H^4_{{-2}}}^{2
}-3\,H^3_{{-2}}H^4_{{-1}}H^3_{{-1}}-5\,H^3_{{-2}}H^4_{{1}} \\
& -H^3_{{-2}}H^4_{{-1}}H^4_{{-2}}+2\,{H^4_{{-2}}}^{2}H^3_{{-1}}-H^4_{{-2}}{%
H^3_{{-1} }}^{2}-2\,H^4_{{-2}}H^3_{{1}}\ , \\
Q_4=&{H^3_{{-2}}}^{3}H^4_{{1}}+{H^4_{{-1}}}^{2}H^3_{{-1}}+8\,H^3_{{4}} H^3_{{%
-2}}-4\,H^3_{{3}}H^3_{{-1}}-3\,H^4_{{2}}H^4_{{-2}}-3\,H^4_{{1}} H^4_{{-1}%
}-4\,H^3_{{2}}{H^3_{{-2}}}^{3}-4\,H^3_{{1}}{H^3_{{-1}}}^{2} \\
& +4\,H^3_{{3}}{H^3_{{-2}}}^{2}-8\,H^3_{{-2}}H^4_{{3}}-7\,{H^3_{{-2}}}^{2}
H^4_{{2}}+2\,H^4_{{2}}H^3_{{-1}}+6\,H^4_{{-2}}H^3_{{3}}+2\,{H^4_{{-2}}}^{2}
H^3_{{1}}-2\,H^4_{{-2}}{H^3_{{-1}}}^{3}+5\,H^4_{{-1}}H^3_{{2}} \\
&. +H^3_{{-2}}H^4_{{-2}}H^4_{{1}}+{H^3_{{-2}}}^{2}H^4_{{-2}}{H^3_{{-1}}}^{2}
+4\,H^3_{{5}}-3\,H^4_{{4}}+6\,{H^3_{{1}}}^{2}+{H^4_{{-2}}}^{2}{H^3_{{-1}}}^{
2}-H^3_{{-2}}H^4_{{-2}}H^4_{{-1}}H^3_{{-1}}-{H^3_{{-2}}}^{3}H^4_{{-1}} H^3_{{%
-1}}\ , \\
& +{H^3_{{-1}}}^{4}+8\,H^3_{{2}}H^3_{{-1}}H^3_{{-2}} +2\,H^3_{{-2}}H^4_{{-1}%
}H^3_{{1}}+2\,{H^3_{{-2}}}^{2}H^4_{{-2}}H^3_{{1}} -6\,H^3_{{-1}}H^3_{{-2}%
}H^4_{{1}}+4\,H^3_{{1}}H^3_{{-1}}{H^3_{{-2}}}^{2} +3\,H^3_{{-2}}H^4_{{-1}}{%
H^3_{{-1}}}^{2} \\
& +2\,H^4_{{-2}}H^3_{{1}}H^3_{{-1}}\ ,
\end{split}%
\end{equation}
\begin{equation}
\begin{split}
Q_3=&-H^4_{{-1}}{H^3_{{-1}}}^{3}+8\,H^3_{{5}}H^3_{{-2}}+4\,H^3_{{4}} H^3_{{-1%
}}+8\,H^3_{{2}}H^3_{{1}}-3\,H^4_{{3}}H^4_{{-2}}-6\,H^4_{{2}}H^4_{{-1}
}-3\,H^4_{{1}}{H^4_{{-2}}}^{2}+6\,{H^4_{{-2}}}^{2}H^3_{{2}} \\
& +4\,{H^3_{{1}}}^{2}H^3_{{-2}} +4\,H^3_{{2}}{H^3_{{-1}}}^{2}+8\,{H^3_{{-2}}}%
^{2} H^3_{{4}}-8\,H^3_{{-2}}H^4_{{4}}-10\,{H^3_{{-2}}}^{2}H^4_{{3}} -4\,H^4_{%
{3}}H^3_{{-1}}+6\,H^4_{{-2}}H^3_{{4}}-4\,H^4_{{1}}{H^3_{{-1}}}^{2} \\
& -4\,{H^3_{{-2}}}^{3}H^4_{{2}}+9\,H^4_{{-1}}H^3_{{3}}-5\,H^3_{{1}}H^4_{{1}}
+H^3_{{-2}}{H^4_{{-1}}}^{2}H^4_{{-2}}-{H^3_{{-2}}}^{2}H^4_{{-1}} H^4_{{-2}%
}H^3_{{-1}}+4\,H^4_{{-2}}H^3_{{1}}H^3_{{-1}}H^3_{{-2}}+4\,H^3_{{6}} \\
& -3\,H^4_{{5}}+{H^3_{{-2}}}^{3}{H^4_{{-1}}}^{2} +H^4_{{-1}}H^4_{{-2}}H^3_{{1%
}}-{H^4_{{-2}}}^{2}H^3_{{-1}}H^4_{{-1}}-{H^4_{{-1}}}^{3}-6\,H^4_{{1} }H^3_{{%
-1}}{H^3_{{-2}}}^{2}+5\,H^4_{{-1}}H^3_{{1}}H^3_{{-1}} +6\,H^3_{{-2}}H^4_{{-1}%
}H^3_{{2}} \\
& +4\,H^3_{{-2}}H^4_{{-2}}H^3_{{3}} +8\,H^3_{{3}}H^3_{{-1}}H^3_{{-2}}-2\,{%
H^3_{{-2}}}^{2}H^4_{{-1}}H^3_{{1}} -7\,H^3_{{-2}}H^4_{{-1}}H^4_{{1}}-6\,H^4_{%
{-2}}H^3_{{2}}H^3_{{-1}} +8\,H^4_{{-2}}H^3_{{2}}{H^3_{{-2}}}^{2} \\
& -3\,H^3_{{-1}}H^3_{{-2}}{H^4_{{-1}}}^{2} -10\,H^3_{{-1}}H^3_{{-2}}H^4_{{2}%
}-2\,H^3_{{-2}}H^4_{{2}}H^4_{{-2}} -3\,{H^3_{{-2}}}^{2}H^4_{{1}}H^4_{{-2}%
}+5\,H^4_{{1}}H^4_{{-2}}H^3_{{-1}} +2\,H^4_{{-1}}H^4_{{-2}}{H^3_{{-1}}}^{2}\
,
\end{split}%
\end{equation}
\begin{equation}
\begin{split}
\\
Q_0=&8\,H^3_{{8}}H^3_{{-2}}+H^4_{{-1}}H^4_{{-2}}H^3_{{2}}H^3_{{-1}} -H^4_{{-2%
}}H^3_{{-1}}H^4_{{1}}H^4_{{-1}}+H^3_{{-2}}H^4_{{2}}H^4_{{-1}}H^4_{{-2 }%
}-13\,H^3_{{-2}}H^4_{{2}}H^4_{{-1}}H^3_{{-1}}+4\,H^3_{{9}}-3\,H^4_{{8 }} \\
& -2\,{H^3_{{3}}}^{2}-3\,{H^4_{{2}}}^{2}+{H^3_{{-2}}}^{3}H^4_{{2}}H^4_{{-1}%
}+H^4_{{-1}}H^3_{{2}}{H^3_{{-1}}}^{2}+H^4_{{-1}}{H^3_{{1}}}^{2}H^3_{{-2}}-{%
H^3_{{-1}}}^{2}H^4_{{1}}H^4_{{-1}}+{H^4_{{-1}}}^{2}H^3_{{1}} {H^3_{{-2}}}^{2}
\\
& +{H^4_{{-1}}}^{2}H^3_{{1}}H^3_{{-1}}-H^4_{{-2}}{H^3_{{\ 2}}}^{2}+4\,{H^3_{{%
1}}}^{3}-12\,H^4_{{1}}H^3_{{2}}H^3_{{-1}}-6\,H^4_{{\ -2}}H^3_{{3}}{H^3_{{-1}}%
}^{2}+3\,{H^4_{{-2}}}^{2}H^3_{{2}}H^4_{{-1}}+3 \,{H^4_{{-1}}}^{2}H^3_{{1}%
}H^4_{{-2}} \\
& -2\,H^4_{{-2}}H^3_{{5}}H^3_{{-1}} -2\,H^4_{{2}}H^3_{{1}}H^3_{{-1}%
}+16\,H^3_{{4}}H^3_{{-2}}H^3_{{1}} +6\,H^4_{{1}}H^3_{{2}}{H^3_{{-2}}}%
^{2}-4\,H^4_{{1}}H^3_{{1}}{H^3_{{-2}}}^{3 }+2\,{H^4_{{-2}}}^{3}H^3_{{1}}H^3_{%
{-1}}-5\,H^3_{{-2}}H^4_{{2}}H^4_{{1 }} \\
& -12\,H^4_{{2}}{H^3_{{-2}}}^{2}H^3_{{1}}+16\,H^3_{{6}}H^3_{{-1}}H^3_{{-2}%
}-12\,H^4_{{3}}{H^3_{{-1}}}^{2}H^3_{{-2}} +5\,H^4_{{-1}}H^3_{{4}}H^3_{{-1}%
}+10\,H^3_{{-2}}H^4_{{-1}}H^3_{{5}} +4\,H^3_{{-2}}H^4_{{-2}}H^3_{{6}} \\
& +14\,H^4_{{-1}}H^3_{{2}}H^3_{{1}}+8\,H^3_{{4}}H^3_{{-2}}{H^3_{{-1 }}}%
^{2}+4\,H^3_{{7}}H^3_{{-1}}+12\,H^3_{{5}}H^3_{{1}} +8\,H^3_{{4}}H^3_{{2}%
}-3\,H^4_{{6}}H^4_{{-2}}-6\,H^4_{{5}}H^4_{{-1}}-3\,H^4_{{3}} H^4_{{1}} \\
& -3\,H^4_{{4}}{H^4_{{-2}}}^{2}-4\,H^4_{{2}}{H^3_{{-1}}}^{3} +3\,{H^4_{{-2}}}%
^{2}{H^3_{{1}}}^{2}+6\,{H^4_{{-2}}}^{2}H^3_{{5}} -3\,H^4_{{2}}{H^4_{{-1}}}%
^{2}-3\,{H^4_{{1}}}^{2}H^4_{{-2}}+4\,H^3_{{5}}{H^3_{{-1}}}^{ 2}+8\,{H^3_{{2}}%
}^{2}H^3_{{-1}} \\
& -4\,{H^3_{{2}}}^{2}{H^3_{{-2}}}^{2}+8\, H^3_{{7}}{H^3_{{-2}}}^{2}-2\,{H^3_{%
{1}}}^{2}{H^3_{{-1}}}^{2}+4\,H^3_{{\ 3}}{H^3_{{-1}}}^{3}-8\,H^3_{{-2}}H^4_{{7%
}}-4\,{H^3_{{-2}}}^{2}{H^4_{{1 }}}^{2}-10\,{H^3_{{-2}}}^{2}H^4_{{6}}-4\,H^4_{%
{6}}H^3_{{-1}} \\
& +6\,H^4_{{-2}}H^3_{{7}} -4\,H^4_{{4}}{H^3_{{-1}}}^{2}-4\,H^4_{{5}}{H^3_{{-2%
}}}^{3 }+9\,H^4_{{-1}}H^3_{{6}}-8\,H^3_{{1}}H^4_{{4}}-4\,H^4_{{3}}H^3_{{2}%
}+3 \,H^3_{{4}}H^4_{{1}}+3\,{H^4_{{1}}}^{2}H^3_{{-1}}+6\,H^4_{{2}}H^3_{{3} }
\\
& +2\,{H^4_{{-2}}}^{3}H^3_{{3}}+6\,{H^4_{{-1}}}^{2}H^3_{{3}}-2\,H^3_{{1 }%
}H^3_{{-2}}H^4_{{1}}H^4_{{-2}}+8\,H^4_{{-2}}H^3_{{4}}H^3_{{-1}}H^3_{{\ -2}%
}+8\,H^4_{{-2}}{H^3_{{-2}}}^{2}H^3_{{3}}H^3_{{-1}} +2\,H^3_{{-1}}H^3_{{-2}%
}H^4_{{3}}H^4_{{-2}} \\
& +8\,H^3_{{2}}H^3_{{-1}}H^3_{{1}}H^3_{{-2}}-6 \,H^4_{{1}}H^3_{{1}}H^3_{{-1}%
}H^3_{{-2}}+16\,H^4_{{-1}}H^3_{{3}}H^3_{{\ -1}}H^3_{{-2}}+2\,H^4_{{-1}}{H^3_{%
{-2}}}^{2}H^3_{{2}}H^3_{{-1}} +8\,H^3_{{-2}}H^4_{{-2}}H^3_{{2}}H^3_{{1}} \\
& +4\,H^3_{{-2}}H^4_{{-2}}H^3_{{2}}{H^3_{{-1}}}^{2}-7\,H^3_{{-1}}{H^3_{{-2}}}%
^{2}H^4_{{1}}H^4_{{-1}}+2\,{H^3_{{-2}}}^{2}H^4_{{-2}}H^4_{{1}}H^4_{{-1}}
-2\,H^4_{{1}}H^3_{{-1}}H^3_{{-2}}{H^4_{{-2}}}^{2}+4\,H^4_{{1}}{H^3_{{-1}}}%
^{2}H^3_{{-2}}H^4_{{-2}} \\
& -2\,H^4_{{1}}H^3_{{-1}}{H^3_{{-2}}}^{3}H^4_{{-2}}+2\,H^4_{{-1}}H^3_{{1 }}{%
H^3_{{-1}}}^{2}H^3_{{-2}}-2\,H^4_{{-1}}H^3_{{1}}H^3_{{-2}}{H^4_{{-2 }}}%
^{2}-4\,{H^3_{{-2}}}^{2}H^4_{{2}}H^4_{{-2}}H^3_{{-1}}-2\,H^4_{{-1}} H^3_{{1}}%
{H^3_{{-2}}}^{3}H^4_{{-2}} \\
& +3\,H^4_{{-2}}H^3_{{2}}{H^3_{{-2}}} ^{2}H^4_{{-1}}+2\,{H^3_{{-2}}}^{2}{%
H^4_{{-2}}}^{2}H^3_{{1}}H^3_{{-1}}- 6\,H^3_{{-2}}H^4_{{2}}H^3_{{2}}-16\,H^3_{%
{1}}H^3_{{-2}}H^4_{{3}} +2\,{H^4_{{-2}}}^{2}H^3_{{3}}H^3_{{-1}}+7\,H^4_{{2}%
}H^4_{{-2}}{H^3_{{-1}}}^{ 2} \\
& -6\,H^4_{{2}}{H^3_{{-1}}}^{2}{H^3_{{-2}}}^{2}-2\,H^4_{{2}}H^4_{{-2}} H^3_{{%
1}}+4\,H^4_{{-2}}{H^3_{{-2}}}^{2}{H^3_{{1}}}^{2} +8\,H^4_{{-2}}{H^3_{{-2}}}%
^{2}H^3_{{5}}-16\,H^3_{{-1}}H^3_{{-2}}H^4_{{5}}-3\,H^4_{{3}} H^4_{{-1}}H^4_{{%
-2}} \\
& +8\,H^3_{{3}}H^3_{{1}}{H^3_{{-2}}}^{2}+8\,H^3_{{5} }H^3_{{-1}}{H^3_{{-2}}}%
^{2}+8\,H^3_{{3}}H^3_{{-2}}H^3_{{2}}-10\,H^3_{{\ -2}}H^4_{{4}}H^4_{{-1}%
}-2\,H^3_{{-2}}H^4_{{5}}H^4_{{-2}}-3\,{H^3_{{-2} }}^{2}H^4_{{4}}H^4_{{-2}} \\
& -7\,{H^3_{{-2}}}^{2}H^4_{{3}}H^4_{{-1}} +2\,H^3_{{2}}H^3_{{-2}}{H^4_{{-1}}}%
^{2}+2\,H^3_{{3}}H^3_{{-2}}H^4_{{1}} -4\,H^4_{{3}}H^3_{{-1}}{H^3_{{-2}}}%
^{3}+10\,H^4_{{-2}}H^3_{{3}}H^3_{{1}}+9 \,H^4_{{1}}H^4_{{-2}}H^3_{{2}} \\
& +9\,H^4_{{-1}}H^4_{{-2}}H^3_{{4}} -4\,H^4_{{-1}}{H^3_{{-2}}}^{3}H^3_{{3}%
}+6\,H^4_{{-1}}{H^3_{{-2}}}^{2}H^3_{{4} }+2\,H^4_{{4}}H^4_{{-2}}H^3_{{-1}%
}-16\,H^4_{{4}}H^3_{{-1}}{H^3_{{-2}}} ^{2}+2\,{H^3_{{-2}}}^{4}H^4_{{1}}H^4_{{%
-1}} \\
& -2\,H^3_{{-2}}{H^4_{{-1}}}^ {2}H^4_{{1}}-4\,{H^4_{{-2}}}^{2}H^3_{{-1}}H^4_{%
{2}}-5\,H^3_{{1}}H^4_{{\ 1}}H^4_{{-1}}-2\,H^4_{{3}}H^3_{{-2}}{H^4_{{-2}}}%
^{2} -5\,H^4_{{3}}H^4_{{-1}}H^3_{{-1}}-2\,H^4_{{3}}{H^3_{{-2}}}^{3}H^4_{{-2}}
\\
& -4\,H^4_{{1}}{H^3_{{-1}}}^{3}H^3_{{-2}}-4\,{H^4_{{-2}}}^{2}{H^3_{{-1}}}%
^{2}H^3_{{1}} +2\,H^4_{{-2}}{H^3_{{-1}}}^{3}H^3_{{1}}+2\,{H^3_{{-2}}}^{2}{%
H^4_{{-2}}}^ {2}H^3_{{3}}+6\,H^4_{{-1}}H^3_{{1}}H^3_{{-2}}H^4_{{-2}}H^3_{{-1}%
}\ .
\end{split}
\notag
\end{equation}
} 

\end{document}